\documentclass[aps,prd,twocolumn,superscriptaddress,nofootinbib,floatfix]{revtex4-2}

\usepackage{amsmath,amssymb,bm}
\usepackage{amsthm}
\usepackage{graphicx}
\usepackage{dcolumn}
\usepackage{booktabs}
\usepackage{xcolor}
\usepackage[colorlinks=true,linkcolor=blue!60!black,citecolor=blue!60!black,urlcolor=blue!60!black]{hyperref}

\makeatletter\@secpenalty=0\makeatother
\newtheorem{lemma}{Lemma}
\newtheorem{prop}{Proposition}
\newcommand{\Hc}{\mathcal{H}}
\newcommand{\LCDM}{$\Lambda$CDM}
\newcommand{\Pl}{P18}
\newcommand{\DK}{DES-Dovekie}
\newcommand{\DESI}{DESI DR2}
\newcommand{\kms}{\,{\rm km\,s^{-1}\,Mpc^{-1}}}

\newcommand{\NdrawsL}{3298}

\newcommand{\NdrawsF}{5966}

\newcommand{\ESSmin}{634}
\newcommand{\khatmax}{0.36}
\newcommand{\dchiA}{-4.1}
\newcommand{\lnBA}{0.72\pm0.03}
\newcommand{\lnBvA}{0.7}

\newcommand{\dchiAEH}{-6.6}
\newcommand{\lnBAEH}{1.49\pm0.04}
\newcommand{\lnBvAEH}{1.5}

\newcommand{\dchiAEHW}{-0.1}
\newcommand{\lnBAEHW}{-1.42\pm0.05}

\newcommand{\lnBcutaA}{0.63}
\newcommand{\SigcutaA}{-0.20\pm0.11}
\newcommand{\lnBcutbA}{0.41}
\newcommand{\SigcutbA}{-0.18\pm0.10}
\newcommand{\lnBcutcA}{-0.35}
\newcommand{\SigcutcA}{-0.09\pm0.09}

\newcommand{\lnBcutbAEH}{1.02}

\newcommand{\lnBcutcAEH}{-0.05}
\newcommand{\SigcutcAEH}{-0.11\pm0.07}
\newcommand{\SigA}{-0.22\pm0.11}
\newcommand{\SigLoA}{-0.39}
\newcommand{\SigHiA}{0.02}
\newcommand{\SeightA}{0.972\pm0.099}

\newcommand{\seightA}{0.953\pm0.099}
\newcommand{\OmA}{0.3119\pm0.0075}
\newcommand{\HzeroA}{67.60\pm0.55}
\newcommand{\etaA}{0.43\pm0.30}
\newcommand{\SeightLA}{0.831\pm0.013}
\newcommand{\OmLA}{0.3154\pm0.0075}
\newcommand{\HzeroLA}{67.35\pm0.54}

\newcommand{\HzeroAE}{67.33\pm0.49}

\newcommand{\HzeroAH}{68.48\pm0.29}

\newcommand{\SigAEH}{-0.24\pm0.09}
\newcommand{\SigLoAEH}{-0.37}
\newcommand{\SigHiAEH}{-0.06}
\newcommand{\SeightAEH}{0.984\pm0.096}
\newcommand{\SeffAEH}{0.868\pm0.025}
\newcommand{\seightAEH}{0.981\pm0.096}
\newcommand{\OmAEH}{0.3018\pm0.0036}
\newcommand{\HzeroAEH}{68.34\pm0.28}
\newcommand{\etaAEH}{0.33\pm0.24}
\newcommand{\SeightLAEH}{0.812\pm0.008}
\newcommand{\OmLAEH}{0.3030\pm0.0036}
\newcommand{\HzeroLAEH}{68.26\pm0.28}
\newcommand{\SigAW}{0.07\pm0.10}

\newcommand{\SigAEHW}{-0.01\pm0.07}
\newcommand{\SigLoAEHW}{-0.14}
\newcommand{\SigHiAEHW}{0.13}
\newcommand{\SeightAEHW}{0.815\pm0.022}

\newcommand{\HzeroLAEHW}{68.32\pm0.27}
\newcommand{\bfSigA}{-0.335}
\newcommand{\bfetaA}{0.11}
\newcommand{\bfseightA}{1.097}
\newcommand{\bfepsA}{-8.5}

\newcommand{\bfSigAEHW}{-0.039}

\newcommand{\dchilowTTA}{-1.56}
\newcommand{\dchilowEEA}{-0.33}
\newcommand{\dchiplikA}{-2.00}
\newcommand{\dchilensA}{-0.25}

\newcommand{\dchilensAH}{-1.24}

\newcommand{\dchidesiAH}{-0.70}

\newcommand{\dchilowTTAEH}{-1.07}
\newcommand{\dchilowEEAEH}{-1.24}
\newcommand{\dchiplikAEH}{-2.67}
\newcommand{\dchilensAEH}{-1.06}
\newcommand{\dchidovAEH}{0.37}
\newcommand{\dchidesiAEH}{-0.89}
\newcommand{\dchikidsAEH}{24.99}
\newcommand{\dchidesyAEH}{52.34}

\newcommand{\lnRLDov}{3.39}
\newcommand{\dDICLDov}{-1.2}

\newcommand{\QDMAPLDESI}{1.85}

\newcommand{\QDMLDESI}{1.91}
\newcommand{\shiftLDESI}{1.88}

\newcommand{\SLDESI}{1.87}

\newcommand{\dDICLDESI}{1.4}

\newcommand{\QDMAPLDovDESI}{2.40}

\newcommand{\shiftLDovDESI}{2.50}

\newcommand{\dDICLDovDESI}{4.0}

\newcommand{\QDMLWL}{1.61}
\newcommand{\shiftLWL}{1.63}
\newcommand{\SLWL}{1.54}

\newcommand{\SeffAL}{0.831\pm0.013}
\newcommand{\QDMAPFDov}{1.11}

\newcommand{\QDMFDov}{1.13}

\newcommand{\lnRFDov}{3.32}
\newcommand{\dDICFDov}{-0.9}
\newcommand{\BMDFDov}{1.24}
\newcommand{\QDMAPFDESI}{1.01}

\newcommand{\QDMFDESI}{1.46}

\newcommand{\lnRFDESI}{6.49}
\newcommand{\dDICFDESI}{-0.5}
\newcommand{\BMDFDESI}{1.28}
\newcommand{\QDMAPFDovDESI}{1.69}

\newcommand{\QDMFDovDESI}{2.12}

\newcommand{\dDICFDovDESI}{2.2}

\newcommand{\QDMAPFWL}{2.59}

\newcommand{\SFWL}{2.34}

\newcommand{\dDICFWL}{4.0}
\newcommand{\SeffAF}{0.876\pm0.028}
\newcommand{\SWL}{0.803\pm0.012}

\newcommand{\OmDov}{0.3308\pm0.0153}
\newcommand{\OmDESI}{0.2978\pm0.0087}
\newcommand{\hrdDESI}{101.52\pm0.74}
\newcommand{\hrdAL}{99.08\pm0.94}

\newcommand{\hrdAF}{99.51\pm0.95}
\newcommand{\OmAF}{0.3119\pm0.0075}

\newcommand{\clsptAEH}{0.841\pm0.014}
\newcommand{\clssptAEH}{1.4}
\newcommand{\clsLsptAEH}{0.6}
\newcommand{\clerassAEH}{0.866\pm0.024}
\newcommand{\clserassAEH}{0.2}
\newcommand{\clsLerassAEH}{3.7}
\newcommand{\clpszAEH}{0.879\pm0.030}
\newcommand{\clspszAEH}{2.8}
\newcommand{\clsLpszAEH}{1.5}

\newcommand{\KzD}{0.30}
\newcommand{\KzK}{0.35}
\newcommand{\NFReras}{0.92}
\newcommand{\NFRpsz}{0.90}
\newcommand{\NFRspt}{1.02}
\newcommand{\NSN}{1820}
\newcommand{\NSNdes}{1623}
\newcommand{\NSNlow}{197}
\newcommand{\SbfFA}{1.113}
\newcommand{\SbfFAEH}{1.114}

\newcommand{\SeffDFAEH}{0.872\pm0.027}

\newcommand{\SeffKFAEH}{0.868\pm0.025}
\newcommand{\SeffKFAEHW}{0.811\pm0.010}

\newcommand{\SeffgDAEH}{0.900}

\newcommand{\SeffgKAEH}{0.895}
\newcommand{\SeffnzDA}{0.917}
\newcommand{\SeffnzDAEH}{0.905}

\newcommand{\SeffnzKAEH}{0.896}

\newcommand{\SeightLAEHx}{0.812\pm0.008}

\newcommand{\SinfA}{0.920\pm0.008}
\newcommand{\SinfAEH}{0.908\pm0.009}

\newcommand{\ageMax}{13.81}
\newcommand{\ageMin}{13.76}
\newcommand{\ageShiftDESI}{0.031}
\newcommand{\binfourD}{3}
\newcommand{\binoneD}{14}
\newcommand{\binoneK}{18}
\newcommand{\binsixK}{4}

\newcommand{\chiinfAEH}{6.5}
\newcommand{\dSeras}{1.4}
\newcommand{\dSpsz}{1.4}
\newcommand{\dSspt}{0.2}

\newcommand{\iswbAEH}{0.47\pm0.21}
\newcommand{\iswbAEHW}{0.97\pm0.12}

\newcommand{\iswbAW}{1.10\pm0.15}

\newcommand{\iswdAEH}{0.62\pm0.14}

\newcommand{\iswrmsF}{15}
\newcommand{\iswrmsL}{25}
\newcommand{\kapdiff}{2}
\newcommand{\liswhi}{0.40}
\newcommand{\liswlo}{0.28}
\newcommand{\npxi}{220}
\newcommand{\nratioc}{2.0}
\newcommand{\nratiocsix}{2.8}
\newcommand{\ppA}{6}

\newcommand{\slopeeras}{6}
\newcommand{\slopepsz}{8}
\newcommand{\slopespt}{10}
\newcommand{\ttlowA}{24}
\newcommand{\ttlowAten}{7}
\newcommand{\zSNdesmin}{0.07}
\newcommand{\zSNlowmax}{0.087}
\newcommand{\zSNmax}{1.14}

\newcommand{\prSeightLP}{0.828\pm0.011}

\newcommand{\prSeightLPEH}{0.811\pm0.008}

\newcommand{\prSeightLPW}{0.816\pm0.008}

\newcommand{\prSeightLPEHW}{0.808\pm0.006}
\newcommand{\prHLPEHW}{68.13\pm0.26}

\newcommand{\prESSFP}{918}
\newcommand{\prkFP}{-0.04}
\newcommand{\prSeightFP}{0.930\pm0.092}

\newcommand{\prSigP}{-0.17\pm0.13}

\newcommand{\prPnegP}{0.89}
\newcommand{\prESSFPEH}{918}
\newcommand{\prkFPEH}{0.11}
\newcommand{\prSeightFPEH}{0.957\pm0.093}

\newcommand{\prSigPEH}{-0.22\pm0.10}

\newcommand{\prESSFPW}{414}
\newcommand{\prkFPW}{0.00}
\newcommand{\prSeightFPW}{0.804\pm0.020}

\newcommand{\prSigPW}{0.07\pm0.09}

\newcommand{\prESSFPEHW}{576}
\newcommand{\prkFPEHW}{0.02}
\newcommand{\prSeightFPEHW}{0.814\pm0.021}

\newcommand{\prSigPEHW}{-0.01\pm0.07}
\newcommand{\prSigLoPEHW}{-0.14}
\newcommand{\prSigHiPEHW}{0.14}

\newcommand{\prlnBP}{0.16\pm0.05}
\newcommand{\prlnBvP}{0.2}
\newcommand{\prlnBPEH}{0.95\pm0.05}
\newcommand{\prlnBvPEH}{0.9}
\newcommand{\prlnBPW}{-1.04\pm0.06}

\newcommand{\prlnBPEHW}{-1.49\pm0.05}
\newcommand{\prlnBvPEHW}{-1.5}
\newcommand{\prWL}{1.6}
\newcommand{\prWF}{1.9}

\begin{document}

\title{Energy exchange confined to the perturbations: the frozen dark sector confronted with Planck, DESI,
supernovae and cosmic shear}

\author{Mart\'in G. Richarte}
\thanks{\href{mailto:martin@df.uba.ar}{martin@df.uba.ar} (corresponding author)}
\affiliation{Centro de Ci\^encias Naturais e Humanas, Universidade Federal do ABC, Santo Andr\'e, SP, Brazil}
\affiliation{Departamento de F\'isica, Facultad de Ciencias Exactas y Naturales, Universidad de Buenos Aires, Buenos Aires, Argentina}
\author{Luiz Filipe Guimar\~aes}
\email{luizfilipe@ufrrj.br}
\affiliation{Departamento de F\'isica, Universidade Federal Rural do Rio de Janeiro, BR 465-07, 23890-000, Serop\'edica - RJ, Brazil}
\affiliation{Departamento de F\'isica, Universidade Estadual de Londrina, Rod. Celso Garcia Cid, Km 380, 86057-970 Londrina, PR, Brazil}
\author{Susana J. Landau}
\email{slandau@df.uba.ar}
\affiliation{CONICET--Universidad de Buenos Aires, Instituto de F\'isica de Buenos Aires (IFIBA). Buenos Aires, Argentina.}
\author{Mat\'ias Leizerovich}
\email{mleize@df.uba.ar}
\affiliation{Universidad de Buenos Aires, Facultad de Ciencias Exactas y Naturales, Departamento de F\'isica. Buenos Aires, Argentina.}
\affiliation{CONICET--Universidad de Buenos Aires, Instituto de F\'isica de Buenos Aires (IFIBA). Buenos Aires, Argentina.}

\date{\today}

\begin{abstract}
We present cosmological constraints on the frozen model, in which dark matter and dark energy exchange energy only
at the level of the perturbations. The expansion history is that of flat $\Lambda$CDM, while on sub-horizon scales
a dimensionless coupling $\Sigma$ enhances ($\Sigma<0$) or suppresses ($\Sigma>0$) the growth of the dark-matter
density contrast. We use Planck 2018 and PR4 CMB and lensing data, DESI DR2 baryon acoustic oscillations (BAO), the DES five-year
supernovae (Dovekie recalibration), and the KiDS-Legacy and DES Year 6 cosmic-shear amplitudes, corrected for the
model's growth. Planck 2018 alone gives $\Sigma=-0.22\pm0.11$, a preference driven by the lensing smoothing of
the acoustic peaks and the low large-scale temperature power. Since BAO and supernovae do not constrain $\Sigma$,
the coupling acts only through the Planck posterior, lowering the Planck--DESI tension from $1.9\sigma$ to
$1.0$--$1.5\sigma$, and that between Planck plus supernovae and DESI from $2.4$--$2.5\sigma$ to $1.7$--$2.1\sigma$.
Combining Planck, DESI and supernovae we find $\Sigma=-0.24\pm0.09$ and $\ln B=1.5$ relative to $\Lambda$CDM, a preference that vanishes when the strongest-coupling region is excluded and weakens with PR4
($\Sigma=\prSigPEH$, $\ln B=\prlnBvPEH$). The enhanced growth, however, raises the
Planck--shear tension from $1.6\sigma$ to $2.3$--$2.6\sigma$. With shear included, $\Sigma=-0.01\pm0.07$ and
$S_8=0.815\pm0.022$, while $\ln B=-1.4$ favors $\Lambda$CDM; PR4 leads to the same conclusion. $H_0$ and the age of the Universe agree with
$\Lambda$CDM for every data combination. Without shear, the
favored coupling predicts half the $\Lambda$CDM integrated Sachs--Wolfe amplitude at $z=0.5$ and an enhanced shear
signal in the lowest tomographic bins. These signatures and the redshift distribution of cluster counts test the
coupling independently of the amplitude of the fluctuations.
\end{abstract}

\maketitle

\section{Introduction}
\label{sec:intro}

The standard \LCDM\ model describes an impressive range of observations with just six parameters, yet the growing
precision of the data has brought to light a few discrepancies between early- and late-time probes. The best known
are the difference between the Hubble constant inferred from the CMB and the one measured with the local distance
ladder, and the somewhat lower amplitude of the matter fluctuations found by weak-lensing surveys; both have been
reviewed extensively~\cite{DiValentino2021,Abdalla2022,CosmoVerse2025}. Assessing such discrepancies requires
care, since different estimators answer different questions and react differently to non-Gaussian posteriors and
to prior volumes. These estimators have been compared in detail in Ref.~\cite{Leizerovich2023} and generalized to
several data sets at once in Ref.~\cite{Leizerovich2025}, and related work has examined how small-scale
inhomogeneities in light propagation~\cite{Kraiselburd2025} or energy diffusion in unimodular
gravity~\cite{Landau2023} could affect the inferred expansion rate. In the second data release of DESI, the BAO
measurements prefer a lower matter density than the CMB within \LCDM, a difference that becomes more significant when type Ia supernovae are included and that has been
read as a hint of dynamical dark energy~\cite{DESI2025DR2,DESI2025DE}. Whether late-time modifications of the
expansion can absorb these discrepancies without spoiling the BAO calibration is itself a subtle
question~\cite{Pedrotti2026}.

An interaction between dark matter and dark energy is one of the most studied extensions of \LCDM, and its history
is almost as long as that of dynamical dark energy~\cite{Wetterich1995,Amendola2000,Damour1990}. Interactions were
first proposed as a way to address the coincidence problem~\cite{Chimento2003,Chimento2006}, for instance through
exchange rates that drive the ratio of the dark densities to a stationary value~\cite{Chimento2003}, and the variety
of couplings explored since then, linear and nonlinear in the densities~\cite{Chimento2010,Chimento2011R}, is
discussed in the review of Ref.~\cite{Velten2014}. Closely related ideas include the diffusion of matter into a dark-energy
field~\cite{Calogero2013} and the decaying vacuum, whose background and structure formation, including the ISW
signal, have been confronted with data~\cite{Velten2013,Velten2015}. Interest in these models grew when it became
clear that an energy flow from dark matter to dark energy can raise the CMB value of $H_0$ and lower the amplitude
of the fluctuations at the same time~\cite{DiValentino2017,DiValentino2020IDE,DiValentino2020NM}, provided that
the perturbations are formulated in a stable way~\cite{Yang2018}, also when the dark energy itself is
dynamical~\cite{Pan2019}. These models have since been tested with
the full shape of the galaxy power spectrum~\cite{Nunes2022}, with forecasts for gravitational-wave standard
sirens~\cite{Yang2019GW} and with the DESI data~\cite{Giare2024,Silva2025,Sabogal2026}, which revealed a close
degeneracy between interacting and evolving dark energy at the background level~\cite{Petri2026}.
Model-independent reconstructions of the interaction kernel~\cite{vonMarttens2021} and forecasts for
narrow-band photometric surveys~\cite{Salzano2021,Figueruelo2021} assess how well the next generation of data
will discriminate among these possibilities. Couplings of dark energy to visible matter, on the other
hand, are being probed directly in the laboratory~\cite{Vagnozzi2021,Yuan2026}. Most of the interacting models
mentioned above exchange energy at the level of the background, which changes the expansion history and is
therefore constrained mainly by geometric probes. A fluid description must also specify how the exchange enters the
perturbations, and some prescriptions lead to large-scale instabilities~\cite{Valiviita2008,Li2014,RicharteXu2016}.

A different class of interactions leaves the background untouched and acts only on the inhomogeneities. The best
known examples transfer momentum: elastic scattering between dark matter and dark
energy~\cite{Simpson2010,BaldiSimpson2015,BaldiSimpson2017} and the pure-momentum couplings of the
Pourtsidou--Skordis--Copeland classification~\cite{Pourtsidou2013,Skordis2015,PourtsidouTram2016,Linton2018},
which have also been described in effective and Lagrangian
frameworks~\cite{KaseTsujikawa2020b,AmendolaTsujikawa2020,BeltranJimenez2021a,Aoki2025}. These couplings act on
the velocity of the dark matter and slow its infall. In the frozen model, proposed by some of us in Ref.~\cite{Richarte2025}, the exchange is
instead an energy transfer confined to the perturbations: the background densities evolve as in the
noninteracting model, while the local dark-matter density gains or loses energy in proportion to the local
density contrast of the dark energy in the frame of the total dark momentum. The dark-matter velocity is
unchanged, but the density responds to the convergence of the flow with a modified, time-dependent efficiency.
Geometric probes are blind to the coupling, redshift-space distortions are only weakly sensitive to it, and the
signal is carried by the density field: the amplitude of the matter fluctuations, the CMB lensing potential, the
integrated Sachs--Wolfe (ISW) effect~\cite{SachsWolfe1967,CrittendenTurok1996} and the abundance of collapsed
objects.

The companion analysis~\cite{Richarte2025} allowed for a dark-energy equation of state $w\neq-1$. Here we fix
$w=-1$, so that the expansion history is exactly that of \LCDM\ and every effect of the coupling is confined to the
growth of structure. We study this case for two reasons. First, since the expansion history is
that of \LCDM, any change in the agreement between the CMB, BAO and supernovae must come from how the coupling
moves the CMB posterior, which isolates the role of the growth of structure in the tensions among these data
sets. Second, the coupling acts on the observables in which the Planck 2018 data depart from \LCDM: the
preference for more lensing smoothing of the acoustic peaks than predicted by
\LCDM~\cite{Planck2018VI,Calabrese2008} and the low amplitude of the temperature spectrum at the largest scales.
A coupling that enhances the late growth of the dark-matter density can absorb both, at the price of a large
present amplitude of the fluctuations, which cosmic shear and cluster counts can test.

In this work we confront the frozen model with Planck 2018 temperature, polarization and lensing data, with the
DESI DR2 BAO measurements and with the DES five-year supernovae in their Dovekie recalibration, and we add the
cosmic-shear measurements of KiDS-Legacy and DES Year 6, compared with the model through an effective amplitude
that accounts for its growth history. Before combining the data sets we quantify their agreement in \LCDM\ and in
the frozen model with several tension estimators based on parameter differences, on the goodness of fit and
on the Bayesian evidence. All posteriors and evidences are computed with exact likelihoods, by adaptive importance
sampling for the combinations that include the CMB and by nested sampling for the external data alone, with the
same priors for every data set. We also repeat the analysis with the Planck PR4 likelihoods, in order to test how the
preference for the coupling depends on the CMB data. On the theoretical side, we formulate the model covariantly,
starting from the Einstein equations, and we show that it is gauge invariant and consistent, that its initial-value
problem is well posed, and how it differs from a vacuum that exchanges energy with the dark matter.
We then derive the joint constraints, compare the model with \LCDM, and follow the
consequences of each data combination for a set of late-time observables: the expansion rate, the Hubble constant,
the age of the Universe, the ISW amplitude, and the amplitude of the fluctuations probed by cosmic shear and
cluster counts, which we compare with independent measurements. Appendix~\ref{app:cov} gives the covariant formulation of the model and the proofs of its consistency, and two
further appendices study in detail how cosmic shear and cluster abundances respond to the coupling, with the published source distributions of the shear surveys and
a halo mass function computed from the linear spectrum of each model.

The paper is organized as follows. Section~\ref{sec:model} presents the field equations, the background and the first-order structure of the model,
its sub-horizon limit with the stability boundary, and its phenomenology, and Sec.~\ref{sec:data} the data. Section~\ref{sec:methods} describes the sampling and the
estimators. Section~\ref{sec:planck} discusses the constraints from Planck alone and the physical boundary of the
model, Sec.~\ref{sec:tensions} the agreement between the data sets, and Sec.~\ref{sec:joint} the joint analysis and
the role of cosmic shear. Section~\ref{sec:pr4} repeats the analysis with Planck PR4. Section~\ref{sec:late} compares the predictions for late-time observables with
independent measurements, and Sec.~\ref{sec:discussion} discusses redshift-space distortions, the relation with
other perturbation-only couplings, and the limitations of the analysis. We conclude in
Sec.~\ref{sec:conclusions}.

\section{The frozen model}
\label{sec:model}

\begin{figure}[tbp]
\includegraphics[width=\columnwidth]{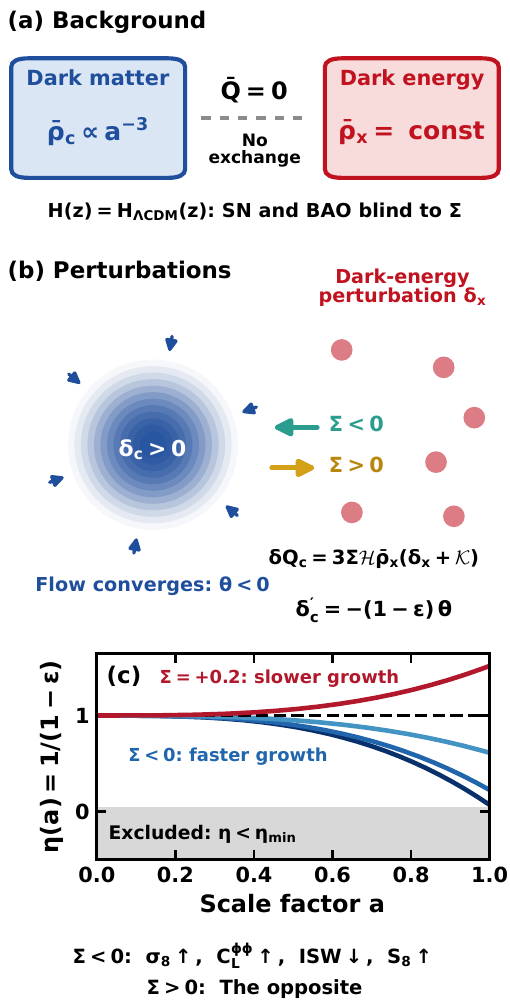}
\caption{Schematic view of the frozen model. (a) In the background the dark matter dilutes as $a^{-3}$ and the
energy density of the dark-energy fluid with $w=-1$ stays constant, with no exchange between them ($\bar Q=0$);
the expansion history is that of \LCDM\ and the geometric probes do not see the coupling. (b) In the perturbations,
energy is exchanged between the dark-matter overdensities and the dark-energy perturbation at the rate
$\delta Q_{\rm c}$ of Eq.~(\ref{eq:dQ}); the velocity field is unchanged, but the dark-matter density responds to the
convergence of the flow with the factor $1-\epsilon$ of Eq.~(\ref{eq:eps}). For $\Sigma<0$ the energy flows into the
overdensities. (c) The factor $\eta=1/(1-\epsilon)$ of Eq.~(\ref{eq:eta}) as a function of the scale factor for
$\Sigma=+0.2$, $0$, $-0.15$, $-0.30$ and $-0.36$: negative couplings accelerate the growth at late times and
approach the stability boundary $\eta\to0$ (shaded region, $\eta<\eta_{\rm min}$). The bottom line summarizes the
observational consequences.}
\label{fig:cartoon}
\end{figure}

The essential idea of the model is summarized in Fig.~\ref{fig:cartoon}. The two dark components do not exchange
energy on average, so the expansion history is that of the noninteracting model; energy flows only between the
local perturbations, from the perturbation of the dark-energy fluid into the dark-matter overdensities or in the
opposite direction, depending on the sign of the coupling. We present the model in four steps: the field equations
and the general form of an energy-momentum exchange, the background, the structure of the first-order
perturbations, and the equations in the gauge used for the numerical solution. The proofs of the statements made
along the way are collected in Appendix~\ref{app:cov}. The construction follows the companion
analysis~\cite{Richarte2025}, with the two refinements described in the footnotes below.

\subsection{Field equations and energy-momentum exchange}
\label{sec:field}

We begin with the field equations, because they determine what a consistent interaction between two components
can be. The physical spacetime is a manifold $M$ with metric $g_{\mu\nu}$ and Levi-Civita connection $\nabla_\mu$.
In this subsection every tensor is a field on $(M,g)$, and indices are raised and lowered with $g_{\mu\nu}$. The
metric obeys the Einstein equations,
\begin{equation}
\mathcal{G}_{\mu\nu}=8\pi G_{\rm N}\sum_{A}T^{(A)}_{\mu\nu},
\label{eq:einstein}
\end{equation}
where $\mathcal{G}_{\mu\nu}$ is the Einstein tensor of $g_{\mu\nu}$, $G_{\rm N}$ is Newton's constant, and $A$ runs over
photons, neutrinos, baryons, cold dark matter (DM, subscript c) and a dark-energy component (subscript x). The
contracted Bianchi identity, $\nabla^\mu\mathcal{G}_{\mu\nu}=0$, holds for every metric. Equation~(\ref{eq:einstein})
therefore requires the conservation of the total energy-momentum tensor, and only of the total. We let the two dark
components exchange energy and momentum,
\begin{equation}
\nabla_\nu T_{\rm c}^{\mu\nu}=Q^\mu,\qquad \nabla_\nu T_{\rm x}^{\mu\nu}=-Q^\mu,
\label{eq:exchange}
\end{equation}
while the other components obey their standard equations.

Let $u^\mu$ be the four-velocity of the DM, $g_{\mu\nu}u^\mu u^\nu=-1$, and
$\gamma_{\mu\nu}=g_{\mu\nu}+u_\mu u_\nu$ the projector onto the rest spaces of the DM observers. The DM is dust, and
we decompose the tensor of the component x with respect to $u^\mu$,
\begin{align}
T_{\rm c}^{\mu\nu}&=\rho_{\rm c}\,u^\mu u^\nu,\nonumber\\
T_{\rm x}^{\mu\nu}&=\rho_{\rm x}\,u^\mu u^\nu+p_{\rm x}\,\gamma^{\mu\nu}+J^\mu u^\nu+u^\mu J^\nu,
\label{eq:Tdecomp}
\end{align}
with $J^\mu u_\mu=0$. Here $\rho_{\rm x}$, $p_{\rm x}$ and $J^\mu$ are the energy density, the pressure and the energy
flux of the component x measured by the DM observers; we assume that it has no anisotropic stress. The exchange
four-vector is decomposed in the same way,
\begin{equation}
Q^\mu=Q\,u^\mu+F^\mu,\qquad F^\mu u_\mu=0,
\label{eq:Qsplit}
\end{equation}
where $Q$ is the rate at which the DM gains energy in its rest frame and $F^\mu$ is the momentum transfer. Projecting
the first of Eqs.~(\ref{eq:exchange}) along $u^\mu$ and orthogonally to it gives the exact relations
\begin{equation}
\dot\rho_{\rm c}+\Theta\,\rho_{\rm c}=Q,\qquad \rho_{\rm c}\,\dot u^\mu=F^\mu,
\label{eq:dmexact}
\end{equation}
where an overdot is the derivative $u^\nu\nabla_\nu$ and $\Theta=\nabla_\mu u^\mu$ is the expansion of the DM
congruence. No approximation has been made so far.

\subsection{Background}
\label{sec:background}

We now specify the solution about which we perturb. The background is a second spacetime, $(\bar M,\bar g)$, with
the spatially flat Friedmann metric $\bar g_{\mu\nu}dx^\mu dx^\nu=a^2(\tau)(-d\tau^2+\delta_{ij}dx^idx^j)$ and its own
Levi-Civita connection $\bar\nabla_\mu$. A bar denotes a quantity of the background, a prime is a derivative with
respect to the conformal time $\tau$, and $\Hc=a'/a$. Homogeneity and isotropy impose $\bar J^\mu=\bar F^\mu=0$ and
$\bar u^\mu=a^{-1}\delta^\mu{}_0$, and they leave the function $\bar Q(\tau)$ free. The frozen model is defined by
the choice
\begin{equation}
\bar Q=0 .
\label{eq:Qbar}
\end{equation}
With a constant equation of state, $\bar p_{\rm x}=w\bar\rho_{\rm x}$, Eqs.~(\ref{eq:exchange}) then reduce to
\begin{equation}
\bar\rho_{\rm c}'+3\Hc\bar\rho_{\rm c}=0,\qquad
\bar\rho_{\rm x}'+3\Hc(1+w)\bar\rho_{\rm x}=0,
\label{eq:bg}
\end{equation}
and Eq.~(\ref{eq:einstein}) to the Friedmann equation, $\Hc^2=(8\pi G_{\rm N}/3)\,a^2\sum_A\bar\rho_A$. The expansion
history is therefore that of the noninteracting model with the same $w$; in this work we take $w=-1$, for which it is
the expansion history of \LCDM\ and all geometric probes are blind to the interaction~\cite{Richarte2025}. Whereas
Ref.~\cite{Richarte2025} kept $w$ as a free parameter, fixing $w=-1$ isolates the effect of the coupling on the growth
of structure; the general case is discussed in Sec.~\ref{sec:discussion}. We denote by
\begin{equation}
r(a)\equiv\frac{\bar\rho_{\rm x}}{\bar\rho_{\rm c}}=r_0\,a^{-3w},\qquad r_0=\frac{\Omega_{\rm x}}{\Omega_{\rm c}},
\label{eq:r}
\end{equation}
the ratio of the two dark densities, which grows as $a^3$ for $w=-1$ and whose present value $r_0$ is about $2.6$. For $w=-1$ the
density $\bar\rho_{\rm x}\equiv\rho_\Lambda$ is a constant and the background tensor of the component is
\begin{equation}
\bar T_{\rm x}{}^\mu{}_\nu=-\rho_\Lambda\,\delta^\mu{}_\nu .
\label{eq:Tbar}
\end{equation}
Equation~(\ref{eq:Tbar}) is all that the choice $w=-1$ states. It is a property of the background tensor, and the
next subsection examines what it implies for the perturbations.

\subsection{Perturbations: What does the background fix and what does it leave free?}
\label{sec:pertstruct}

\begin{figure}[tbp]
\includegraphics[width=\columnwidth]{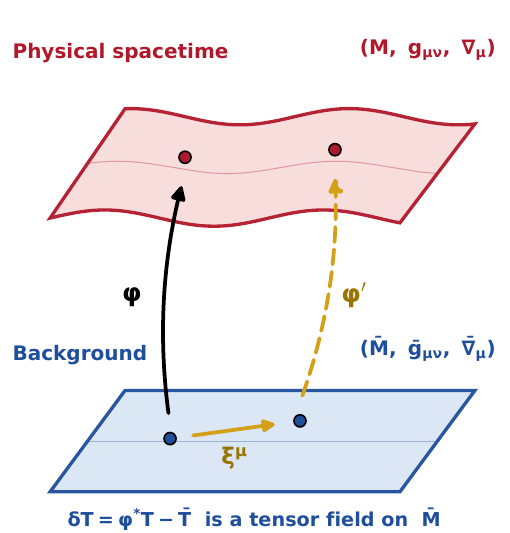}
\caption{The two spacetimes of perturbation theory. The exact equations of Sec.~\ref{sec:field} hold on the physical
spacetime $(M,g_{\mu\nu})$, with covariant derivative $\nabla_\mu$. The background $(\bar M,\bar g_{\mu\nu})$ is the
Friedmann solution of Sec.~\ref{sec:background}, with covariant derivative $\bar\nabla_\mu$. A map $\varphi$
identifies the points of the two manifolds, and the perturbation $\delta T=\varphi^*T-\bar T$ of any tensor is a
field on $\bar M$: first-order equations contain $\bar g_{\mu\nu}$ and $\bar\nabla_\mu$ only. A second map
$\varphi'$ differs from $\varphi$ by a diffeomorphism of $\bar M$ generated by $\xi^\mu$, which changes $\delta T$ by
the Lie derivative $\pounds_\xi\bar T$ (a gauge transformation).}
\label{fig:spacetimes}
\end{figure}

Perturbation theory compares the physical spacetime with the background, as sketched in
Fig.~\ref{fig:spacetimes}. Given a one-parameter family of solutions
and a diffeomorphism $\varphi:\bar M\to M$ that identifies their points, the perturbation of a tensor field $T$ is
$\delta T=\varphi^*T-\bar T$ to first order in the parameter~\cite{StewartWalker1974,Wald1984}. Thus $\delta T$ is a
tensor field on the background manifold $\bar M$: in every first-order equation indices are moved with
$\bar g_{\mu\nu}$ and the covariant derivative is $\bar\nabla_\mu$. A change of the identification map, generated by
a vector field $\xi^\mu$ on $\bar M$, is a gauge transformation,
\begin{equation}
\delta T\ \longrightarrow\ \delta T+\pounds_\xi\bar T ,
\label{eq:gauge}
\end{equation}
where $\pounds_\xi$ is the Lie derivative. By the Stewart--Walker lemma~\cite{StewartWalker1974}, $\delta T$ is gauge
invariant if and only if $\bar T$ vanishes, is a constant scalar, or is a combination with constant coefficients of
Kronecker deltas.

Applied to Eqs.~(\ref{eq:Qbar}) and (\ref{eq:Tbar}), the lemma has three consequences, proved in
Appendix~\ref{app:cov}. First, $\delta Q^\mu$ and $\delta T_{\rm x}{}^\mu{}_\nu$ are gauge invariant: no choice of
coordinates can remove them or generate them, and covariance places no condition on their values. Second, the
first-order form of the exchange law for the component x contains no metric perturbation,
\begin{equation}
\bar\nabla_\mu\,\delta T_{\rm x}{}^\mu{}_\nu=-\delta Q_\nu .
\label{eq:L}
\end{equation}
Third, $\delta T_{\rm x}{}^\mu{}_\nu$ is an arbitrary symmetric tensor on $\bar M$,
\begin{equation}
\delta T_{\rm x}{}^\mu{}_\nu=\delta\rho_{\rm x}\,\bar u^\mu\bar u_\nu+\delta p_{\rm x}\,\bar\gamma^\mu{}_\nu
+J^\mu\bar u_\nu+\bar u^\mu J_\nu ,
\label{eq:dTx}
\end{equation}
whose components do not depend on the timelike frame used to define them.

These statements fix the logical status of the choice $w=-1$. Equation~(\ref{eq:Tbar}) leaves $\delta\rho_{\rm x}$,
$\delta p_{\rm x}$ and $J^\mu$ free. A vacuum term, $T_{\rm x}^{\mu\nu}=-Vg^{\mu\nu}$, is the particular case
\begin{equation}
\delta p_{\rm x}=-\delta\rho_{\rm x},\qquad J^\mu=0 ,
\label{eq:vacuum}
\end{equation}
for which Eq.~(\ref{eq:L}) gives $\delta Q_\nu=\partial_\nu\delta V$: the exchange derives from a single scalar,
and for geodesic DM it vanishes at first order whenever $\bar Q=0$~\cite{Wands2012,Martinelli2019}. A
perturbation-only exchange is thus impossible for a vacuum coupled to geodesic DM (Appendix~\ref{app:cov}).
Conditions (\ref{eq:vacuum}) are, however, hypotheses about the perturbations that the equation of state does not
imply, and the frozen model does not make them. Its component x is a dark-energy fluid with $w=-1$ on the
background, whose first-order pressure perturbation is not minus its density perturbation and which carries an
energy flux. Both are generated by the interaction and vanish with it, so that for $\Sigma=0$ the component is the
cosmological constant.

\begin{figure*}[tbp]
\includegraphics[width=0.86\textwidth]{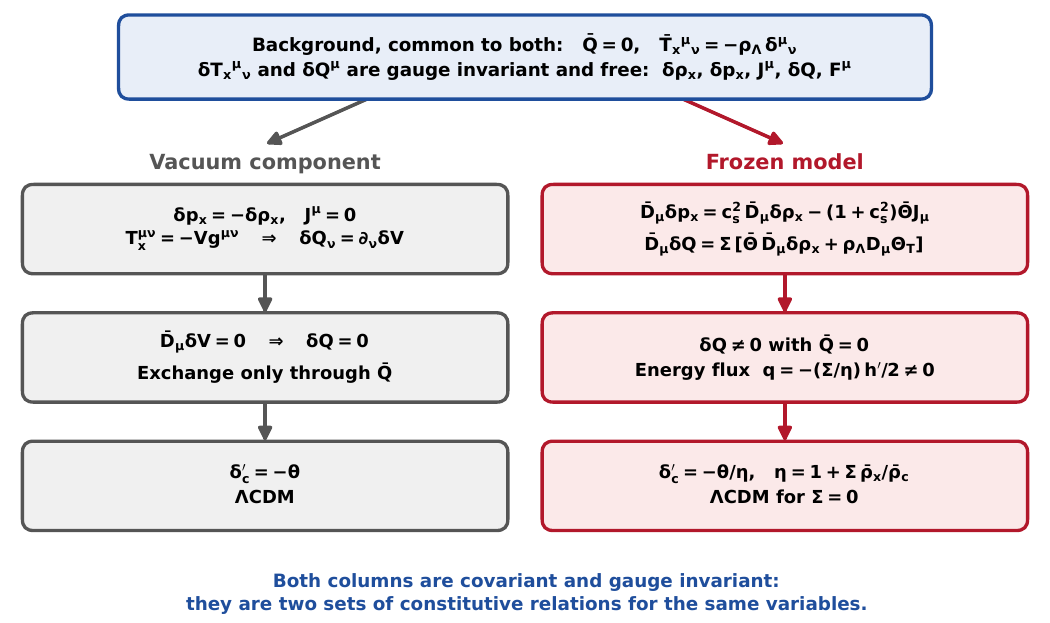}
\caption{The same background with two first-order closures. Both start from $\bar Q=0$ and
$\bar T_{\rm x}{}^\mu{}_\nu=-\rho_\Lambda\delta^\mu{}_\nu$, which leave the perturbations of the component free. Left:
a vacuum component, $T_{\rm x}^{\mu\nu}=-Vg^{\mu\nu}$, has $\delta p_{\rm x}=-\delta\rho_{\rm x}$ and no energy flux; its
exchange derives from the single scalar $\delta V$, and for geodesic dark matter ($F^\mu=0$) it vanishes at first
order, so the growth of structure is that of \LCDM. Right: the frozen model closes the pressure perturbation with a
sound speed [Eq.~(\ref{eq:closurecov})] and specifies the energy transfer by Eq.~(\ref{eq:exchcov}); the component
carries an energy flux proportional to the coupling [Eq.~(\ref{eq:qqs})], the exchange is nonzero with $\bar Q=0$,
and the growth of the dark-matter density is modified by the factor $1/\eta$ [Eq.~(\ref{eq:eps})]. The two columns
are different sets of constitutive relations for the same gauge-invariant variables, and they coincide for
$\Sigma=0$.}
\label{fig:flow}
\end{figure*}

Figure~\ref{fig:flow} summarizes the comparison between the two closures, which is completed below with the
definition of the frozen model.

Projecting Eq.~(\ref{eq:L}) along $\bar u^\mu$ and orthogonally to it, we obtain
\begin{align}
\delta\dot\rho_{\rm x}+\bar\Theta\,(\delta\rho_{\rm x}+\delta p_{\rm x})+\bar D_\mu J^\mu&=-\delta Q,
\label{eq:encov}\\
\dot J_\mu+\tfrac43\bar\Theta\,J_\mu+\bar D_\mu\delta p_{\rm x}&=-F_\mu ,
\label{eq:momcov}
\end{align}
where, acting on first-order quantities, the overdot is $\bar u^\nu\bar\nabla_\nu$, the spatial derivative is
$\bar D_\mu=\bar\gamma_\mu{}^\nu\bar\nabla_\nu$, and $\bar\Theta=3\Hc/a$. The five gauge-invariant scalar functions
contained in $\delta\rho_{\rm x}$, $\delta p_{\rm x}$, $J^\mu$, $\delta Q$ and $F^\mu$ are related by these two
equations only. The remaining freedom must be fixed by constitutive relations, as in any multicomponent system: the
Einstein equations constrain the total energy-momentum and leave the intrinsic stress and transfer perturbations of
each component to its physics~\cite{KodamaSasaki1984,MalikWands2009}. The frozen model is defined by three such
relations. There is no momentum transfer in the DM frame,
\begin{equation}
F^\mu=0 ,
\label{eq:F0}
\end{equation}
so that the DM remains geodesic by Eq.~(\ref{eq:dmexact}). The pressure perturbation is closed with a rest-frame
sound speed $c_s$,
\begin{equation}
\bar D_\mu\delta p_{\rm x}=c_s^2\,\bar D_\mu\delta\rho_{\rm x}-(1+c_s^2)\,\bar\Theta\,J_\mu ,
\label{eq:closurecov}
\end{equation}
which is the limit $w\to-1$ of the usual closure and, written with $J_\mu$, does not refer to a rest frame of the
component. Finally, the energy transfer is
\begin{align}
\bar D_\mu\delta Q&=\Sigma\left[\bar\Theta\,\bar D_\mu\delta\rho_{\rm x}+\rho_\Lambda\,D_\mu\Theta_T\right],\nonumber\\
\Theta_T&\equiv\Theta+\frac{\bar D_\nu J^\nu}{\bar\rho_{\rm c}} ,
\label{eq:exchcov}
\end{align}
where $\Sigma$ is a dimensionless constant, $\Theta_T$ is the expansion of the frame of the total dark momentum, and
$D_\mu\Theta=\gamma_\mu{}^\nu\nabla_\nu\Theta$ is the gradient of the expansion of the physical DM congruence
projected onto its rest spaces; it vanishes on the background and is therefore a gauge-invariant first-order
vector~\cite{EllisBruni1989}. Equation~(\ref{eq:exchcov}) is the first-order part of the covariant coupling
$Q\propto\rho_{\rm x}\Theta$ that renders the interactions $Q\propto H\rho$ gauge invariant~\cite{Gavela2010}, with
vanishing homogeneous part.

All three relations are tensorial and are built from quantities that vanish on the background, so the
model is covariant and gauge invariant by construction. It is also consistent: Appendix~\ref{app:cov} shows that,
for any Friedmann background and any first-order closure that conserves the total energy-momentum, the linearized
Einstein equations preserve their constraints. A background therefore does not select its perturbative closure.
The frozen closure was studied on backgrounds with $w\neq-1$ in Ref.~\cite{Richarte2025} and is studied on the
\LCDM\ background here, and on this single background the coupling $\Sigma$ labels a one-parameter family of
consistent first-order theories, of which $\Sigma=0$ is \LCDM\ itself.

\subsection{Equations in the synchronous gauge}
\label{sec:sync}

For the numerical solution we write the equations for a Fourier mode of wave number $k$ in the synchronous gauge
comoving with the DM~\cite{Ma1995}, in which the scalar metric perturbation enters the matter equations through its
trace $h$. We restore a general constant $w$, which makes the connection with Ref.~\cite{Richarte2025}. The density
contrast of the fluid is $\delta_{\rm x}=\delta\rho_{\rm x}/\bar\rho_{\rm x}$, and we define the scalar $q$ by
\begin{equation}
\bar\rho_{\rm x}\,q\equiv ik^j\,\delta T_{\rm x}{}^0{}_j ,
\label{eq:qdef}
\end{equation}
so that $\bar\rho_{\rm x}q=a\,\bar D_\mu J^\mu$ is the divergence of the energy flux. The variable $q$ is a component
of the energy-momentum tensor. For $w\neq-1$ one may attribute to the component a velocity divergence
$\theta_{\rm x}\equiv q/(1+w)$. At $w=-1$ the background tensor (\ref{eq:Tbar}) selects no rest frame and
$\theta_{\rm x}$ is not defined, while $q$ remains finite~\cite{KunzSapone2006}; $\theta_{\rm x}$ enters none of the
equations below. In this gauge $\Theta_T=(3\Hc/a)(1+\mathcal{K})$, and Eq.~(\ref{eq:exchcov}) becomes the energy
transfer per unit conformal time,
\begin{equation}
\delta Q_{\rm c}=-\delta Q_{\rm x}=3\Sigma\,\Hc\,\bar\rho_{\rm x}\left(\delta_{\rm x}+\mathcal{K}\right),\qquad
\mathcal{K}=\frac{h'/2+\theta_T}{3\Hc},
\label{eq:dQ}
\end{equation}
where $\theta_T$ is the velocity divergence of the frame of the total dark momentum,
\begin{equation}
\theta_T=\frac{\bar\rho_{\rm x}\,q}{\bar\rho_{\rm c}+(1+w)\bar\rho_{\rm x}}
=\frac{q}{1+w+r^{-1}} .
\label{eq:thetaT}
\end{equation}
The combination $\delta_{\rm x}+\mathcal{K}$ is the relative perturbation of $\rho_{\rm x}\Theta_T$\footnote{The
earlier implementation of the model used the fluid velocity divergence rescaled by the factor $1+R_I$ of
Ref.~\cite{Richarte2025} in $\mathcal{K}$; the
momentum-weighted velocity of Eq.~(\ref{eq:thetaT}) is the natural frame-independent choice and is adopted here
throughout.}. Since the density ratio $r$ is finite and positive, $\theta_T$ is regular at $w=-1$, where it equals
$r\,q$. The DM equations are
\begin{align}
\delta_{\rm c}'&=-\theta_{\rm c}-\frac{h'}{2}+3\Sigma\,\Hc\,r\left(\delta_{\rm x}+\mathcal{K}\right),
\label{eq:dc}\\
\theta_{\rm c}'&=-\Hc\,\theta_{\rm c},
\label{eq:thc}
\end{align}
and, since the DM is initially at rest in this gauge, $\theta_{\rm c}=0$ at all times, in agreement with
Eq.~(\ref{eq:F0}). The fluid, with rest-frame sound speed $c_s$ and adiabatic sound speed $c_a^2=w$, obeys
\begin{align}
\delta_{\rm x}'&=-q-(1+w)\frac{h'}{2}-3(c_s^2-w)\Hc\,\delta_{\rm x}-9(c_s^2-c_a^2)\Hc^2\frac{q}{k^2}
\nonumber\\
&\quad-3\Sigma\,\Hc\left(\delta_{\rm x}+\mathcal{K}\right),
\label{eq:dx}\\
q'&=-(1-3c_s^2)\Hc\,q+c_s^2k^2\delta_{\rm x},
\label{eq:q}
\end{align}
which at $w=-1$ are Eqs.~(\ref{eq:encov}) and (\ref{eq:momcov}) with the closure (\ref{eq:closurecov}). We set
$c_s^2=1$. The system is regular at $w=-1$: written in the variables $(\delta_{\rm x},q)$, its coefficients are
analytic functions of $w$ there, and so is its solution (Appendix~\ref{app:cov}). The baryons, photons and neutrinos
follow the standard equations, and the Einstein equations are unchanged: the total density and momentum
perturbations source the metric as usual.

\paragraph{Initial conditions.} Deep in the radiation era, the adiabatic growing mode has
$\delta_{\rm x}=D_{\rm x}h$ with
\begin{equation}
D_{\rm x}=-\frac{(4-3c_s^2)(1+w+\Sigma)}{8+6c_s^2-12w+3\Sigma(4-3c_s^2)},
\label{eq:ic}
\end{equation}
and $q=c_s^2D_{\rm x}h\,k(k\tau)/(4-3c_s^2)$, with $h=-\tfrac32\delta_\gamma$ in that regime\footnote{The numerator
$1+w+\Sigma$ of Eq.~(\ref{eq:ic}) corrects a sign in the corresponding expression of Ref.~\cite{Richarte2025};
it follows from the requirement that the fluid perturbation be a power-law solution of
Eqs.~(\ref{eq:dx})--(\ref{eq:q}) with the transfer of Eq.~(\ref{eq:dQ}).}. For $w=-1$ and $\Sigma=0$ the fluid
is unperturbed and the model reduces exactly to \LCDM.

\subsection{Sub-horizon limit, propagation speed and stability boundary}
\label{sec:qs}

On sub-horizon scales the fluid with $c_s^2=1$ does not cluster: its density contrast is suppressed by
$(\Hc/k)^2$, and Eq.~(\ref{eq:dx}) becomes a constraint for its energy flux,
$q\simeq-(1+w)h'/2-3\Sigma\Hc\mathcal{K}$. With $\theta_T=rq/[1+(1+w)r]$ this gives
\begin{equation}
3\Hc\,\mathcal{K}=\frac{1}{1+(1+w+\Sigma)\,r}\,\frac{h'}{2}.
\end{equation}
Inserting this into Eq.~(\ref{eq:dc}) and using that on these scales the Newtonian-gauge velocity divergence of
the DM is $\theta=h'/2$, we find the continuity equation
\begin{equation}
\delta_{\rm c}'=-\left[1-\epsilon(a)\right]\theta,\qquad
\epsilon(a)=\frac{\Sigma\,r}{1+(1+w+\Sigma)\,r},
\label{eq:eps}
\end{equation}
while the baryons obey $\delta_{\rm b}'=-\theta$ and both components fall in the same gravitational
potential.
The DM density therefore responds to the convergence of the flow with the factor $1-\epsilon=1/\eta(a)$, where we define the regularity parameter
\begin{equation}
\eta(a)\equiv1+(1+w+\Sigma)\,r(a)=1+\Sigma r_0a^3\quad(w=-1).
\label{eq:eta}
\end{equation}
For $\Sigma>0$ the exchange damps the growth of the DM density; for $\Sigma<0$ it enhances it, and
$\epsilon\to-\infty$ when $\eta\to0$. At $w=-1$ the energy flux in this limit is
\begin{equation}
q=-\frac{\Sigma}{\eta}\,\frac{h'}{2},
\label{eq:qqs}
\end{equation}
which is finite, proportional to the coupling, and nonzero whenever the metric evolves. An unperturbed component,
$\delta_{\rm x}=q=0$, is not a solution for $\Sigma\neq0$: by Eqs.~(\ref{eq:dx}) and (\ref{eq:dQ}) it would require
$h'=0$, which the Einstein equations allow only in the absence of density perturbations
(Appendix~\ref{app:cov}).

The parameter $\eta$ has a direct physical meaning. Combining Eqs.~(\ref{eq:dx}) and (\ref{eq:q}), the term $k^2q$
appears with the coefficient $c_s^2\eta$, so the perturbations of the component x propagate in the DM frame with
the speed
\begin{equation}
c_{\rm eff}^2=c_s^2\,\eta .
\label{eq:ceff}
\end{equation}
For $\eta>0$ the system is hyperbolic and the initial-value problem is well posed; Eq.~(\ref{eq:qqs}) is its
quasi-static solution. For $\eta<0$ the system has a gradient instability with rate $k\sqrt{c_s^2|\eta|}$. The
condition $\eta>0$ is therefore the stability boundary of the model. For constant $\Sigma<0$, $\eta$ decreases
monotonically and vanishes at
\begin{equation}
a_*=(-\Sigma\,r_0)^{-1/3}=(1-\eta_0)^{-1/3},
\label{eq:astar}
\end{equation}
where $\eta_0\equiv\eta(1)$; the evolution is stable up to today only if $\eta_0>0$, that is,
$\Sigma>-1/r_0$, about $-0.38$, and then $a_*>1$. The initial-value problem is well posed on $0<a<a_*$, an interval
that contains the whole past light cone of the observer, so every prediction that can be compared with observations
is determined by the solution on $a\leq1$. A constant negative coupling is thus an effective description of the
observed epoch, which is not meant to be extrapolated beyond $a_*$; in a complete model the product $\Sigma\,r$ is expected to
saturate. In this respect it is similar to other effective parametrizations whose extrapolation is
singular at a finite future time, such as a constant $w<-1$~\cite{Caldwell2003}. The closer $\eta_0$ is to zero, the
sooner the description ceases to be valid: $\eta_0=0.05$, $0.1$, $0.2$ and $0.5$ correspond to $a_*=1.017$, $1.036$,
$1.077$ and $1.26$. For $\Sigma>0$, on the other hand, $\eta>1$ at all times: no critical scale factor exists, the system is hyperbolic
at all times and the response of the dark matter is bounded. With $c_s^2=1$ the speed (\ref{eq:ceff}) then exceeds
unity, and subluminal propagation requires $c_s^2\leq1/\eta_0$, a choice to which the observables are insensitive.
The two signs of the coupling are compared in Appendix~\ref{app:signs}, which also gives the posterior
distribution of $a_*$. Following the convention of the companion analyses, we require
$\eta_0\ge\eta_{\rm min}=0.05$ ($\epsilon_0\ge-19$) and treat the excluded region as a region of vanishing
likelihood; we show in Sec.~\ref{sec:planck} how the results depend on $\eta_{\rm min}$.

In this limit the DM mass is not conserved: the exchange converts dark energy into DM where matter
converges ($\Sigma<0$) or the opposite ($\Sigma>0$). The velocity field is unchanged, so redshift-space
distortions, which measure $\theta$, are much less sensitive to $\Sigma$ than the density
field~\cite{Richarte2025}; quantities that respond to the density, such as $\sigma_8$, the lensing potential, the
integrated Sachs--Wolfe (ISW) effect and the abundance of halos and voids, carry the signal.

\subsection{Phenomenology}

\begin{figure}[tbp]
\includegraphics[width=\columnwidth]{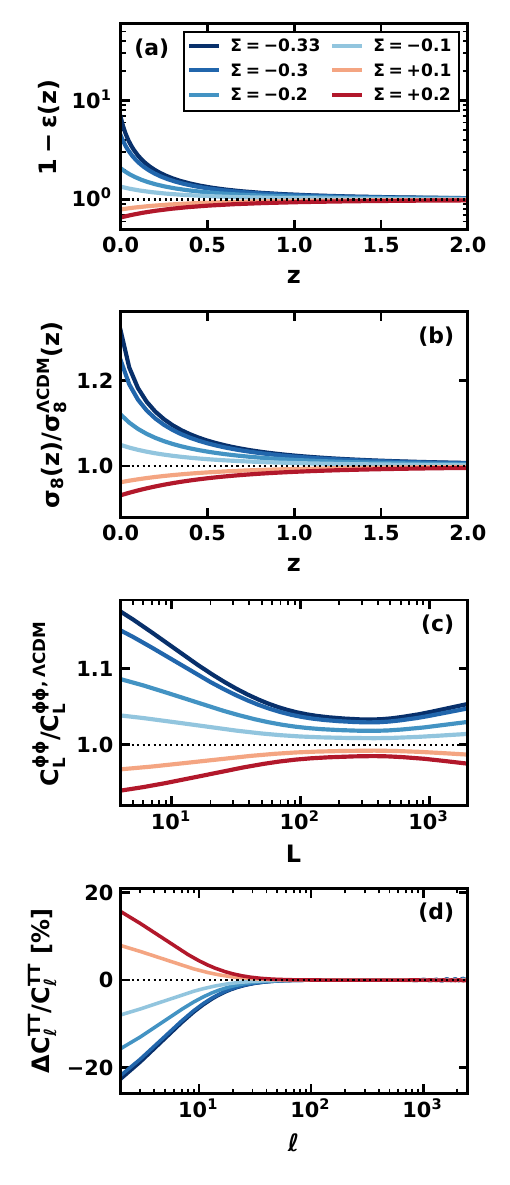}
\caption{Effect of the coupling at fixed $H_0$, $\omega_{\rm b}$, $\omega_{\rm c}$, $A_{\rm s}$, $n_{\rm s}$ and
$\tau$. (a) Factor $1-\epsilon$ that multiplies the DM velocity divergence in the continuity equation,
Eq.~(\ref{eq:eps}). (b) Ratio of $\sigma_8(z)$ to its \LCDM\ value. (c) Ratio of the CMB lensing potential power
spectrum. (d) Relative change of the lensed temperature power spectrum, dominated at low multipoles by the late
ISW effect. For $\Sigma=-0.33$ the model is at $\eta_0=0.15$.}
\label{fig:pheno}
\end{figure}

Figure~\ref{fig:pheno} illustrates the effect of $\Sigma$ at fixed values of the other parameters. Because
$\epsilon\propto a^3$ at early times, the modification is confined to $z\lesssim1$, where it changes
$\sigma_8(z)$ by up to $30\%$ for $\Sigma$ close to the boundary (panel b). The enhanced growth for $\Sigma<0$
raises the CMB lensing potential by $2$--$6\%$ on the scales of the Planck lensing reconstruction (panel c),
which also increases the lensing smoothing of the acoustic peaks. At the same time it slows the decay of the
gravitational potentials at late times, and the late ISW contribution to the large-angle temperature anisotropy
drops by up to $20\%$ at $\ell\lesssim10$ (panel d); for $\Sigma>0$ both effects have the opposite sign.

\section{Data}
\label{sec:data}

We use four groups of observations. The CMB and the two geometric probes enter all the fits, cosmic shear enters
the combinations labeled ``shear,'' and the cluster abundances are used only as a consistency check. Because the
frozen model and \LCDM\ share the same expansion history, we show each data set, in the form in which it enters the
likelihood, together with the best fits of both models; the figures of this section thus show where the coupling
acts and where it is invisible.

\subsection{Planck 2018 (\Pl)}

We use the Planck 2018 low-$\ell$ temperature likelihood ($2\le\ell<30$), the low-$\ell$ $E$-mode polarization
likelihood, the high-$\ell$ TT, TE and EE likelihood in its foreground-marginalized compressed form, and the
CMB lensing reconstruction in its conservative range of lensing multipoles,
$8\le L\le400$~\cite{Planck2018V,Planck2018VIII,Planck2018VI}; as usual, we denote the multipoles of the lensing
potential by $L$ and those of the CMB spectra by $\ell$. The
overall calibration $A_{\rm P}$ is sampled with a Gaussian prior $1.0000\pm0.0025$. This combination constrains
$\Sigma$ through the three effects of Fig.~\ref{fig:pheno}: the late ISW contribution at low multipoles, the
lensing smoothing of the acoustic peaks, and the amplitude of the reconstructed lensing potential.

\begin{figure}[tbp]
\includegraphics[width=\columnwidth]{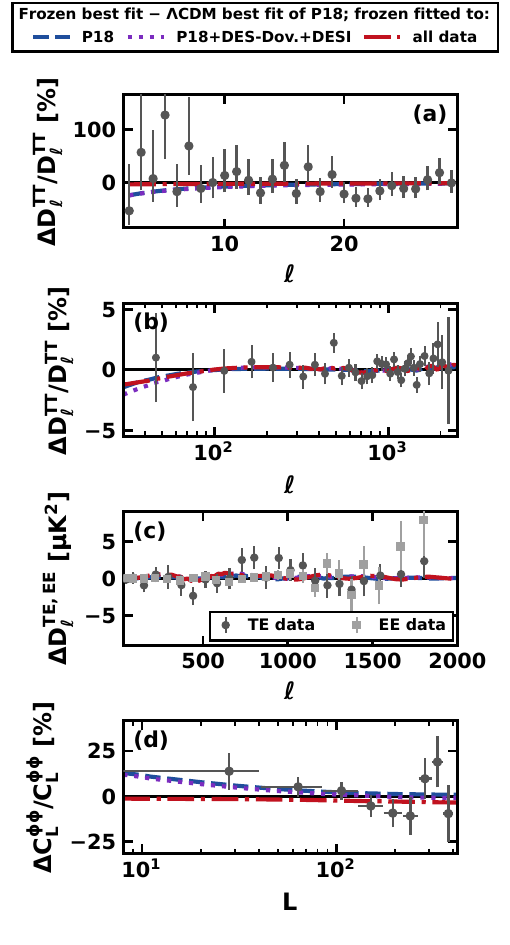}
\caption{Planck 2018 data used in this work, shown as differences with respect to the \LCDM\ best fit of \Pl\
(horizontal line). The curves are the differences between the best fits of the frozen model to \Pl\ (dashed blue),
to \Pl+\DK+\DESI\ (dotted purple) and to all data (dash-dotted red) and that \LCDM\ best fit; they are relative
differences in panels (a), (b) and (d) and absolute differences in panel (c). (a) Low-$\ell$ temperature spectrum:
maximum-likelihood values of the individual multipoles with their $68\%$ intervals (the upper intervals at $\ell=3$
and $5$ extend beyond the panel). (b) High-$\ell$ TT bandpowers, rebinned by six. (c) TE and EE bandpowers, rebinned
by eight; the curves show TE, since the EE differences are below $0.1\,\mu{\rm K}^2$. (d) Lensing-potential
bandpowers in the range $8\le L\le400$, with the horizontal bars marking the bin widths. The two fits without shear
nearly coincide.}
\label{fig:planckdata}
\end{figure}

Figure~\ref{fig:planckdata} shows these data together with the best fits of \LCDM\ and of the frozen model. The
low-$\ell$ points are the maximum-likelihood values of the Gibbs-sampled temperature spectrum~\cite{Planck2018V}, and the
high-$\ell$ points are the foreground-marginalized bandpowers, to which the models are compared after the same
binning that enters the likelihood. At the best fit of \Pl\ the coupling lowers the temperature power by
$\ttlowA\%$ at $\ell=2$ and by $\ttlowAten\%$ at $\ell=10$, the multipoles where the observed spectrum lies below
the \LCDM\ prediction (panel a), while the high-$\ell$ spectra change by less than the size of the error bars (panel b) and the polarization spectra by less than $1\,\mu{\rm K}^2$ (panel c). The lensing bandpowers, on the other hand,
are almost unaffected: the frozen best fit raises the lensing power by $\ppA\%$ in the first bandpower, centered near
$L=28$, which is also the bandpower that lies highest above \LCDM, and by less than $1\%$ at $L\gtrsim200$
(panel d).

\begin{figure}[tbp]
\includegraphics[width=\columnwidth]{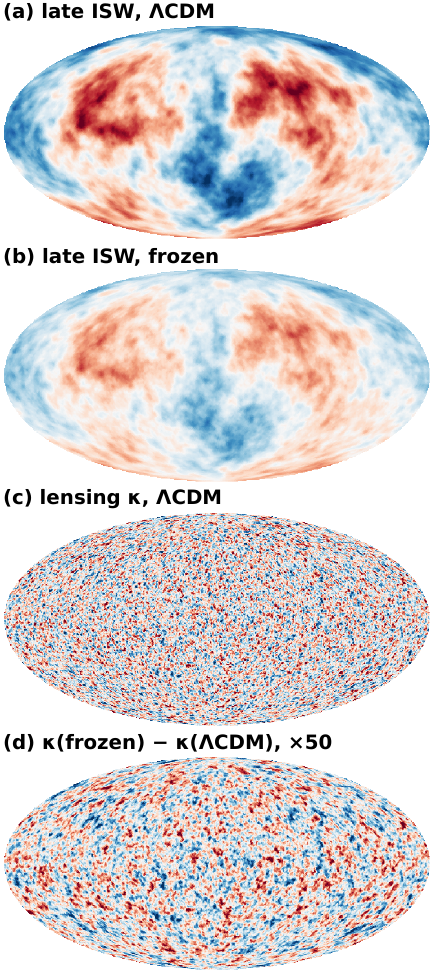}
\caption{Simulated full-sky maps (Mollweide projection) for the best fits of \LCDM\ and of the frozen model to
\Pl+\DK+\DESI, drawn with identical random phases. (a, b) Late-ISW contribution to the temperature anisotropy, on a
common color scale ($\pm63\,\mu$K). (c) Lensing convergence of \LCDM, band-limited to $L\le400$ and smoothed with
a $1^\circ$ Gaussian beam. (d) Difference between the convergence of the frozen model and that of \LCDM,
multiplied by 50 to make it visible on the color scale of panel (c). These are simulations built from the model
spectra, not the Planck maps.}
\label{fig:simmaps}
\end{figure}

The coupling thus changes the late ISW signal much more than the lensing potential, and this difference can be
seen directly on the sky. Figure~\ref{fig:simmaps} shows Gaussian realizations of the late-ISW temperature map and of the lensing convergence for the best fits of \LCDM\
and of the frozen model to \Pl+\DK+\DESI, generated with identical random phases so that the maps differ only
through the power spectra. The rms of the late-ISW map drops from $\iswrmsL$ to $\iswrmsF\,\mu$K, and the late-ISW
power at $2\le\ell\le20$ is $\liswlo$--$\liswhi$ times the \LCDM\ value, whereas the convergence maps,
band-limited to the multipoles of the Planck reconstruction, differ by about $\kapdiff\%$ of their rms. The
reconstruction therefore leaves room for a large change of the growth at $z\lesssim0.5$, where its kernel is small,
while the large-scale temperature is sensitive to it.

\subsection{DES-Dovekie}

The Dark Energy Survey five-year supernova sample~\cite{DES2024}, recalibrated with the Dovekie photometric
calibration~\cite{Popovic2025}, contains $\NSN$ type Ia supernovae: $\NSNdes$ photometrically classified
supernovae from the DES deep fields at $\zSNdesmin<z<\zSNmax$ and $\NSNlow$ spectroscopically confirmed
supernovae from external low-redshift samples at $z<\zSNlowmax$ (Table~\ref{tab:sn}). We use the distance moduli
with their full statistical and systematic covariance and marginalize analytically over the absolute magnitude;
the probability that each DES supernova is not of type Ia is included in its uncertainty. The data measure the
shape of the luminosity distance, and therefore $\Omega_{\rm m}$.

\begin{table}[tbp]
\caption{Composition of the DES-Dovekie Hubble diagram: number of supernovae, redshift range and median redshift of each sample (redshifts corrected to the CMB frame and for peculiar velocities). Data from Refs.~\cite{DES2024,Popovic2025}.}
\label{tab:sn}
\begin{ruledtabular}
\begin{tabular}{lrcc}
Sample & $N$ & $z_{\rm min}$--$z_{\rm max}$ & median $z$\\
\hline
DES 5-year & 1623 & 0.073--1.144 & 0.519\\
Foundation & 117 & 0.025--0.087 & 0.043\\
CfA3K & 31 & 0.025--0.085 & 0.034\\
CfA4p2 & 22 & 0.025--0.071 & 0.038\\
CfA3S & 16 & 0.025--0.059 & 0.034\\
CSP DR3 & 8 & 0.029--0.057 & 0.041\\
CfA4p3 & 3 & 0.033--0.053 & 0.039\\
\hline
Total & 1820 & 0.025--1.144 & 0.479\\
\end{tabular}
\end{ruledtabular}
\end{table}

\begin{figure}[tbp]
\includegraphics[width=\columnwidth]{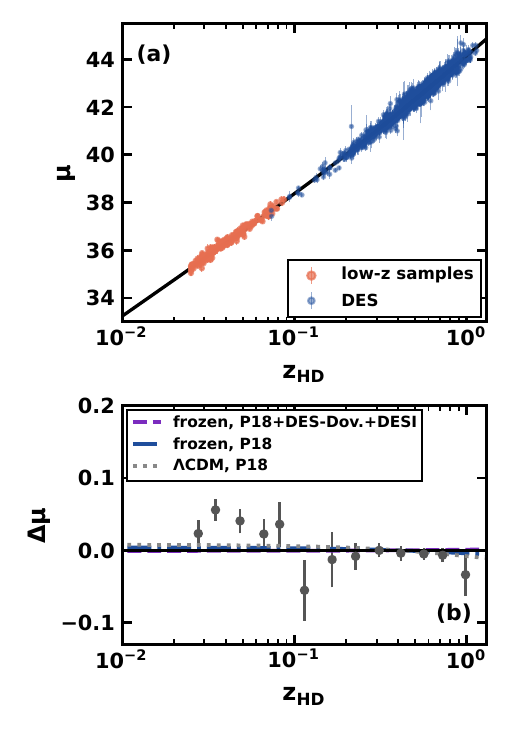}
\caption{DES-Dovekie supernovae. (a) Hubble diagram of the DES (blue) and low-redshift (orange) supernovae, with
the absolute magnitude fixed to its best-fit value for the \LCDM\ best fit of \Pl+\DK+\DESI\ (black line); points
with uncertainties larger than one magnitude, which correspond to a low probability of being of type Ia, are not
shown. (b) Inverse-variance weighted residuals in redshift bins (diagonal errors only), and the difference between
other best fits and the reference, after marginalization over the absolute magnitude; the full covariance used in
the fits, which includes the calibration systematics that correlate the low-redshift points, is not reflected in
these error bars. Because the
two models share the same background, all curves in panel (b) differ only through the parameters preferred by each
data combination.}
\label{fig:sn}
\end{figure}

Figure~\ref{fig:sn} shows the Hubble diagram and the binned residuals with respect to the \LCDM\ best fit of
\Pl+\DK+\DESI. Because the frozen model has the same expansion history as \LCDM, the two models would give identical
Hubble diagrams if fitted to the supernovae alone; the small difference we find stems only from the shift of the
cosmological parameters induced by the other data sets in the fit, and it stays below $0.01$~mag at all redshifts
once the absolute magnitude is marginalized over.

\subsection{\DESI}

\begin{figure}[tbp]
\includegraphics[width=\columnwidth]{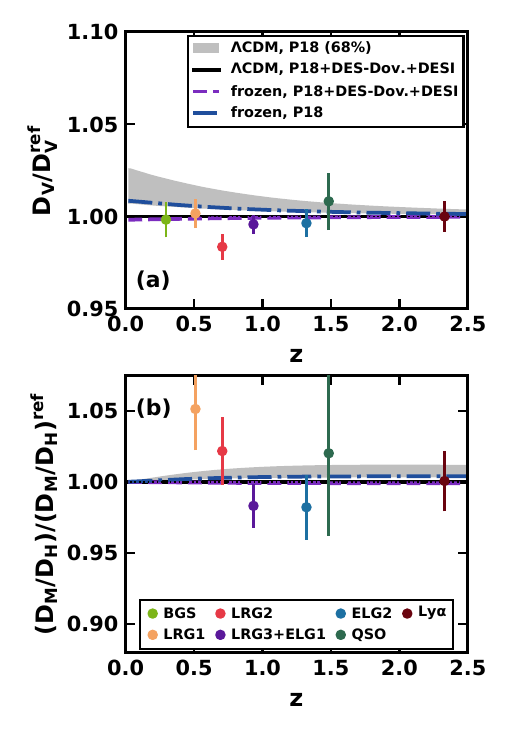}
\caption{DESI DR2 BAO distances relative to the \LCDM\ best fit of \Pl+\DK+\DESI: (a) angle-averaged distance
$D_V/r_{\rm d}$ and (b) ratio of the transverse and line-of-sight distances, $D_M/D_H$. The gray band is the $68\%$ region of \LCDM\ constrained by \Pl\ alone; the lines are the best fits
of \LCDM\ and of the frozen model to \Pl+\DK+\DESI\ (they overlap) and of the frozen model to \Pl\ alone. The
largest residuals with respect to the \LCDM\ best fit of \Pl+\DK+\DESI\ are $-2.4\sigma$ for the second LRG bin
in $D_V$ and $+1.8\sigma$ for the first LRG bin in $D_M/D_H$; neither model can change them, since the coupling
does not enter the background.}
\label{fig:bao}
\end{figure}

We use the DESI DR2 measurements of the transverse and line-of-sight BAO distances, $D_M/r_{\rm d}$ and
$D_H/r_{\rm d}$, and of the angle-averaged distance $D_V/r_{\rm d}$ for the bright galaxy sample, for all tracers
in $0.1<z<4.2$, with their covariance~\cite{DESI2025DR2}. Uncalibrated BAO constrain
$\Omega_{\rm m}$ and $hr_{\rm d}$. Figure~\ref{fig:bao} compares the measurements with the best fits of the two
models. The curves of \LCDM\ and of the frozen model fitted to \Pl+\DK+\DESI\ are indistinguishable, as they must
be for two models with the same background; the best fit of the frozen model to \Pl\ alone, on the other hand, lies at
the lower edge of the band allowed by \Pl\ in \LCDM\ (gray), closer to the DESI points, which is the geometric
counterpart of the reduction of the tension discussed in Sec.~\ref{sec:tensions}.

\subsection{Cosmic shear}

We use the amplitude of the matter fluctuations measured by cosmic shear in the Kilo-Degree Survey (KiDS-Legacy),
$S_8=0.815^{+0.016}_{-0.021}$~\cite{KiDSLegacy2025}, and in the Dark Energy Survey Year 6,
$S_8=0.798^{+0.014}_{-0.015}$~\cite{DESY6shear2026}, with $S_8\equiv\sigma_8\sqrt{\Omega_{\rm m}/0.3}$. These are
the most recent results of the two collaborations; the DES value, obtained with the nonlinear alignment model of
intrinsic alignments, supersedes the Year 3 measurement, $S_8=0.759^{+0.025}_{-0.023}$~\cite{Amon2022,Secco2022}.
These values are derived under \LCDM\ growth, while in the frozen model the growth history at $z\lesssim1$ is
different. We therefore compare them with the effective amplitude
\begin{equation}
S_8^{\rm eff}=S_8\left[\frac{\int dz\,K(z)\,G^2(z)}{\int dz\,K(z)}\right]^{1/2},
\label{eq:s8eff}
\end{equation}
with
\begin{equation}
G(z)=\frac{\sigma_8(z)/\sigma_8(0)}{\sigma_8^{\Lambda}(z)/\sigma_8^{\Lambda}(0)},
\end{equation}
where $\sigma_8^{\Lambda}$ refers to \LCDM\ with the same parameters and $\Sigma=0$, and $K(z)$ is the redshift
kernel of the shear power spectrum at $\ell=300$. Because the growth modification is scale independent on the
scales probed, the shear power of the frozen model is close to that of a \LCDM\ model with amplitude
$S_8^{\rm eff}$; Appendix~\ref{app:wl} derives this result, tests it against the published source distributions
of both surveys and against a simulated \LCDM\ analysis of the DES Year 3 correlation functions, and shows how the
tomographic signal of the two models differs. The two surveys cover almost disjoint areas of the sky and are treated as independent split-normal likelihoods,
$-2\ln\mathcal{L}=(S_8^{\rm eff}-\mu)^2/\sigma_\pm^2$, where $\mu$ is the published central value and $\sigma_+$
($\sigma_-$) the published upper (lower) error, used above (below) $\mu$. Their product peaks at
$S_8=0.803\pm0.012$, which we quote as the combined shear measurement. Section~\ref{sec:joint} shows that the results
are stable when DES Year 6 is replaced by DES Year 3, or when the two likelihoods are replaced by the joint
KiDS-Legacy and DES Year 3 analysis of Ref.~\cite{Stolzner2025}, $S_8=0.818^{+0.012}_{-0.014}$. Since $G(z)<1$ at $z>0$ for $\Sigma<0$, $S_8^{\rm eff}$ is smaller than $S_8$: the shear kernels peak
at $z=0.3$--$0.4$, where part of the late enhancement of the growth has not yet happened.

\subsection{Cluster abundances}

As a check that does not enter the fits, we compare the predictions with three determinations of the amplitude
from cluster counts: SPT clusters with weak-lensing mass calibration, $S_8=0.795\pm0.029$~\cite{Bocquet2024};
eROSITA eRASS1, $S_8=0.86\pm0.01$~\cite{Ghirardini2024}; and the Planck PSZ2 sample,
$S_8=0.78\pm0.02$~\cite{Aymerich2024}. We use Eq.~(\ref{eq:s8eff}) with kernels that follow the redshift
distribution of each sample. Cluster counts depend on the halo mass function, which has been calibrated only for
models in which the DM mass is conserved; Appendix~\ref{app:clusters} quantifies how well the effective
amplitude describes the counts of the frozen model with a mass function of \LCDM\ form, and shows that the
redshift evolution of the counts carries additional information.

The three samples are complementary, as Table~\ref{tab:clusters} summarizes: SPT selects
massive clusters through their Sunyaev--Zeldovich signal up to $z=1.78$ in the southern sky, eRASS1 selects
X-ray clusters at $0.1<z<0.8$ over the western Galactic hemisphere, and the Planck sample covers most of the sky
outside the Galactic plane with the highest signal-to-noise detections. The masses are calibrated with weak lensing in the first two and with X-ray and weak-lensing data in the third.
We list these properties because the selection, the sky coverage and the mass calibration differ among the
samples, and these differences are possible sources of the spread of their amplitudes discussed in Sec.~\ref{sec:late}.

\begin{table}[tbp]
\caption{Cluster samples used for comparison: number of clusters in the cosmological sample, redshift range, sky
coverage, mass calibration and published amplitude.}
\label{tab:clusters}
\begin{ruledtabular}
\footnotesize
\begin{tabular}{lccc}
 & SPT~\cite{Bocquet2024} & eRASS1~\cite{Ghirardini2024} & PSZ2~\cite{Aymerich2024}\\
\hline
$N_{\rm cl}$ & 1005 & 5259 & 439\\
$z$ range & 0.25--1.78 & 0.1--0.8 & ---\\
Area & 5200 deg$^2$ & 12791 deg$^2$ & 65\% of sky\\
Selection & SZ & X-ray & SZ, S/N$>6$\\
Masses & DES, HST WL & DES, KiDS, HSC WL & X-ray + WL\\
$S_8$ & $0.795\pm0.029$ & $0.86\pm0.01$ & $0.78\pm0.02$\\
\end{tabular}
\end{ruledtabular}
\end{table}

\subsection{Theory settings}

We assume one massive neutrino with $m_\nu=0.06$~eV and $N_{\rm ur}=2.0328$, take the primordial helium fraction
from big-bang nucleosynthesis, and use the revised halo-model fit of Ref.~\cite{Takahashi2012} for the nonlinear
correction to the CMB lensing potential, a prescription calibrated on \LCDM\ simulations.

\section{Statistical methodology}
\label{sec:methods}

Comparing models whose posteriors are strongly non-Gaussian, and combining data sets whose agreement is itself
in question, calls for some care in the statistical treatment. In this section we describe the parameters and
priors, the sampling strategy, which uses exact likelihoods throughout, and the estimators with which we quantify
the agreement between data sets and compare the frozen model with \LCDM.

\subsection{Parameters and priors}

The frozen model has eight free parameters,
$\bm\theta=\{H_0,\omega_{\rm b},\omega_{\rm c},\ln(10^{10}A_{\rm s}),n_{\rm s},\tau,\Sigma,A_{\rm P}\}$, and \LCDM\
the same seven without $\Sigma$. The priors (Table~\ref{tab:priors}) are uniform, except for $A_{\rm P}$, and
identical for every data set and for both models; this is necessary for the evidence-based estimators of
Sec.~\ref{sec:estimators} to be comparable. The prior range of $\omega_{\rm b}$ is the domain of validity of the BBN table that sets the primordial helium
abundance $Y_{\rm He}$ consistently with $\omega_{\rm b}$, which matters mainly for the recombination history probed
by the CMB. Unlike many BAO analyses, which calibrate $r_{\rm d}$ with a BBN prior on $\omega_{\rm b}$, we impose no
prior from the observed light-element abundances, so $\omega_{\rm b}$ is constrained by the data alone. Points beyond the physical boundary $\eta_0<\eta_{\rm min}$
are assigned zero likelihood; the prior volume of $\Sigma$ is therefore the full interval
$[-0.5,0.3]$, and the boundary acts as part of the data model, as in a Markov chain that rejects these points.

\begin{table}[tbp]
\caption{Priors. All priors are uniform in the indicated range, except for the Planck calibration, which is
Gaussian. They are identical for all data sets and, for the parameters in common, for \LCDM\ and the frozen
model. Points with $\eta_0=1+\Sigma r_0<\eta_{\rm min}=0.05$ have zero likelihood.}
\label{tab:priors}
\begin{ruledtabular}
\begin{tabular}{lc}
Parameter & Prior \\
\hline
$H_0$ [$\kms$] & $[40,100]$ \\
$\omega_{\rm b}$ & $[0.0073,0.03287]$ \\
$\omega_{\rm c}$ & $[0.001,0.99]$ \\
$\ln(10^{10}A_{\rm s})$ & $[1.61,3.91]$ \\
$n_{\rm s}$ & $[0.8,1.2]$ \\
$\tau$ & $[0.01,0.8]$ \\
$\Sigma$ (frozen) & $[-0.5,0.3]$ \\
$A_{\rm P}$ & $\mathcal{N}(1,0.0025)$ \\
\end{tabular}
\end{ruledtabular}
\end{table}

\subsection{Sampling}
\label{sec:sampling}

Each evaluation of the \Pl\ likelihood requires the full solution of the Boltzmann hierarchy, about $6$~s. We
sample the posteriors that involve \Pl\ by adaptive importance sampling with exact likelihoods, in the form of
population Monte Carlo with deterministic-mixture weights~\cite{Veach1995,Owen2000,Cornuet2012}. In round $r$ we
draw $n_r$ points from a mixture $q_r$ of multivariate Student-$t$ distributions with five degrees of freedom,
and at every draw we evaluate all likelihoods at once: \Pl, \DK, \DESI\ and the shear and cluster amplitudes.
After $R$ rounds, every draw $\bm\theta_i$ carries, for each target $T$, the weight
\begin{equation}
w_T(\bm\theta_i)=\frac{\pi(\bm\theta_i)\,\mathcal{L}_T(\bm\theta_i)}{\sum_r(n_r/n)\,q_r(\bm\theta_i)},\qquad
n=\sum_rn_r,
\label{eq:dmw}
\end{equation}
and the evidence is $Z_T=n^{-1}\sum_iw_T(\bm\theta_i)$. Because all likelihoods are computed at every draw, a single
set of draws serves all targets: \Pl, \Pl+\DK, \Pl+\DESI, \Pl+\DK+\DESI, \Pl+shear and \Pl+\DK+\DESI+shear. The
weights are exact for any choice of the $q_r$, which affect only the efficiency, and the effective sample size
$N_{\rm eff}=(\sum w)^2/\sum w^2$ and the Pareto-smoothing diagnostic $\hat k$~\cite{Vehtari2015} monitor their
reliability.

The first round of each model draws from the Laplace approximations of the \LCDM\ posteriors of the four
Planck-based combinations, with covariances inflated by $1.5^2$; for the frozen model the $\Sigma$ direction is
added with the regression of the other parameters on $\Sigma$ measured in a preliminary Markov chain. Later rounds
use one component per target, fitted to its weighted posterior and inflated by $1.15$--$1.35$, plus a broad
component with $10\%$ of the weight. In the frozen model the posterior of $\Sigma$ is strongly skewed and limited
by the physical boundary; in the variable $\ln\eta_0$, which maps the allowed region onto the real line, it is
close to Gaussian, and from the third round on the proposal components are Student-$t$ distributions in the space
where $\Sigma$ is replaced by $\ln\eta_0$, with the corresponding Jacobian in Eq.~(\ref{eq:dmw}). After the first
two rounds ($1878$ draws) the effective sample size of \Pl\ was $155$; the third round, the first one in
$\ln\eta_0$, raised it to $648$ with $1196$ further draws. The last round of each model puts most of
its weight on the combinations with cosmic shear. The sampling is summarized in Appendix~\ref{app:sampling}.

For \DK\ and \DESI\ alone the likelihoods depend only on $\{H_0,\omega_{\rm b},\omega_{\rm c}\}$, and on $\Sigma$
through the boundary, and we use nested sampling~\cite{Skilling2006,Lange2023} with $2000$ live points and about
$2\times10^4$ effective samples. To make this affordable we compute the expansion history from the densities of
photons and neutrinos tabulated with the same Boltzmann solver, and the drag-epoch sound horizon from a bicubic
interpolation in $(\omega_{\rm b},\ln\omega_{\rm c})$; the log-likelihoods agree with the full computation to
$2\times10^{-6}$ (\DK) and $4\times10^{-3}$ (\DESI).

The best fits (maxima of $\mathcal{L}\pi$) are found with a quadratic surrogate of $-2\ln(\mathcal{L}\pi)$ fitted to
all exact evaluations within $\Delta\chi^2<16$ of the minimum, in coordinates whitened with the posterior
covariance (with $\ln(\eta_0-\eta_{\rm min})$ in place of $\Sigma$, so that every step stays inside the boundary),
followed by an exact evaluation at the minimum of the surrogate within a trust radius, and iterated until the exact
$\chi^2$ improves by less than $0.03$, which takes one to five iterations. For \DK\ and \DESI\ alone we use a
simplex search.

\subsection{Tension estimators}
\label{sec:estimators}

We quantify the agreement between \Pl\ (denoted $A$) and an external data set $X$ with the same set of
estimators in both models~\cite{Raveri2019,Raveri2020,Raveri2021,Marshall2006,Handley2019a,Handley2019b,
Spiegelhalter2002,Joudaki2017,Lemos2021}. Each of them answers a slightly different question, and their strengths
and limitations for non-Gaussian posteriors and for several data sets at once have been compared in detail in
Refs.~\cite{Leizerovich2023,Leizerovich2025}; we report all of them side by side, so that no conclusion rests on a
single estimator. Their definitions are collected in Appendix~\ref{app:estimators}:
\begin{itemize}
\item the difference in means $Q_{\rm DM}$ and the parameter-shift probability in the parameters constrained by
$X$: $\Omega_{\rm m}$ for \DK, $(\Omega_{\rm m},hr_{\rm d})$ for \DESI, and $S_8^{\rm eff}$ for cosmic shear;
\item the update difference in means $Q_{\rm UDM}$;
\item the goodness-of-fit loss $Q_{\rm DMAP}=\chi^2_{AX}-\chi^2_A-\chi^2_X$ at the best fits, distributed as
$\chi^2$ with $N_{\rm eff}^A+N_{\rm eff}^X-N_{\rm eff}^{AX}$ degrees of freedom, where $N_{\rm eff}$ counts the
parameters constrained beyond the prior;
\item the evidence ratio $\ln R=\ln Z_{AX}-\ln Z_A-\ln Z_X$, the information ratio $\ln I$, the suspiciousness
$\ln S=\ln R-\ln I$ and the Bayesian model dimensionality $d$, with the probability to exceed obtained from
$d-2\ln S\sim\chi^2_d$;
\item the change of the deviance information criterion, $\Delta{\rm DIC}={\rm DIC}_{AX}-{\rm DIC}_A-{\rm DIC}_X$.
\end{itemize}
Probabilities are converted to an equivalent number of Gaussian standard deviations with the two-tailed
convention. For the frozen model the prior used to count $N_{\rm eff}$ is restricted to the physical region, so
that the boundary is not counted as a constraint of the external data. For cosmic shear, which enters as a
Gaussian constraint on a single amplitude, the shear-alone posterior is known analytically:
$\chi^2_{X}$ is the minimum over a common amplitude of the two surveys,
$\langle\ln\mathcal{L}_X\rangle=\ln\mathcal{L}_{X,\max}-\tfrac12$ and $d_X=1$, so $Q_{\rm DMAP}$, $\ln S$ and
$\Delta{\rm DIC}$ follow without sampling; the evidence of the shear data alone is not defined independently of
the prior on $A_{\rm s}$, and we do not quote $\ln R$ for this pair.

\subsection{Model comparison}

We compare the frozen model with \LCDM\ with the Bayes factor $\ln B=\ln Z_{\rm frozen}-\ln Z_{\Lambda{\rm CDM}}$
for each data combination and with the difference of the best-fit $\chi^2$. Since \LCDM\ is the point
$\Sigma=0$ of the frozen model, $\ln B$ includes the Occam penalty of the prior range of $\Sigma$; with the uniform
prior adopted here, it changes by $\ln(\Delta\Sigma/0.8)$ for a different prior width $\Delta\Sigma$ that contains
the posterior.

\section{Planck alone and the physical boundary}
\label{sec:planck}

\begin{figure}[tbp]
\includegraphics[width=\columnwidth]{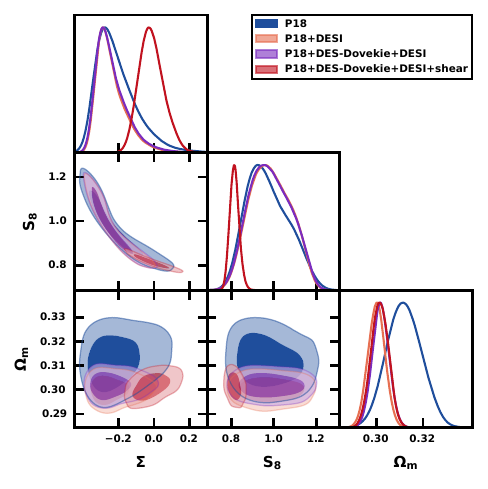}
\caption{Marginalized posteriors (68\% and 95\% regions) of $\Sigma$, $S_8$ and $\Omega_{\rm m}$ in the frozen
model for \Pl, \Pl+\DESI, \Pl+\DK+\DESI\ and \Pl+\DK+\DESI+shear. Without cosmic shear the coupling is pushed
toward the physical boundary and $S_8$ is close to unity; the shear data bring it back to values close to zero.}
\label{fig:triangle}
\end{figure}

\begin{figure}[tbp]
\includegraphics[width=\columnwidth]{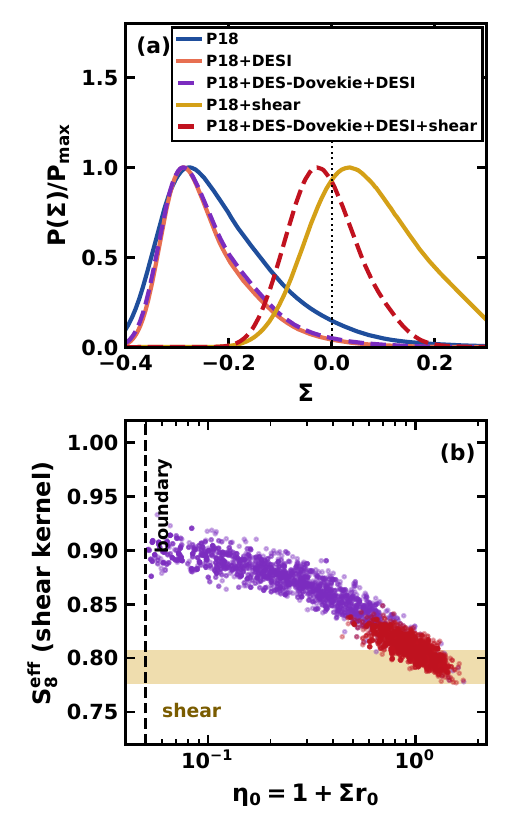}
\caption{(a) Marginalized posteriors of $\Sigma$, normalized to their maxima. (b) Samples of the
\Pl+\DK+\DESI\ posterior (purple) and of the combination with cosmic shear (red) in the plane of the regularity
parameter $\eta_0$ and of the amplitude $S_8^{\rm eff}$ seen by the shear kernel; the band is the $68\%$ interval of the combined KiDS-Legacy and DES Year 6 cosmic-shear measurement, $S_8=\SWL$, and the dashed line the physical boundary $\eta_0=0.05$.}
\label{fig:sigma}
\end{figure}

\Pl\ alone constrains the coupling to $\Sigma=\SigA$, with the $95\%$ interval $[\SigLoA,\SigHiA]$
(Table~\ref{tab:constraints} and Figs.~\ref{fig:triangle} and \ref{fig:sigma}). The posterior is strongly
asymmetric: it rises steeply toward negative $\Sigma$, peaks near $\Sigma=-0.29$, close to the physical
boundary, and has a long tail toward positive values. In terms of the regularity parameter,
$\eta_0=\etaA$. The best fit sits at $\Sigma=\bfSigA$, that is $\eta_0=\bfetaA$ and $\epsilon_0=\bfepsA$, deep in
the region where the exchange enhances the late growth of the DM density, and it improves the fit to \Pl\ by
$\Delta\chi^2=\dchiA$ with respect to \LCDM. The amplitude of the fluctuations follows the coupling: the
posterior of $\sigma_8$ is $\seightA$ and that of $S_8$ is $\SeightA$, against $\SeightLA$ in \LCDM, and the
best fit has $\sigma_8=\bfseightA$. The parameters shared with \LCDM\ move by less than half a standard
deviation: $H_0=\HzeroA\kms$ and $\Omega_{\rm m}=\OmA$, against $\HzeroLA\kms$ and $\OmLA$ in \LCDM.

\begin{figure}[tbp]
\includegraphics[width=\columnwidth]{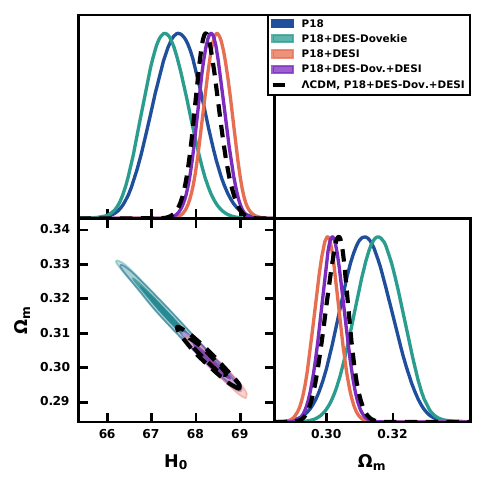}
\caption{Marginalized posteriors of $H_0$ (in ${\rm km\,s^{-1}\,Mpc^{-1}}$) and $\Omega_{\rm m}$ in the frozen
model for \Pl, \Pl+\DK, \Pl+\DESI\ and \Pl+\DK+\DESI, compared with \LCDM\ for \Pl+\DK+\DESI\ (dashed). All
posteriors lie along the CMB degeneracy direction; the supernovae pull toward the upper left and DESI toward
the lower right of this line, and the frozen model moves the \Pl\ posterior slightly toward DESI.}
\label{fig:h0om}
\end{figure}

\begin{figure}[tbp]
\includegraphics[width=\columnwidth]{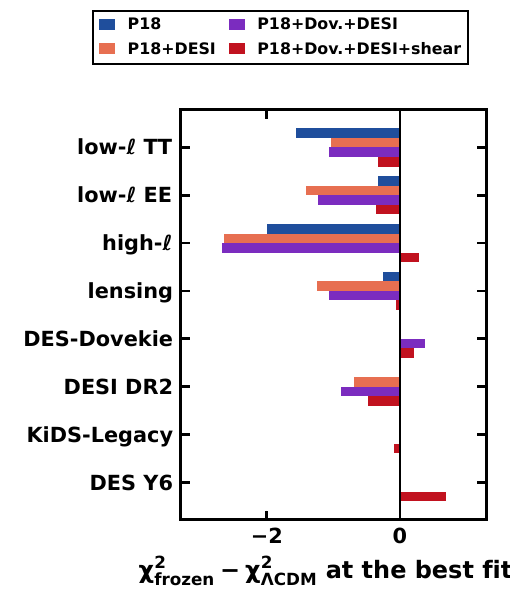}
\caption{Difference of the $\chi^2$ of each likelihood between the best fits of the frozen model and of \LCDM,
for the data combinations that enter each fit.}
\label{fig:chi2}
\end{figure}

\paragraph{Origin of the preference.} Figure~\ref{fig:chi2} decomposes the improvement of the best fit. All four
Planck likelihoods contribute with the same sign: the high-$\ell$ spectra by $\dchiplikA$, the low-$\ell$
temperature by $\dchilowTTA$, the low-$\ell$ polarization by $\dchilowEEA$ and the lensing reconstruction by
$\dchilensA$. These are the two effects of Fig.~\ref{fig:pheno}. The enhanced late growth increases the lensing
smoothing of the acoustic peaks, which the high-$\ell$ Planck 2018 spectra prefer to be larger than in \LCDM, the
well-known lensing anomaly~\cite{Planck2018VI,Calabrese2008}; and it suppresses the late ISW contribution at
$\ell\lesssim10$, where the observed temperature spectrum is low. The low-$\ell$ polarization improves slightly because the optical depth moves to lower values,
$\tau=0.052$ against $0.054$ in \LCDM, once the amplitude of the peaks is no longer tied to the lensing smoothing. The lensing
reconstruction itself is only marginally affected, because the growth enhancement acts at $z\lesssim0.5$, where its
kernel is small; this is why $\sigma_8$ can reach values above unity without being excluded by the CMB lensing
data. Since the lensing anomaly is weaker in the Planck PR4 likelihoods~\cite{Rosenberg2022,Tristram2024}, the
preference for $\Sigma<0$ is expected to be weaker with those data, which Sec.~\ref{sec:pr4} confirms.

\paragraph{Dependence on the boundary.} The posterior does not vanish at the boundary, so the lower limit on
$\Sigma$ is set by the choice of $\eta_{\rm min}$ rather than by the data. Raising the cut removes the region of
largest growth: for \Pl\ we find $\Sigma=\SigcutaA$, $\SigcutbA$ and $\SigcutcA$ for $\eta_{\rm min}=0.1$, $0.2$
and $0.5$, and the Bayes factor with respect to \LCDM\ drops from $\ln B=\lnBA$ to $\lnBcutaA$, $\lnBcutbA$ and
$\lnBcutcA$ (Table~\ref{tab:models}). The best fits, at $\eta_0$ close to $0.1$, correspond to $\epsilon_0$ between
$-11$ and $-8$: the DM density responds to the convergence of the flow ten times more efficiently than the baryons. This
is a regime in which the linear description of the exchange and the nonlinear corrections calibrated on \LCDM\
cannot be taken for granted, and we consider that the region $\eta_0\lesssim0.2$ should be interpreted with caution even when it is formally allowed. By
Eq.~(\ref{eq:astar}), it is also the region in which the stability boundary is reached shortly after the present
epoch, at $a_*<1.08$.

\section{Agreement between the data sets}
\label{sec:tensions}

\begin{table*}[tbp]
\caption{Agreement between the data sets in \LCDM\ and in the frozen model. The first five rows give the
equivalent number of Gaussian standard deviations of the goodness-of-fit loss $Q_{\rm DMAP}$, of the difference
in means $Q_{\rm DM}$, of the update difference in means from \Pl\ to the joint posterior $Q_{\rm UDM}$, of the
parameter shift and of the suspiciousness; the shift parameters are $\Omega_{\rm m}$ (\DK),
$(\Omega_{\rm m},hr_{\rm d})$ (\DESI) and $S_8^{\rm eff}$ (shear). The last rows give the log-evidence ratio,
the log-suspiciousness, the Bayesian model dimensionality of the tension and the change of the deviance
information criterion; $\ln R>0$, $\ln S>0$ and $\Delta{\rm DIC}<0$ indicate agreement.}
\label{tab:tension}
\begin{ruledtabular}
\begin{tabular}{lcccccccc}
 & \multicolumn{2}{c}{\Pl\ vs \DK} & \multicolumn{2}{c}{\Pl\ vs \DESI} & \multicolumn{2}{c}{\Pl+\DK\ vs \DESI} & \multicolumn{2}{c}{\Pl\ vs shear} \\
 & \LCDM & frozen & \LCDM & frozen & \LCDM & frozen & \LCDM & frozen \\
\hline
$Q_{\rm DMAP}$ [$\sigma$] & $0.91$ & $1.11$ & $1.85$ & $1.01$ & $2.40$ & $1.69$ & $1.62$ & $2.59$ \\
$Q_{\rm DM}$ [$\sigma$] & $0.91$ & $1.13$ & $1.91$ & $1.46$ & $2.46$ & $2.12$ & $1.61$ & $2.38$ \\
$Q_{\rm UDM}$ [$\sigma$] & $0.98$ & $1.15$ & $1.54$ & $1.37$ & $1.79$ & $1.70$ & -- & -- \\
shift [$\sigma$] & $0.89$ & $1.13$ & $1.88$ & $1.37$ & $2.50$ & $2.03$ & $1.63$ & $2.29$ \\
$S$ [$\sigma$] & $0.91$ & $1.11$ & $1.87$ & $1.35$ & $2.44$ & $2.05$ & $1.54$ & $2.34$ \\
\hline
$\ln R$ & $3.39$ & $3.32$ & $5.34$ & $6.49$ & $4.04$ & $5.25$ & -- & -- \\
$\ln S$ & $0.08$ & $-0.17$ & $-1.71$ & $-0.51$ & $-3.04$ & $-1.82$ & $-0.70$ & $-2.65$ \\
$d$ & $0.99$ & $1.24$ & $1.83$ & $1.28$ & $1.74$ & $1.31$ & $1.04$ & $1.53$ \\
$\Delta{\rm DIC}$ & $-1.2$ & $-0.9$ & $1.4$ & $-0.5$ & $4.0$ & $2.2$ & $0.2$ & $4.0$ \\
\end{tabular}
\end{ruledtabular}
\end{table*}

\begin{figure}[tbp]
\includegraphics[width=\columnwidth]{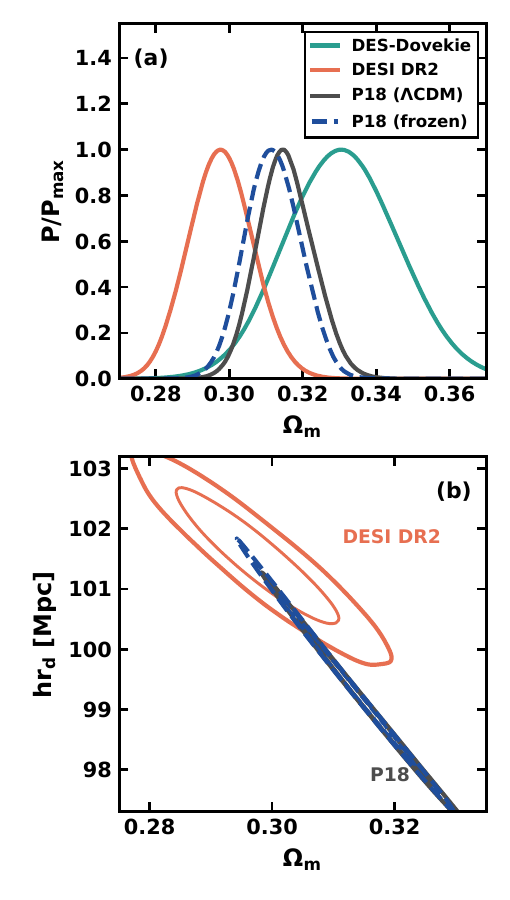}
\caption{(a) Marginalized posteriors of $\Omega_{\rm m}$ for \DK\ and \DESI\ alone and for \Pl\ in \LCDM\ and in
the frozen model. (b) 68\% and 95\% regions in the plane $(\Omega_{\rm m},hr_{\rm d})$ measured by uncalibrated
BAO, for \DESI\ and for \Pl\ in \LCDM\ (solid) and in the frozen model (dashed).}
\label{fig:geometry}
\end{figure}

\begin{figure}[tbp]
\includegraphics[width=\columnwidth]{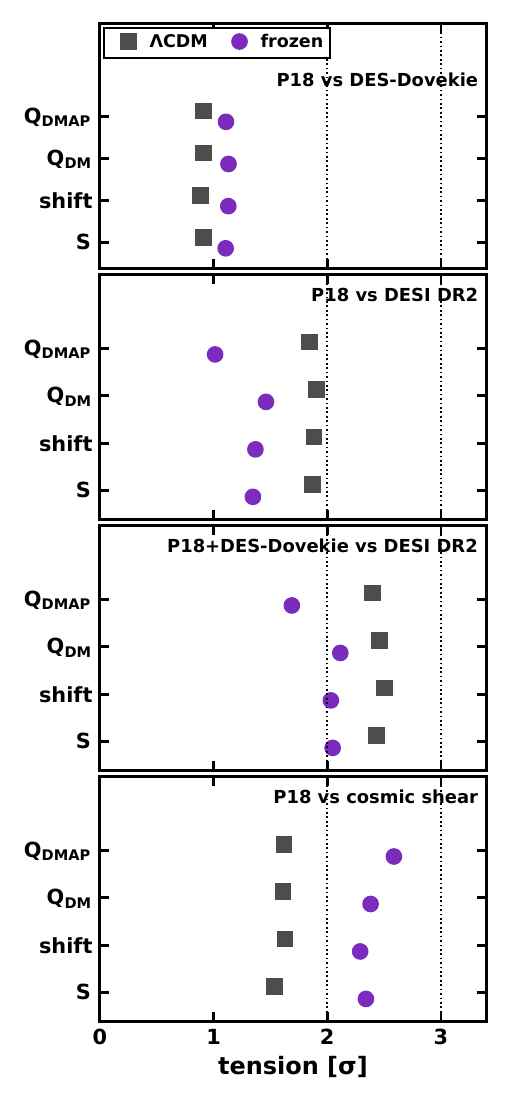}
\caption{Tension estimators, in equivalent Gaussian standard deviations, for \LCDM\ (squares) and the frozen model
(circles): goodness-of-fit loss $Q_{\rm DMAP}$, difference in means $Q_{\rm DM}$, parameter shift and
suspiciousness $S$. The dotted lines mark $2\sigma$ and $3\sigma$.}
\label{fig:tension}
\end{figure}

Before combining the data sets we test their agreement with the estimators of Sec.~\ref{sec:estimators}. The
results are collected in Table~\ref{tab:tension} and summarized in Fig.~\ref{fig:tension}. Because the frozen
model has the expansion history of \LCDM, the posteriors of \DK\ and \DESI\ alone are the same in the two models,
apart from the volume of the $\Sigma$ prior, which they do not constrain; any change in the level of tension comes
from how the coupling moves the Planck posterior in the parameters that the external data measure
(Fig.~\ref{fig:geometry}).

\paragraph{\Pl\ versus \DK.} The supernovae prefer a higher matter density, $\Omega_{\rm m}=\OmDov$, than \Pl. In
\LCDM\ the two data sets agree: all estimators give $0.9\sigma$, $\ln R=\lnRLDov$ and $\Delta{\rm DIC}=\dDICLDov$.
In the frozen model \Pl\ moves to a slightly lower matter density, $\Omega_{\rm m}=\OmAF$, and the estimators rise
to $\QDMAPFDov$--$\QDMFDov\sigma$; the agreement is still good, with $\Delta{\rm DIC}=\dDICFDov$.

\paragraph{\Pl\ versus \DESI.} DESI measures $\Omega_{\rm m}=\OmDESI$ and $hr_{\rm d}=\hrdDESI$~Mpc. In \LCDM\ the
moderate discrepancy with \Pl\ ($hr_{\rm d}=\hrdAL$~Mpc) is reproduced by all estimators:
$Q_{\rm DMAP}$, $Q_{\rm DM}$, the shift and the suspiciousness give $\QDMAPLDESI$, $\QDMLDESI$, $\shiftLDESI$ and
$\SLDESI$ standard deviations, and the deviance information criterion increases by $\dDICLDESI$ when the two are
combined. The frozen model reduces it to $\QDMAPFDESI$--$\QDMFDESI\sigma$, with $\Delta{\rm DIC}=\dDICFDESI$: the
coupling lowers $\Omega_{\rm m}$ and raises $hr_{\rm d}$ of \Pl\ to $\hrdAF$~Mpc along the direction of the DESI
measurement. The reduction is largest in $Q_{\rm DMAP}$, because the best fit of \Pl+\DESI\ also improves the
lensing likelihood ($\dchilensAH$) and the DESI $\chi^2$ ($\dchidesiAH$) at the same time, while the estimators based
on parameter differences decrease by about $0.5\sigma$.

\paragraph{\Pl+\DK\ versus \DESI.} Adding the supernovae to \Pl\ widens the gap with DESI, as found in previous analyses~\cite{DESI2025DE}: in \LCDM\ the four estimators give $\QDMAPLDovDESI$--$\shiftLDovDESI\sigma$ and $\Delta{\rm
DIC}=\dDICLDovDESI$. The frozen model reduces this to $\QDMAPFDovDESI$--$\QDMFDovDESI\sigma$ and
$\Delta{\rm DIC}=\dDICFDovDESI$. In both models the three data sets can be combined, but in \LCDM\ the combination is
a compromise between discordant measurements of the matter density.

\paragraph{\Pl\ versus cosmic shear.} The effect of the coupling on the amplitude of the fluctuations goes in the
opposite direction. In \LCDM, \Pl\ predicts $S_8^{\rm eff}=S_8=\SeffAL$ against $\SWL$ for the combined shear data,
a $\QDMLWL\sigma$ difference, and the other estimators agree ($\SLWL$--$\shiftLWL\sigma$). In the frozen model, \Pl\
alone predicts $S_8^{\rm eff}=\SeffAF$, and the tension rises to $\SFWL$--$\QDMAPFWL\sigma$, with
$\Delta{\rm DIC}=\dDICFWL$. The same coupling that relieves the disagreement with DESI therefore increases the one
with cosmic shear.

\paragraph{Consistency of the estimators.} The estimators agree with each other to $0.2\sigma$ in \LCDM\ for
every pair. In the frozen model the spread is larger, up to $0.45\sigma$ for \Pl\ versus \DESI, because the
posteriors are not Gaussian and $Q_{\rm DMAP}$ uses the best fit at the edge of the allowed region. The Bayesian
model dimensionality of the \DK\ and \DESI\ comparisons is $d=\BMDFDov$ and $\BMDFDESI$ in the frozen model, close to
the number of parameters that these data constrain, and the suspiciousness agrees with the parameter-based
estimators. The evidence ratio is positive in all cases, $\ln R=\lnRFDov$--$\lnRFDESI$, but it grows with the prior
volume of the parameters that the external data do not constrain, which is why the suspiciousness is the relevant
evidence-based estimator. None of the pairs exceeds $3\sigma$ in either model, so all combinations are allowed.

\section{Joint constraints}
\label{sec:joint}

\begin{table*}[tbp]
\caption{Marginalized constraints (mean and standard deviation) on the parameters of the frozen model for each
data combination. The posterior of $\Sigma$ is asymmetric for the combinations without shear; its $95\%$ intervals
are given in the text. $S_8^{\rm eff}$ is the amplitude seen by the KiDS-Legacy shear kernel,
Eq.~(\ref{eq:s8eff}). $H_0$ in $\kms$. Columns: \Pl\ alone and \Pl\ combined with DES-Dovekie (Dov.), \DESI,
both, cosmic shear, or all of them. The last three rows give the corresponding \LCDM\ values.}
\label{tab:constraints}
\begin{ruledtabular}
\footnotesize
\begin{tabular}{lcccccc}
Parameter & \Pl & $+$Dov. & $+$DESI & $+$Dov.$+$DESI & $+$shear & $+$all \\
\hline
$\Sigma$ & $-0.216\pm0.113$ & $-0.204\pm0.121$ & $-0.244\pm0.084$ & $-0.242\pm0.088$ & $0.066\pm0.099$ & $-0.015\pm0.069$ \\
$\eta_0$ & $0.43\pm0.30$ & $0.47\pm0.31$ & $0.32\pm0.23$ & $0.33\pm0.24$ & $1.17\pm0.26$ & $0.96\pm0.19$ \\
$\sigma_8$ & $0.953\pm0.099$ & $0.941\pm0.099$ & $0.985\pm0.095$ & $0.981\pm0.096$ & $0.791\pm0.026$ & $0.813\pm0.023$ \\
$S_8$ & $0.972\pm0.099$ & $0.965\pm0.099$ & $0.985\pm0.094$ & $0.984\pm0.096$ & $0.804\pm0.021$ & $0.815\pm0.022$ \\
$S_8^{\rm eff}$ & $0.876\pm0.028$ & $0.878\pm0.029$ & $0.866\pm0.025$ & $0.868\pm0.025$ & $0.812\pm0.010$ & $0.811\pm0.010$ \\
$\Omega_{\rm m}$ & $0.3119\pm0.0075$ & $0.3156\pm0.0067$ & $0.3001\pm0.0037$ & $0.3018\pm0.0036$ & $0.3107\pm0.0077$ & $0.3015\pm0.0037$ \\
$H_0$ & $67.60\pm0.55$ & $67.33\pm0.49$ & $68.48\pm0.29$ & $68.34\pm0.28$ & $67.69\pm0.57$ & $68.36\pm0.29$ \\
$hr_{\rm d}$ [Mpc] & $99.51\pm0.95$ & $99.04\pm0.84$ & $101.05\pm0.49$ & $100.82\pm0.47$ & $99.67\pm0.99$ & $100.87\pm0.49$ \\
$n_{\rm s}$ & $0.9658\pm0.0042$ & $0.9644\pm0.0039$ & $0.9707\pm0.0034$ & $0.9699\pm0.0034$ & $0.9663\pm0.0041$ & $0.9698\pm0.0032$ \\
$\ln(10^{10}A_{\rm s})$ & $3.038\pm0.014$ & $3.036\pm0.014$ & $3.044\pm0.014$ & $3.043\pm0.014$ & $3.043\pm0.015$ & $3.047\pm0.015$ \\
$\tau$ & $0.0520\pm0.0073$ & $0.0507\pm0.0070$ & $0.0568\pm0.0070$ & $0.0560\pm0.0070$ & $0.0547\pm0.0074$ & $0.0579\pm0.0072$ \\
\hline
$S_8$ (\LCDM) & $0.831\pm0.013$ & $0.835\pm0.012$ & $0.809\pm0.009$ & $0.812\pm0.008$ & $0.816\pm0.008$ & $0.809\pm0.007$ \\
$\Omega_{\rm m}$ (\LCDM) & $0.3154\pm0.0075$ & $0.3183\pm0.0069$ & $0.3012\pm0.0038$ & $0.3030\pm0.0036$ & $0.3076\pm0.0052$ & $0.3022\pm0.0033$ \\
$H_0$ (\LCDM) & $67.35\pm0.54$ & $67.14\pm0.49$ & $68.39\pm0.30$ & $68.26\pm0.28$ & $67.91\pm0.40$ & $68.32\pm0.27$ \\
\end{tabular}
\end{ruledtabular}
\end{table*}

\subsection{Planck, DES-Dovekie and DESI}

Since none of the comparisons of Sec.~\ref{sec:tensions} exceeds $3\sigma$, we combine the three data sets. For
\Pl+\DK+\DESI\ the coupling is $\Sigma=\SigAEH$, with the $95\%$ interval $[\SigLoAEH,\SigHiAEH]$, which excludes
$\Sigma=0$ (Table~\ref{tab:constraints}). The geometric data fix the matter density and the expansion rate,
$\Omega_{\rm m}=\OmAEH$ and $H_0=\HzeroAEH\kms$, which coincide with \LCDM\ ($\OmLAEH$ and $\HzeroLAEH\kms$), as they
must for a model with the same background. Figure~\ref{fig:h0om} shows how the four combinations without shear
arrange themselves along the degeneracy direction of the CMB, on which $\Omega_{\rm m}h^3$ is nearly constant. The
supernovae prefer a higher matter density and a lower expansion rate, $H_0=\HzeroAE\kms$ in the frozen model,
and DESI the opposite, $H_0=\HzeroAH\kms$; the combination of the three settles on an intermediate point that
is the same in both models. None of these values approaches the local distance-ladder measurements, a point to
which we return in Sec.~\ref{sec:late}. The coupling is again pushed toward the boundary, $\eta_0=\etaAEH$,
because DESI selects the region of lower $\Omega_{\rm m}$ in which \Pl\ prefers more lensing, and the amplitude of
the fluctuations becomes large: $\sigma_8=\seightAEH$ and $S_8=\SeightAEH$, against $S_8=\SeightLAEH$ in \LCDM.

The best fit improves the $\chi^2$ by $\dchiAEH$ with respect to \LCDM, distributed over all the CMB likelihoods
(high-$\ell$ $\dchiplikAEH$, low-$\ell$ TT $\dchilowTTAEH$, low-$\ell$ EE $\dchilowEEAEH$, lensing $\dchilensAEH$) and
DESI ($\dchidesiAEH$), with a small cost for the supernovae ($+\dchidovAEH$). The Bayes factor is $\ln B=\lnBAEH$
(Table~\ref{tab:models}), a weak preference on the Jeffreys scale once the Occam penalty of the prior range of
$\Sigma$ is included. This preference rests on the region close to the boundary: with $\eta_{\rm min}=0.2$ it
drops to $\ln B=\lnBcutbAEH$ and with $\eta_{\rm min}=0.5$ it vanishes, $\ln B=\lnBcutcAEH$, while the coupling
moves to $\Sigma=\SigcutcAEH$.

\begin{table}[tbp]
\caption{Model comparison. Best-fit $\chi^2$ of the data in \LCDM\ and in the frozen model, their difference,
and the Bayes factor $\ln B=\ln Z_{\rm frozen}-\ln Z_{\Lambda{\rm CDM}}$, whose Monte Carlo error is $0.03$--$0.05$.
The last three columns give $\ln B$ when the physical cut is raised to $\eta_{\rm min}=0.1$, $0.2$ and $0.5$.
Data: \Pl\ alone and \Pl\ combined with DES-Dovekie (Dov.), \DESI, cosmic shear, or all of them.}
\label{tab:models}
\begin{ruledtabular}
\footnotesize
\begin{tabular}{lcccccc}
Data & $\chi^2_{\Lambda}$ & $\Delta\chi^2$ & $\ln B$ & \multicolumn{3}{c}{$\ln B(\eta_{\rm min})$} \\
 & & & & $0.1$ & $0.2$ & $0.5$ \\
\hline
\Pl & $1011.1$ & $-4.1$ & $0.72$ & $0.63$ & $0.41$ & $-0.35$ \\
$+$Dov. & $2643.4$ & $-3.4$ & $0.52$ & $0.45$ & $0.27$ & $-0.41$ \\
$+$DESI & $1026.9$ & $-7.0$ & $1.64$ & $1.48$ & $1.15$ & $0.01$ \\
$+$Dov.$+$DESI & $2661.9$ & $-6.6$ & $1.49$ & $1.33$ & $1.02$ & $-0.05$ \\
$+$shear & $1014.2$ & $-0.1$ & $-1.15$ & $-1.15$ & $-1.15$ & $-1.15$ \\
$+$all & $2662.7$ & $-0.1$ & $-1.42$ & $-1.42$ & $-1.42$ & $-1.43$ \\
\end{tabular}
\end{ruledtabular}
\end{table}

\subsection{Adding cosmic shear}

The amplitude favored by \Pl+\DK+\DESI\ is larger than the one measured by the shear surveys. The amplitude relevant for the
shear kernels, $S_8^{\rm eff}=\SeffAEH$, is smaller than $S_8$ because the growth enhancement is concentrated at
$z\lesssim0.5$ (Fig.~\ref{fig:sigma}b), but it is still $2.4\sigma$ above the combined shear measurement, $\SWL$.
At the best fit of \Pl+\DK+\DESI\ the two shear likelihoods would add $\Delta\chi^2=\dchikidsAEH$ (KiDS-Legacy) and
$\dchidesyAEH$ (DES Year 6) with respect to \LCDM.

With cosmic shear included, the coupling is consistent with zero: $\Sigma=\SigAW$ for \Pl+shear and
$\Sigma=\SigAEHW$ for the full combination, with the $95\%$ interval $[\SigLoAEHW,\SigHiAEHW]$, and
$S_8=\SeightAEHW$. The region near the boundary is excluded, so these constraints no longer depend on
$\eta_{\rm min}$. The best fit of the full combination has $\Sigma=\bfSigAEHW$ and the same $\chi^2$ as \LCDM\
($\Delta\chi^2=\dchiAEHW$), and the Bayes factor, $\ln B=\lnBAEHW$, now favors \LCDM: the data constrain the
coupling but do not require it. The reduction of the tension with DESI found in Sec.~\ref{sec:tensions}
is therefore not retained once the shear amplitude is taken into account, since both effects come from the same
enhancement of the late growth.

This conclusion does not depend on the choice of the shear data. Table~\ref{tab:altshear} repeats the analysis of the
full combination with two alternatives: DES Year 3 in place of DES Year 6, and the joint analysis of KiDS-Legacy and DES Year 3 of Ref.~\cite{Stolzner2025}, in which the two
surveys are analyzed with a common pipeline and which gives a higher amplitude than the product of the two
individual results. Because every likelihood is evaluated at every draw (Sec.~\ref{sec:sampling}), each alternative
is an exact reweighting of the same samples, with effective sample sizes between $550$ and $760$. The coupling moves
by less than one standard deviation, from $\Sigma=0.01\pm0.08$ to $-0.08\pm0.07$; it is consistent with zero in the
three cases, and the Bayes factor favors \LCDM\ in all of them.

\begin{table}[tbp]
\caption{Dependence of the results of the full combination, \Pl+\DK+\DESI+shear, on the cosmic-shear data: mean and
standard deviation of the coupling and of $S_8$ in the frozen model, and Bayes factor of the frozen model relative to
\LCDM. The first row is the baseline of this work, and the last row uses the joint analysis of
Ref.~\cite{Stolzner2025}. The $95\%$ intervals of the coupling are $[\SigLoAEHW,\SigHiAEHW]$, $[-0.14,0.17]$ and
$[-0.19,0.06]$, and in \LCDM\ the three choices give $S_8=0.809\pm0.007$, $0.808\pm0.007$ and $0.814\pm0.007$.}
\label{tab:altshear}
\begin{ruledtabular}
\begin{tabular}{lccc}
Shear data & $\Sigma$ & $S_8$ & $\ln B$\\
\hline
KiDS-Legacy, DES Y6 & $-0.01\pm0.07$ & $0.815\pm0.022$ & $-1.4$\\
KiDS-Legacy, DES Y3 & $0.01\pm0.08$ & $0.808\pm0.025$ & $-1.3$\\
Joint analysis & $-0.08\pm0.07$ & $0.842\pm0.026$ & $-0.8$\\
\end{tabular}
\end{ruledtabular}
\end{table}

\section{Update with Planck PR4}
\label{sec:pr4}

The analysis of the previous sections uses the Planck 2018 likelihoods. In this section we repeat it with the latest
Planck data release, PR4, in order to test how the preference for a negative coupling depends on the CMB likelihood.
The data, the treatment of the nuisance parameters and the sampling are described first, and the results are then
compared with those of Planck 2018.

\subsection{Data and nuisance parameters}

The low-multipole temperature and $E$-mode likelihoods are those of Sec.~\ref{sec:data}, as in the PR4 analyses. At
high multipoles we use the PR4 temperature and polarization spectra in the unbinned form of
Ref.~\cite{Rosenberg2022}: TT at $143\times143$ ($30\le\ell\le2000$), $217\times217$ and $143\times217$
($500\le\ell\le2500$), and TE and EE ($30\le\ell\le2000$), a total of $9915$ data points. The lensing reconstruction
is that of PR4~\cite{Carron2022}, in the same conservative range of lensing multipoles. We denote this combination by
PR4.

In contrast with the compressed likelihood of Sec.~\ref{sec:data}, this likelihood is not marginalized over the
foregrounds. Besides the overall calibration $A_{\rm P}$, it has eight nuisance parameters: the amplitudes $A_i$ and
the tilts $\gamma_i$ of a power-law residual, $A_i(\ell/1500)^{\gamma_i}$, in each of the three temperature spectra,
with uniform priors $A_i\in[0,50]\,\mu{\rm K}^2$ and $\gamma_i\in[0,5]$, and the polarization calibrations
$c_{TE}$ and $c_{EE}$, with Gaussian priors $1\pm0.01$. We marginalize over the eight parameters at every
cosmological point. The model vector is linear in $A_i$, $1/c_{TE}$ and $1/c_{EE}$ at fixed tilts, so the likelihood
is Gaussian in these five quantities. We integrate the three amplitudes analytically, the two calibrations on a
$21\times21$ grid centered on their posterior with a spacing of $0.6$ standard deviations, and the three tilts on a
grid of $34\times32\times36$ nodes that covers their priors, except for the tilt of the $143\times143$ spectrum,
whose grid ends at $\gamma=3$ because its posterior vanishes above that value. The result agrees with the direct evaluation of the
likelihood to $10^{-4}$ in $\chi^2$ and is stable to $4\times10^{-4}$ in $\ln\mathcal{L}$ when the number of nodes is
reduced by a third. The prior bounds of the amplitudes are more than five standard deviations away for every
point within $\Delta\ln\mathcal{L}=40$ of the maximum.

\subsection{Sampling}

The posteriors are obtained with the adaptive importance sampling of Sec.~\ref{sec:sampling}, in two stages. In the
first stage, all the draws of the Planck 2018 analysis ($3298$ for \LCDM\ and $5966$ for the frozen model) are evaluated with the PR4 likelihoods, which is an exact reweighting. The PR4 posterior is narrower
than that of Planck 2018 and shifted with respect to it, so the effective sample size of this stage is about ten
times smaller than in the baseline analysis. In the second stage we therefore draw new points ($2000$ for \LCDM\ and $3500$ for the
frozen model) from mixtures of Student-$t$ distributions fitted to the PR4-weighted samples of each data
combination, and we evaluate all likelihoods at them. The final weights
use the deterministic mixture of the proposals of all rounds and stages, Eq.~(\ref{eq:dmw}), and the evidence is the
mean weight. With the two stages the effective sample size of the frozen model lies between $\prESSFPW$ and $\prESSFP$ and the
Pareto diagnostic is below $0.2$ for all combinations (Table~\ref{tab:pr4}), so the estimates are as reliable as
those of the baseline analysis. The
unbinned likelihood is sensitive to the numerical precision of the Boltzmann solution: the same point computed on two
different machines differs by $0.2$ in $\ln\mathcal{L}$ (root mean square), a noise that is independent of the
parameters and lowers the effective sample size by a few percent.

\subsection{Results}

\begin{table*}[tbp]
\caption{Constraints with Planck PR4 in place of Planck 2018: mean and standard deviation of the coupling and of
$S_8$ in the frozen model, $S_8$ in \LCDM, Bayes factor of the frozen model relative to \LCDM, and effective sample
size $N_{\rm eff}$ and Pareto diagnostic $\hat k$ of the frozen-model samples. }
\label{tab:pr4}
\begin{ruledtabular}
\begin{tabular}{lcccccc}
Data & $\Sigma$ & $S_8$ & $S_8$ (\LCDM) & $\ln B$ & $N_{\rm eff}$ & $\hat k$\\
\hline

PR4 & $\prSigP$ & $\prSeightFP$ & $\prSeightLP$ & $\prlnBP$ & $\prESSFP$ & $\prkFP$\\
PR4+\DK+\DESI & $\prSigPEH$ & $\prSeightFPEH$ & $\prSeightLPEH$ & $\prlnBPEH$ & $\prESSFPEH$ & $\prkFPEH$\\
PR4+shear & $\prSigPW$ & $\prSeightFPW$ & $\prSeightLPW$ & $\prlnBPW$ & $\prESSFPW$ & $\prkFPW$\\
PR4+all & $\prSigPEHW$ & $\prSeightFPEHW$ & $\prSeightLPEHW$ & $\prlnBPEHW$ & $\prESSFPEHW$ & $\prkFPEHW$\\

\end{tabular}
\end{ruledtabular}
\end{table*}

Table~\ref{tab:pr4} summarizes the constraints. In \LCDM\ the PR4 results are close to those of Planck 2018:
$S_8=\prSeightLP$ for PR4 alone, against $\SeightLA$, and $H_0=\prHLPEHW\kms$ for the full combination, against
$\HzeroLAEHW\kms$.

In the frozen model the preference for a negative coupling is weaker than with Planck 2018. PR4 alone gives
$\Sigma=\prSigP$, against $\SigA$, and the posterior probability of $\Sigma<0$ is $\prPnegP$. The Bayes factor
relative to \LCDM\ decreases from $\ln B=\lnBvA$ to $\prlnBP$, so that PR4 alone does not distinguish between the
two models. With the supernovae and the BAO, which constrain the coupling only through the other parameters,
$\Sigma=\prSigPEH$ and $\ln B=\prlnBPEH$, against $\SigAEH$ and $\lnBvAEH$ with Planck 2018. This is the behavior
expected from the origin of the preference (Sec.~\ref{sec:planck}): the smoothing of the acoustic peaks in excess of
the \LCDM\ prediction, which a negative coupling reproduces through the enhanced lensing, is less pronounced in the
PR4 spectra~\cite{Rosenberg2022}.

The amplitude seen by the shear surveys remains larger than the measured one. PR4 alone differs from the combined shear measurement by
$\prWL\sigma$ in \LCDM\ and by $\prWF\sigma$ in the frozen model, and with the shear data included the coupling
returns to zero: $\Sigma=\prSigPW$ for PR4+shear and $\Sigma=\prSigPEHW$ for the full combination, with the $95\%$
interval $[\prSigLoPEHW,\prSigHiPEHW]$, $S_8=\prSeightFPEHW$ and $\ln B=\prlnBPEHW$. These values coincide with those of
Sec.~\ref{sec:joint} within a small fraction of their errors. The change of the CMB likelihood therefore weakens
the preference for a negative coupling without shear and leaves the results with shear unchanged: once cosmic shear
is included, the coupling is consistent with zero for both releases.

\section{Late-time observables}
\label{sec:late}

The constraints of the previous sections determine, for each data combination, a prediction for observables that
were not used in the fits. Comparing these predictions with independent measurements is a useful complement to the
tension estimators: it shows directly which features of the data each model can or cannot reproduce, and it
identifies the observables that could decide between them in the future. In the figures of this section the
predictions are shown as boxes that span the $68\%$ credible interval, with whiskers at $95\%$ and a line at the
median, for the six data combinations and for both models; the measurements are shown with their $68\%$ errors.

\subsection{Expansion history}

\begin{figure}[tbp]
\includegraphics[width=\columnwidth]{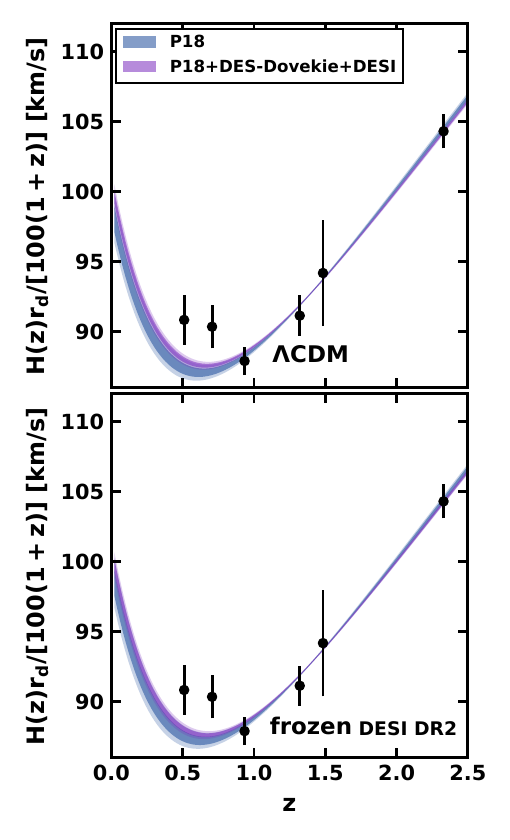}
\caption{Posterior of $H(z)r_{\rm d}/[100(1+z)]$ (68\% and 95\% bands) for \Pl\ and \Pl+\DK+\DESI\ in \LCDM\
(top) and in the frozen model (bottom), compared with the DESI DR2 measurements of $D_H/r_{\rm d}$ (points).}
\label{fig:hz}
\end{figure}

Because the two models share the same background, their predictions for the expansion rate differ only through the
values of $H_0$, $\omega_{\rm b}$ and $\omega_{\rm c}$ preferred by each combination. Figure~\ref{fig:hz} shows
the posterior of $H(z)r_{\rm d}$, the combination measured by the radial BAO, scaled by $1/[100(1+z)]$ for
clarity. The bands of the two models are nearly identical: for \Pl\ the frozen model is shifted upward by $0.2$--$0.4\%$ at
$z\lesssim0.5$, a consequence of its slightly lower matter density, and for \Pl+\DK+\DESI\ the two models coincide
to within $0.2\%$. The DESI points at $z=0.51$ and $0.71$ lie about $1.5\sigma$ above both
predictions, while the others agree well, in line with the moderate tension discussed in
Sec.~\ref{sec:tensions}. Beyond these small parameter shifts, a model that changes only the growth cannot modify the
expansion history, and the frozen model leaves untouched the features that motivate a dynamical dark energy in the
DESI data.

\subsection{Hubble constant}

\begin{figure}[tbp]
\includegraphics[width=\columnwidth]{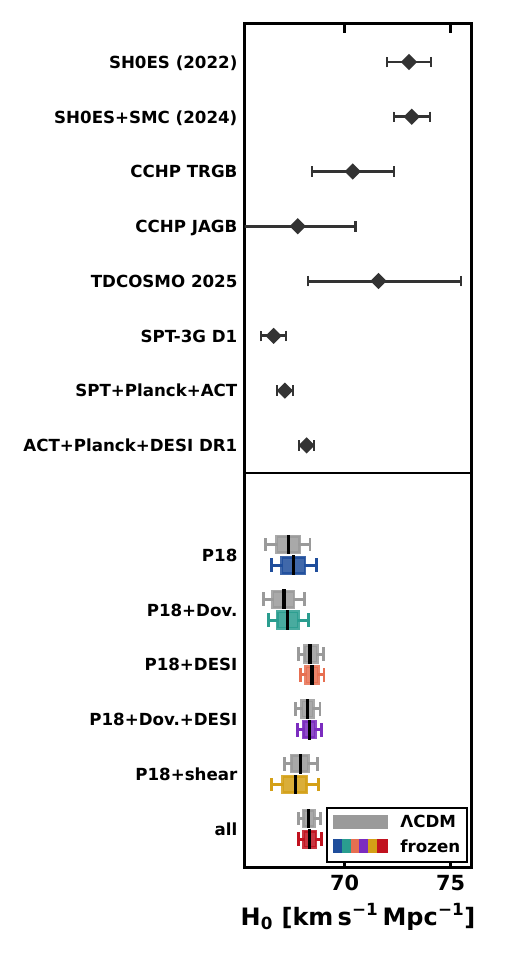}
\caption{Hubble constant. Top: measurements from the distance ladder~\cite{Riess2022,Breuval2024}, from the
Chicago-Carnegie program~\cite{Freedman2025}, from strong-lensing time delays~\cite{TDCOSMO2025}, and from recent
CMB analyses in \LCDM~\cite{Camphuis2025,Louis2025}, with their $68\%$ errors. Bottom: posteriors of \LCDM\ (gray)
and of the frozen model (colored by data combination, as in Fig.~\ref{fig:sigma}) for the six data combinations; boxes enclose the $68\%$ intervals, whiskers the
$95\%$ intervals, and the vertical line marks the median.}
\label{fig:h0box}
\end{figure}

The same reasoning applies to the Hubble constant, shown in Fig.~\ref{fig:h0box} together with a selection of
recent measurements. Both models predict $H_0=67.1$--$68.5\kms$ for every data combination, and the frozen model
moves the \Pl\ value by only $+0.25\kms$ with respect to \LCDM. The distance-ladder measurements of the SH0ES team,
$73.04\pm1.04\kms$~\cite{Riess2022} and $73.17\pm0.86\kms$ with the additional anchor of the Small Magellanic
Cloud~\cite{Breuval2024}, therefore remain $4.4\sigma$ and $5.3$--$5.4\sigma$ above the prediction of \Pl+\DK+\DESI\ in
both models. The tip-of-the-red-giant-branch and J-region asymptotic giant branch calibrations of the Chicago-Carnegie
program with the James Webb Space Telescope, $70.39\pm1.94$ and $67.80\pm2.72\kms$~\cite{Freedman2025}, and the
strong-lensing time delays of TDCOSMO, $71.6^{+3.9}_{-3.3}\kms$~\cite{TDCOSMO2025}, are compatible with both
models within $1.5\sigma$. Among the CMB-based determinations, SPT-3G alone, $66.66\pm0.60\kms$, and its
combination with Planck and ACT, $67.19\pm0.38\kms$~\cite{Camphuis2025}, agree with our \Pl\ predictions, while the
combination of ACT, Planck and CMB lensing with DESI DR1, $68.22\pm0.36\kms$~\cite{Louis2025}, agrees with our
\Pl+\DK+\DESI\ prediction; in \LCDM-like backgrounds the BAO raise the CMB inference of $H_0$ by about
$1\kms$. The small effect of the coupling on $H_0$ is consistent with the way it reduces the tension with DESI: the
sound horizon depends only on $\omega_{\rm b}$ and $\omega_{\rm c}$, which the coupling leaves untouched, and the
shift of the \Pl\ posterior toward DESI, from $hr_{\rm d}=\hrdAL$ to $\hrdAF$~Mpc, is a small displacement along the
CMB degeneracy direction, with a slightly larger $h$ and a slightly lower matter density (Fig.~\ref{fig:h0om}). The frozen model was not designed to address the
Hubble tension, and it does not; any model that leaves the expansion history and the sound horizon untouched
inherits the \LCDM\ value of $H_0$ from the CMB, whatever it does to the growth of structure. Its interest lies in
the growth sector, to which we now turn after a brief look at the age of the Universe.

\subsection{Age of the Universe}

\begin{figure}[tbp]
\includegraphics[width=\columnwidth]{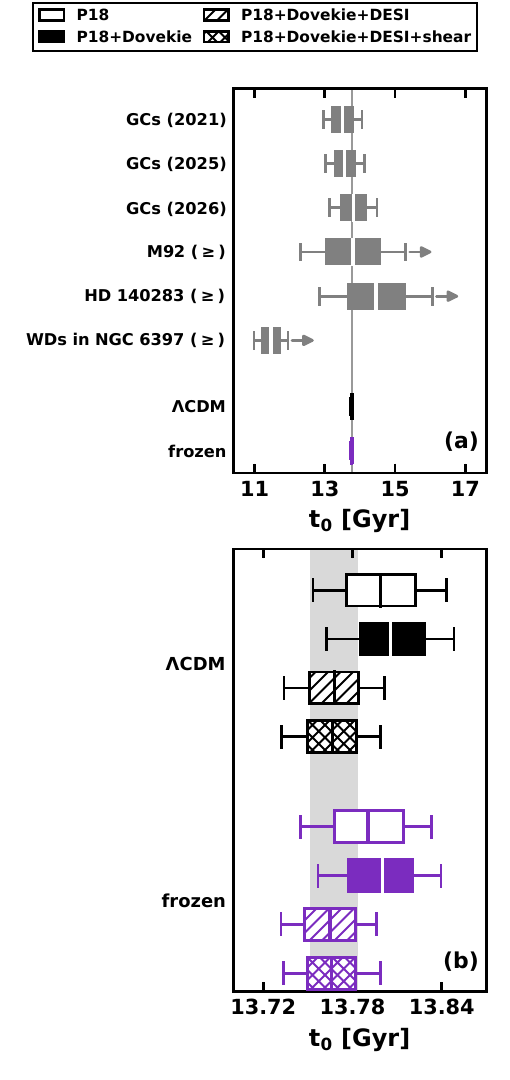}
\caption{Age of the Universe. Boxes enclose the $68\%$ intervals, whiskers the $95\%$ intervals, and the vertical
line marks the median. (a) Ages of the Universe inferred from globular clusters~\cite{Valcin2021,Valcin2025,Valcin2026}
and ages of the cluster M92~\cite{Ying2023}, of the subgiant HD~140283~\cite{Bond2013} and of the white dwarfs of
NGC~6397~\cite{Hansen2007}, which give lower bounds (gray; statistical and systematic errors added in quadrature),
compared with the posteriors of \LCDM\ and of the frozen model for \Pl+\DK+\DESI. (b) Posteriors for \Pl\ (open
boxes), \Pl+\DK\ (filled), \Pl+\DK+\DESI\ (hatched) and the full combination with shear (cross-hatched). The gray
band is the $68\%$ interval of \LCDM\ for \Pl+\DK+\DESI.}
\label{fig:agebox}
\end{figure}

The age of the Universe is another quantity fixed by the background, and its comparison with the oldest objects is
a classic consistency test of the cosmological model. Figure~\ref{fig:agebox} and Table~\ref{tab:age} summarize the
results. All models and data combinations give an age of about $13.8$~Gyr, with an uncertainty of $0.02$~Gyr and
posterior means between $\ageMin$ and $\ageMax$~Gyr; adding DESI lowers the age by $\ageShiftDESI$~Gyr through the larger $H_0$, and the
frozen model and \LCDM\ differ by less than $0.01$~Gyr for the same data. The ages of the Universe inferred from
the oldest globular clusters, $13.5^{+0.28}_{-0.27}$~Gyr~\cite{Valcin2021}, $13.57^{+0.28}_{-0.27}$~Gyr with Gaia
distances~\cite{Valcin2025} and $13.81\pm0.34$~Gyr with multiple stellar populations~\cite{Valcin2026}
(statistical and systematic errors added in quadrature), agree with the prediction to better than $1.1\sigma$. The
ages of individual objects provide lower bounds: the globular cluster M92, $13.80\pm0.75$~Gyr~\cite{Ying2023}, and
the white-dwarf cooling sequence of NGC~6397, $11.47\pm0.24$~Gyr~\cite{Hansen2007}, are compatible,
while the metal-poor subgiant HD~140283, $14.46\pm0.8$~Gyr~\cite{Bond2013}, lies $0.8$--$0.9\sigma$ above the
prediction, a mild excess well within its uncertainty. The table
also shows that the frozen model leaves the linear structure at $z=10$ essentially untouched, because the exchange
becomes effective only at late times; any change in the abundance of the first halos is due to the shift of the
parameters shared with \LCDM.

\begin{table*}[tbp]
\caption{Best fits of \LCDM\ and of the frozen model: Hubble constant (in ${\rm km\,s^{-1}\,Mpc^{-1}}$), matter density, amplitude of the fluctuations, age of the Universe $t_0$, the dimensionless product $H_0t_0$, and linear structure at high redshift: the age at $z=10$, the rms linear fluctuation $\sigma_{10}$ of the total matter at $z=10$ in spheres containing $10^{10}\,M_\odot$, relative to the \LCDM\ best fit of \Pl, and the factors $\mathcal{B}_{10}$ and $\mathcal{B}_{11}$ by which the abundance of collapsed halos of $10^{10}$ and $10^{11}\,M_\odot$ at $z=10$ changes in the Press--Schechter approximation, with the linear collapse threshold of \LCDM. The best fits illustrate how the coupling changes these derived
quantities; the marginalized constraints on the parameters are given in Table~\ref{tab:constraints}, the posterior
of $t_0$ is shown in Fig.~\ref{fig:agebox}, and the column $\langle t_0\rangle$ gives its posterior mean and
standard deviation.}
\label{tab:age}
\begin{ruledtabular}
\begin{tabular}{llccccccccccc}
Model & Data & $H_0$ & $\Omega_{\rm m}$ & $\sigma_8$ & $S_8$ & $t_0$ [Gyr] & $\langle t_0\rangle$ [Gyr] & $H_0t_0$ & $t(z{=}10)$ [Myr] & $\sigma_{10}$ & $\mathcal{B}_{10}$ & $\mathcal{B}_{11}$\\
\hline
\LCDM & \Pl & 67.34 & 0.315 & 0.810 & 0.831 & 13.799 & $13.80\pm0.02$ & 0.950 & 470 & 1.000 & 1.00 & 1.00\\
\LCDM & \Pl+\DK & 67.13 & 0.318 & 0.811 & 0.835 & 13.806 & $13.81\pm0.02$ & 0.948 & 470 & 1.001 & 1.02 & 1.03\\
\LCDM & \Pl+\DESI & 68.41 & 0.301 & 0.807 & 0.809 & 13.763 & $13.76\pm0.02$ & 0.963 & 474 & 0.994 & 0.92 & 0.85\\
\LCDM & \Pl+\DK+\DESI & 68.28 & 0.303 & 0.808 & 0.811 & 13.768 & $13.77\pm0.02$ & 0.961 & 474 & 0.995 & 0.93 & 0.87\\
\LCDM & \Pl+shear & 67.92 & 0.307 & 0.805 & 0.815 & 13.780 & $13.78\pm0.02$ & 0.957 & 472 & 0.992 & 0.90 & 0.83\\
\LCDM & all & 68.34 & 0.302 & 0.806 & 0.809 & 13.766 & $13.77\pm0.02$ & 0.962 & 474 & 0.993 & 0.90 & 0.82\\
\hline
frozen & \Pl & 67.79 & 0.309 & 1.097 & 1.113 & 13.784 & $13.79\pm0.02$ & 0.956 & 472 & 0.991 & 0.89 & 0.81\\
frozen & \Pl+\DK & 67.47 & 0.314 & 1.080 & 1.104 & 13.795 & $13.80\pm0.02$ & 0.952 & 471 & 0.993 & 0.92 & 0.86\\
frozen & \Pl+\DESI & 68.53 & 0.299 & 1.115 & 1.114 & 13.759 & $13.76\pm0.02$ & 0.964 & 474 & 0.987 & 0.83 & 0.71\\
frozen & \Pl+\DK+\DESI & 68.39 & 0.301 & 1.112 & 1.114 & 13.764 & $13.77\pm0.02$ & 0.963 & 474 & 0.987 & 0.84 & 0.73\\
frozen & \Pl+shear & 67.76 & 0.310 & 0.795 & 0.808 & 13.786 & $13.79\pm0.02$ & 0.955 & 472 & 0.995 & 0.93 & 0.88\\
frozen & all & 68.40 & 0.301 & 0.819 & 0.821 & 13.764 & $13.77\pm0.02$ & 0.963 & 474 & 0.990 & 0.87 & 0.78\\
\end{tabular}
\end{ruledtabular}
\end{table*}

\subsection{Integrated Sachs--Wolfe effect}

\begin{figure}[tbp]
\includegraphics[width=\columnwidth]{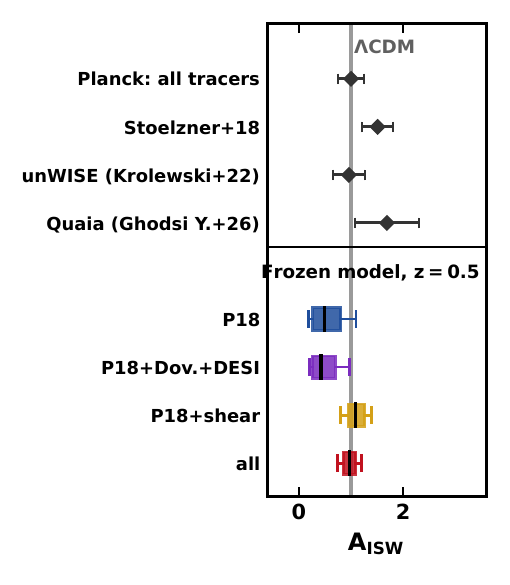}
\caption{ISW amplitude relative to \LCDM. Top: four representative measurements from the cross-correlation of the
CMB temperature with tracers of the large-scale
structure~\cite{Planck2016ISW,Stolzner2018,Krolewski2022,Ghodsi2026}, with their $68\%$ errors; the complete list is
given in the text. Bottom: posterior predictive of the bias-independent amplitude, Eq.~(\ref{eq:aisw}), in the frozen
model at $z=0.5$ for four data combinations (boxes: $68\%$; whiskers: $95\%$). The vertical gray line is \LCDM, for
which the amplitude is unity by construction. Without shear data the frozen model predicts about half of the \LCDM\
signal; with shear data the predictions return to unity. The measurements share tracers and sky areas and are
therefore correlated with each other.}
\label{fig:iswbox}
\end{figure}

The enhanced growth for $\Sigma<0$ slows the decay of the gravitational potentials, and the ISW signal of the
large-scale structure is reduced. Figure~\ref{fig:iswbox} compares the posterior predictive of the
bias-independent ISW amplitude, Eq.~(\ref{eq:aisw}), at $z=0.5$, with the amplitudes measured by
cross-correlating the CMB temperature with tracers of the large-scale structure. These measurements are
normalized to the \LCDM\ prediction and calibrate the galaxy bias with the auto-correlation of the tracers, so to a
good approximation they measure the same ratio of decay rates~\cite{Zhang2006,Dong2022}. For \Pl+\DK+\DESI\ the
frozen model predicts $A_{\rm ISW}=\iswbAEH$ at $z=0.5$ and $\iswdAEH$ at $z=1.5$, and a sign change at
$z\lesssim0.15$ for the values of $\Sigma$ closest to the boundary; with the shear data the prediction is
$A_{\rm ISW}=\iswbAEHW$, indistinguishable from \LCDM, while \Pl+shear alone, which allows slightly positive
couplings, predicts $\iswbAW$.

Most of these measurements are consistent with \LCDM: the combination of all tracers in the Planck 2015 analysis
gives $1.00\pm0.25$~\cite{Planck2016ISW}, the unWISE galaxies $0.96\pm0.30$~\cite{Krolewski2022}, the DESI
Legacy Imaging Survey $0.98\pm0.35$~\cite{Hang2021}, the RACS radio galaxies
$0.94^{+0.42}_{-0.41}$~\cite{BahrKalus2022} and the Quaia quasars $1.69\pm0.61$~\cite{Ghodsi2026}, while earlier
combined analyses found somewhat larger values, $1.38\pm0.32$~\cite{Giannantonio2012} and
$1.51\pm0.30$~\cite{Stolzner2018}. Taken one at a time, they lie $1.0$--$1.6\sigma$ (the recent single-tracer
measurements and Planck) and $2.4$--$2.8\sigma$ (the earlier combinations) above the prediction of
\Pl+\DK+\DESI\ in the frozen model, and the Quaia measurement lies $1.7\sigma$ above the prediction at $z=1.5$.
Since the tracers overlap, these numbers cannot be combined naively, but they consistently point in the same
direction as cosmic shear: the strong-coupling region favored by the CMB and BAO alone predicts less ISW
signal than is measured. The excess ISW imprint reported for the largest supervoids, $A_{\rm ISW}=4.1\pm2.0$~\cite{Kovacs2019},
is not accounted for by the model either: negative values of $\Sigma$ reduce the imprint, and the positive values
allowed by the shear data raise it by at most about $40\%$. A tomographic measurement of the bias-independent ratio
with a precision of $0.2$, within the reach of the combination of CMB maps with DESI and Euclid samples, would test
the region of strong coupling independently of the shear amplitude and of the modeling of nonlinear scales.

\subsection{Amplitude of the fluctuations}

\begin{figure}[tbp]
\includegraphics[width=\columnwidth]{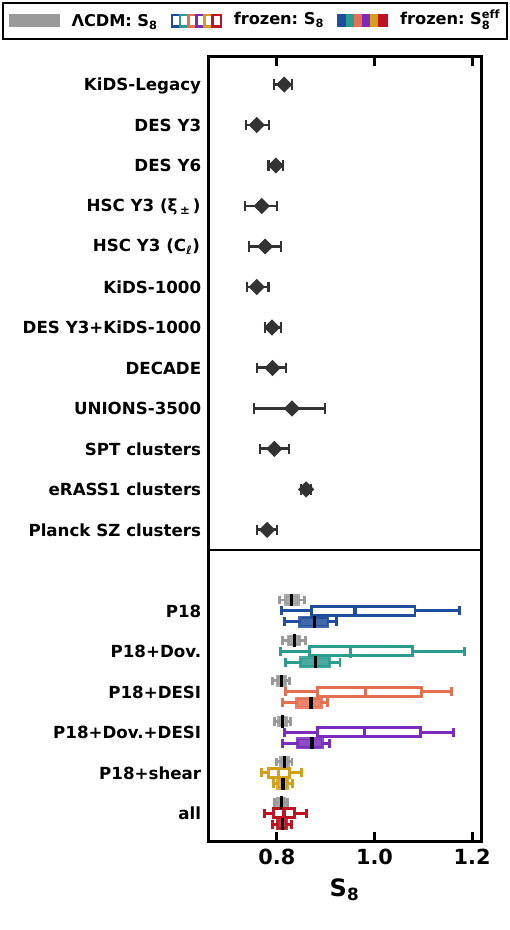}
\caption{Amplitude of the matter fluctuations. Top: measurements from cosmic
shear~\cite{KiDSLegacy2025,Amon2022,DESY6shear2026,Li2023HSC,Dalal2023,Asgari2021,DESKiDS2023,Anbajagane2025,Goh2026}
and from cluster abundances~\cite{Bocquet2024,Ghirardini2024,Aymerich2024}, with their $68\%$ errors. Bottom:
posteriors of $S_8$ in \LCDM\ (gray), of $S_8$ in the frozen model (open boxes) and of the effective amplitude
$S_8^{\rm eff}$ for the KiDS-Legacy kernel (filled boxes), for the six data combinations (boxes: $68\%$; whiskers: $95\%$); the colors of the frozen boxes identify the data combination, as in Fig.~\ref{fig:sigma}. The open boxes show how large the present amplitude of the frozen model becomes without shear data, and the
filled ones what a shear survey analyzed under \LCDM\ would measure; only the latter is to be compared with the
shear measurements.}
\label{fig:s8box}
\end{figure}

Figure~\ref{fig:s8box} gathers the measurements of $S_8$ from cosmic shear and from cluster abundances and compares
them with the predictions. For the frozen model we show both the true $S_8$ of the model and the effective amplitude
seen by a shear survey; the difference between the two, about $0.1$ for the combinations without shear, is the
growth enhancement at $z\lesssim0.3$ that the shear kernels only partly see. The cosmic-shear surveys are in good
mutual agreement, with central values between $0.76$ and $0.83$: KiDS-Legacy, $0.815^{+0.016}_{-0.021}$~\cite{KiDSLegacy2025},
DES Year 3, $0.759^{+0.025}_{-0.023}$~\cite{Amon2022}, and its successor DES Year 6, $0.798^{+0.014}_{-0.015}$~\cite{DESY6shear2026},
HSC Year 3, $0.769^{+0.031}_{-0.034}$ and $0.776^{+0.032}_{-0.033}$ in real and harmonic
space~\cite{Li2023HSC,Dalal2023}, KiDS-1000, $0.759^{+0.024}_{-0.021}$~\cite{Asgari2021}, the joint analysis of
DES Year 3 and KiDS-1000, $0.790^{+0.018}_{-0.014}$~\cite{DESKiDS2023}, DECADE,
$0.791^{+0.027}_{-0.032}$~\cite{Anbajagane2025}, and UNIONS, $0.831^{+0.067}_{-0.078}$~\cite{Goh2026}.

For \Pl+\DK+\DESI, \LCDM\ predicts $S_8=\SeightLAEHx$, within about $1\sigma$ of most surveys and $2.0$--$2.1\sigma$
above DES Year 3 and KiDS-1000. The frozen model predicts an effective amplitude of $\SeffKFAEH$ for the KiDS kernel
and $\SeffDFAEH$ for the DES Year 6 kernel, which lies $1.8\sigma$ above KiDS-Legacy, $2.1$--$2.5\sigma$ above DECADE,
HSC, DES Year 6 and the DES--KiDS combination, and about $3\sigma$ above DES Year 3 and KiDS-1000. The surveys that
were not used in our fits are therefore consistent with the conclusion of Sec.~\ref{sec:joint}. When the shear
data are included, the frozen model returns to $S_8^{\rm eff}=\SeffKFAEHW$ and agrees with all surveys as well as
\LCDM\ does.

\subsection{Cluster abundances}

The three cluster samples, which cover different parts of the sky and different redshift ranges
(Sec.~\ref{sec:data}), give different indications (Fig.~\ref{fig:s8box}). For \Pl+\DK+\DESI\ in
\LCDM\ the predicted amplitude is $\clsLsptAEH\sigma$ from SPT, $\clsLerassAEH\sigma$ from eRASS1 and
$\clsLpszAEH\sigma$ from PSZ2. The frozen model, which raises the amplitude at the low redshifts of the cluster
samples, moves the prediction to $\clsptAEH$ (SPT kernel), $\clerassAEH$ (eRASS1) and $\clpszAEH$ (PSZ2): it
removes the tension with eRASS1 ($\clserassAEH\sigma$) and increases those with SPT ($\clssptAEH\sigma$) and PSZ2
($\clspszAEH\sigma$). After the shear data are included the predictions return to the \LCDM\ values. Given the
spread among the cluster samples themselves, which is under active study by the cluster
community~\cite{Bocquet2024,Ghirardini2024,Aymerich2024}, and the fact that the halo mass function of a model in
which the DM mass is not conserved has not yet been calibrated, we do not include the clusters in the fits. We also note that,
within \LCDM, the prediction of \Pl+\DK+\DESI\ differs from eRASS1 by $3.7\sigma$, a discrepancy comparable to
that of the frozen model with PSZ2, so the cluster samples alone cannot currently discriminate between the two
models; a joint analysis of the samples with a common mass calibration would help in this respect, and the redshift
distribution of the counts would be a sharper test (Appendix~\ref{app:clusters}).

\section{Discussion}
\label{sec:discussion}

The results of the previous sections can be summarized as follows. The geometric data do not see the coupling,
the CMB prefers an enhanced late growth, and the probes of the present amplitude of the fluctuations disfavor it.
In this section we compare the model with other perturbation-only couplings, discuss which observables could break the remaining degeneracies, and list the
limitations of our analysis.

\subsection{Redshift-space distortions and other perturbation-only couplings}

Redshift-space distortions measure the divergence of the velocity field, which the exchange does not modify
directly (Sec.~\ref{sec:qs}). At fixed primordial amplitude and $z=0.38$, the value of $f\sigma_8$ built from the
velocity divergence changes by $+2.0\%$ for $\Sigma=-0.3$ and by $-0.6\%$ for $\Sigma=0.1$, three times less than
$\sigma_8(z)$ ($+6.9\%$ and $-1.8\%$), and well below the precision of individual RSD measurements. The region of
strong coupling favored by \Pl+\DK+\DESI\ is therefore only mildly constrained by current RSD data, which were not included here. The relevant combination is $f_v\sigma_8$, with $f_v$ obtained from the velocity
divergence, and not $f_\delta\sigma_8$, built from the logarithmic growth rate of the density: in \LCDM\ the two
coincide, but in the frozen model they differ precisely because the exchange changes the density at fixed velocity,
so a comparison of $f_\delta\sigma_8$ with RSD data would overestimate their sensitivity to $\Sigma$. On the scales
used by RSD analyses the growth of the model is scale independent to good accuracy (Appendix~\ref{app:wl}), so no
additional dependence on $k$ arises. This is the main difference with the momentum-transfer
models~\cite{Simpson2010,Pourtsidou2013,BaldiSimpson2017,Linton2022}, which slow the infall of the DM, suppress the
growth of the velocity and the density alike, and are typically constrained by RSD and favored by the low shear
amplitude. In the frozen model the two observables decouple: the coupling acts on the density at fixed velocity,
and the CMB prefers $\Sigma<0$, the sign that enhances the growth. The same decoupling also distinguishes
the model from interactions that exchange energy in the background, whose signatures in the expansion history are
the main target of the analyses of
Refs.~\cite{DiValentino2017,DiValentino2020IDE,Nunes2022,Giare2024,Silva2025,Petri2026}: here the geometric data
can only fix the parameters shared with \LCDM, and the growth observables carry all the information on the
coupling.

\subsection{Breaking the degeneracies}

Our analysis points to three observables that can test the region of strong coupling. The first is
the ISW effect discussed in Sec.~\ref{sec:late}: the bias-independent ratio of the ISW--galaxy and lensing--galaxy
cross-spectra measures the decay rate of the potentials directly, and the frozen model predicts a reduction by a
factor of about two at $z=0.5$ for the couplings favored without shear. The second is the tomographic
structure of cosmic shear (Appendix~\ref{app:wl}): the model enhances the signal of the lowest source bins relative
to the highest ones in a way that no rescaling of the \LCDM\ amplitude can reproduce. The third is the redshift
distribution of cluster counts (Appendix~\ref{app:clusters}), which the model steepens at low redshift. All three
are within the reach of the combination of the next CMB experiments with the Euclid, Rubin and DESI samples, and
two of them do not depend on the absolute calibration of the amplitude of the fluctuations, which makes them
complementary to the $S_8$ comparison that currently limits the model. A further handle is the cross-correlation of
CMB lensing with galaxy samples at $z\lesssim0.5$, which isolates the redshift range where the growth departs from
\LCDM\ and is free of the calibration uncertainties of cosmic shear; and the abundance and density profiles of
cosmic voids respond to the growth of underdense regions and would provide another test.

\subsection{Limitations}

Five aspects of the analysis deserve further work. (i) The cosmic-shear data enter as compressed amplitudes,
corrected for the growth history of the model through an effective kernel. Appendix~\ref{app:wl} shows that a
\LCDM\ analysis of the DES Year 3 correlation functions of the frozen model recovers this effective amplitude to
better than $0.5\%$, but a full analysis would require the nonlinear power spectrum of the model and the response of
baryonic feedback and intrinsic alignments to the modified growth. (ii) The nonlinear correction to the CMB lensing
potential and to the shear spectra uses a fit calibrated on \LCDM, which may not describe the model near the
boundary, where $\epsilon_0$ is about $-10$ and the nonlinear evolution with a varying DM mass is unknown. (iii) The
preference for $\Sigma<0$ depends on the CMB likelihood. It is driven by the lensing smoothing of the high-$\ell$
spectra and by the low-$\ell$ temperature deficit, and the former is weaker in the PR4
analyses~\cite{Rosenberg2022,Tristram2024}. Section~\ref{sec:pr4} confirms it with one of the two PR4 likelihoods; the
other one, and the ground-based spectra, remain to be tested. (iv) We fixed $w=-1$. For $w\neq-1$ the background changes, the geometric probes constrain $w$
directly, and the boundary moves to $1+(1+w+\Sigma)r_0>0$. (v) The cluster comparison relies on a halo mass function
calibrated for a conserved DM mass; a dedicated study of collapse with energy exchange is needed before cluster
counts and void abundances can test the model on their own.

\section{Conclusions}
\label{sec:conclusions}

In summary, we have confronted the frozen model, in which dark matter and dark energy exchange energy only at the
level of the perturbations, with Planck 2018 and Planck PR4, DESI DR2, the DES five-year supernovae in their Dovekie recalibration and the
cosmic-shear amplitudes of KiDS-Legacy and DES Year 6. On sub-horizon scales the exchange changes the efficiency
with which the dark-matter density responds to the convergence of the flow by the factor $1/\eta(a)$, where
$\eta=1+\Sigma r$ and $r$ is the ratio of the dark-energy density to the dark-matter density, while it leaves the expansion history and the velocity field untouched. We have shown that the model is covariant,
gauge invariant and consistent, that the dark-energy fluid with $w=-1$ carries an energy flux proportional to the
coupling and is therefore not a vacuum term, and that $\eta$ sets the propagation speed of its perturbations. The
evolution is hyperbolic only for $\eta>0$. For a negative coupling this stability boundary is reached at a finite
scale factor in the future, outside the domain of dependence of any observation, and the data without cosmic shear
pull the model toward it; for a positive coupling it does not exist. All posteriors and
evidences were computed with exact likelihoods, and the agreement between the data sets was assessed with the same
estimators in \LCDM\ and in the frozen model. Our main findings are the following.

\begin{enumerate}
\item Planck 2018 alone prefers a negative coupling, $\Sigma=\SigA$, with a posterior that rises toward the
boundary and a best fit at $\eta_0=\bfetaA$ that improves the $\chi^2$ by $\dchiA$. The preference is driven by the
lensing smoothing of the high-$\ell$ spectra and by the low-$\ell$ temperature deficit, which the enhanced late
growth reproduces at the price of $\sigma_8$ close to unity; the CMB lensing reconstruction is barely affected.
\item Since the background is that of \LCDM, the coupling changes the agreement between data sets only through the
Planck posterior. It reduces the tension between \Pl\ and \DESI\ from $\QDMAPLDESI$--$\QDMLDESI\sigma$ to
$\QDMAPFDESI$--$\QDMFDESI\sigma$, and that between \Pl+\DK\ and \DESI\ from $\QDMAPLDovDESI$--$\shiftLDovDESI\sigma$ to
$\QDMAPFDovDESI$--$\QDMFDovDESI\sigma$; the agreement with \DK\ remains good ($\QDMAPFDov$--$\QDMFDov\sigma$).
\item The joint fit of \Pl, \DK\ and \DESI\ gives $\Sigma=\SigAEH$, with $\Sigma=0$ outside the $95\%$ interval,
$\Delta\chi^2=\dchiAEH$ and $\ln B=\lnBAEH$. This weak preference rests on the region next to the boundary: it
disappears when the cut is raised to $\eta_{\rm min}=0.5$ ($\ln B=\lnBcutcAEH$).
\item The same enhancement of the growth raises the amplitude seen by cosmic shear, and the tension between \Pl\ and
the shear data grows from $\QDMLWL\sigma$ in \LCDM\ to $\SFWL$--$\QDMAPFWL\sigma$. With the shear data included the
coupling is consistent with zero, $\Sigma=\SigAEHW$, the best fit coincides with \LCDM\ and $\ln B=\lnBAEHW$
favors \LCDM. In its present form, the model does not accommodate at the same time the relief of the BAO tension and the
measured shear amplitude, because both respond to the same enhancement of the late growth. These
results do not depend on the choice of the shear data among the most recent measurements.
\item With Planck PR4 in place of Planck 2018 the preference of the CMB for a negative coupling is weaker:
$\Sigma=\prSigP$ and $\ln B=\prlnBvP$ for PR4 alone, and $\Sigma=\prSigPEH$ with $\ln B=\prlnBvPEH$ when supernovae and
BAO are added. With cosmic shear the conclusion is unchanged, $\Sigma=\prSigPEHW$ and $\ln B=\prlnBvPEHW$. The preference found
with Planck 2018 is thus sensitive to the CMB likelihood, and the combination with cosmic shear brings the coupling
back to values consistent with zero in both releases.
\item The late-time observables that depend only on the background are those of \LCDM: for every data combination
the expansion rate, the Hubble constant and the age of the Universe agree between the two models within a fraction
of their errors, so the model neither eases nor worsens the discrepancy with the local distance ladder, and the
age remains compatible with the oldest stars and globular clusters.
\item The region favored without shear predicts a reduced ISW signal, about half of the \LCDM\ value at $z=0.5$ for the
bias-independent ratio of the ISW--galaxy and lensing--galaxy cross-spectra, and a sign change at very low
redshift. With the published source distributions of DES Year 3 and KiDS-Legacy, the model also enhances the shear
signal of the lowest tomographic bins relative to the highest ones (Appendix~\ref{app:wl}), and a simulated \LCDM\ analysis of the DES Year 3
correlation functions confirms that the effective amplitude we use is what such an analysis measures. Cluster
counts give mixed indications at the level of the amplitude, but their redshift distribution is steeper in
the frozen model than in a \LCDM\ model with the same parameters and effective amplitude.
\end{enumerate}

The next steps are to include the ACT DR6 and SPT-3G spectra~\cite{Louis2025,Camphuis2025}, which measure the
damping tail and the lensing of the CMB independently of Planck, and to study in detail the positive couplings, for
which the initial-value problem is well posed at all times. Extending the model to $w\neq-1$ would make the
background informative. On the side of structure formation, the model needs a full cosmic-shear likelihood with its
nonlinear evolution and tomographic signature, and a halo mass function for a dark matter whose mass is not
conserved, which would turn the redshift distribution of cluster counts into a sharp test. Finally, a microscopic
(Lagrangian) realization of the exchange will tell whether the region close to the boundary, where the data pull
the model, is physically acceptable. The methods used here, from the uniform treatment of tensions to the
validation of compressed growth probes, can be applied to other couplings that leave the expansion history
untouched. Upcoming CMB lensing maps, the full DESI spectroscopic samples and the first cosmological analyses of
Euclid and of the Rubin Observatory will measure the growth of structure at the redshifts where the frozen model
departs from \LCDM, and will therefore test directly the region of strong coupling favored by the CMB.

\begin{acknowledgments}
M.G.R. is partially supported by UFABC and by WFSA. L.F.G. thanks Funda\c{c}\~ao Arauc\'aria for financial
support. S.J.L. and M.L. are supported by grant PIP 11220250100565CO CONICET and grant 20020170100129BA UBACYT. L.F.G. thanks Jo\~ao Victor Silva Rebou\c{c}as for useful discussions about the DES-Dovekie data. We acknowledge the use of
the CHE cluster, managed and funded by the COSMO/CBPF/MCTI, with financial support from FINEP and FAPERJ, and
operating at the RSDC--Datacenter para Ci\^encia Alfredo Marques de Oliveira/CBPF.
\end{acknowledgments}

\section*{Data Availability}
Some of the data that support the findings of this article are openly
available~\cite{dataPLA,dataCamSpecPR4,dataPR4lensing,dataDESSN,dataDESI,dataKiDS,dataDESY3,dataDESY6}. The Planck 2018 temperature, polarization and lensing
likelihoods~\cite{Planck2018V,Planck2018VIII,Planck2018VI} are distributed by the Planck Legacy Archive at
\url{https://pla.esac.esa.int}~\cite{dataPLA}, which also hosts the PR4 (NPIPE) maps. The CamSpec PR4 spectra and
likelihood~\cite{Rosenberg2022} are available at
\url{https://github.com/CobayaSampler/planck_native_data}~\cite{dataCamSpecPR4}, and the PR4 lensing
likelihood~\cite{Carron2022} at \url{https://github.com/carronj/planck_PR4_lensing}~\cite{dataPR4lensing}. The distance moduli of the DES five-year supernova
sample~\cite{DES2024} in the Dovekie recalibration~\cite{Popovic2025}, together with their covariance, are available at \url{https://github.com/des-science/DES-SN5YR}~\cite{dataDESSN}. The DESI DR2 BAO
measurements and their covariance~\cite{DESI2025DR2} are available through the DESI public data releases at
\url{https://data.desi.lbl.gov/doc/releases/}~\cite{dataDESI}. The source redshift distributions of
KiDS-Legacy~\cite{KiDSLegacy2025} are available at
\url{https://kids.strw.leidenuniv.nl/DR5/legacy_wl.php}~\cite{dataKiDS}, and those of DES Year 3, together with its
cosmic-shear correlation functions and their covariance~\cite{Amon2022,Secco2022}, at
\url{https://des.ncsa.illinois.edu/releases/y3a2}~\cite{dataDESY3}. The cosmic-shear data vector, covariance and
source redshift distributions of DES Year~6~\cite{DESY6shear2026} are available through the DES Year~6 cosmology
release at \url{https://www.darkenergysurvey.org/des-y6-cosmology-results-papers/}~\cite{dataDESY6}. The cosmic-shear and cluster-count amplitudes,
Hubble-constant measurements, stellar ages and ISW amplitudes used in Secs.~\ref{sec:data} and \ref{sec:late} are
quoted from the cited publications. The posterior samples, best-fit parameters, tension estimates and numerical
codes produced in this work cannot be made publicly available at present because they are being used in ongoing work
by the first author. The observational data are publicly available, as listed above, and the codes may be made available upon
reasonable request to the first author.

\appendix

\section{Covariant formulation and consistency of the first-order theory}
\label{app:cov}

This appendix gives the proofs of the statements made in Sec.~\ref{sec:model}. We first specify on which space each
object is defined, then establish the first-order structure of a component with $w=-1$, its form in the
Kodama--Sasaki formalism, the consistency of the theory, its regularity at $w=-1$, its relation with a vacuum term,
and the propagation speed of the perturbations. The last subsection verifies the results numerically.

\subsection{Spaces, metrics and derivatives}

Two spacetimes enter the construction, and it is useful to state explicitly where each object lives.

(i) \emph{The physical spacetime} $(M,g_{\mu\nu})$, with Levi-Civita connection $\nabla_\mu$. The Einstein equations
(\ref{eq:einstein}), the exchange laws (\ref{eq:exchange}), the decompositions (\ref{eq:Tdecomp}) and
(\ref{eq:Qsplit}), the four-velocity $u^\mu$, the projector $\gamma_{\mu\nu}=g_{\mu\nu}+u_\mu u_\nu$, the expansion
$\Theta=\nabla_\mu u^\mu$, the overdot $u^\nu\nabla_\nu$ and the projected derivative
$D_\mu=\gamma_\mu{}^\nu\nabla_\nu$ are defined on $M$, and their indices are moved with $g_{\mu\nu}$.

(ii) \emph{The background} $(\bar M,\bar g_{\mu\nu})$, with Levi-Civita connection $\bar\nabla_\mu$, four-velocity
$\bar u^\mu=a^{-1}\delta^\mu{}_0$, projector $\bar\gamma_{\mu\nu}=\bar g_{\mu\nu}+\bar u_\mu\bar u_\nu$, projected
derivative $\bar D_\mu=\bar\gamma_\mu{}^\nu\bar\nabla_\nu$ and expansion $\bar\Theta=3\Hc/a$; here
$\bar\nabla_\mu\bar u_\nu=\tfrac13\bar\Theta\,\bar\gamma_{\mu\nu}$.

(iii) \emph{The perturbations.} With an identification map $\varphi:\bar M\to M$, the perturbation
$\delta T=\varphi^*T-\bar T$ of any tensor field is a tensor field on $\bar M$. First-order equations are therefore
equations on $\bar M$: their indices are moved with $\bar g_{\mu\nu}$ and they contain $\bar\nabla_\mu$ only. When
a derivative defined on $M$ acts on a quantity that vanishes on the background, it may be replaced by its
background counterpart, since the difference is of second order: $D_\mu X=\bar D_\mu X$ and
$\dot X=\bar u^\nu\bar\nabla_\nu X$ for first-order $X$. The gradient $D_\mu\Theta$ is the one object in the
definition of the model that must be built on $M$, because $\bar\Theta$ depends on time; it vanishes on the
background because $\bar\Theta$ is homogeneous on the rest spaces of $\bar u^\mu$.

\subsection{First-order structure}

\begin{lemma}[Stewart--Walker~\cite{StewartWalker1974}]\label{lem:SW}
The perturbation $\delta T$ is invariant under the gauge transformations (\ref{eq:gauge}) if and only if
$\pounds_\xi\bar T=0$ for every $\xi^\mu$, that is, if and only if $\bar T$ (i) vanishes, or (ii) is a constant
scalar, or (iii) is a linear combination with constant coefficients of products of Kronecker deltas.
\end{lemma}

\begin{lemma}\label{lem:struct}
Let $\bar Q^\mu=0$ and $\bar T_{\rm x}{}^\mu{}_\nu=-\rho_\Lambda\delta^\mu{}_\nu$ with $\rho_\Lambda$ constant. Then
(a) $\delta T_{\rm x}{}^\mu{}_\nu$ and $\delta Q^\mu$ are gauge invariant; (b) the first-order exchange law is
Eq.~(\ref{eq:L}), without metric perturbations; (c) $\delta T_{\rm x}{}^\mu{}_\nu$ is an arbitrary symmetric tensor
on $\bar M$, and the components $\delta\rho_{\rm x}$, $\delta p_{\rm x}$ and $J^\mu$ of Eq.~(\ref{eq:dTx}), as well
as a possible anisotropic stress, are gauge invariant and independent of the timelike frame used to define them.
\end{lemma}
\begin{proof}
(a) Apply Lemma~\ref{lem:SW}, case (iii) to $\bar T_{\rm x}{}^\mu{}_\nu$ and case (i) to $\bar Q^\mu$. (b) For any
tensor,
\begin{align}
\delta(\nabla_\mu T^\mu{}_\nu)&=\bar\nabla_\mu\delta T^\mu{}_\nu
+\delta\Gamma^\mu_{\mu\sigma}\bar T^\sigma{}_\nu-\delta\Gamma^\sigma_{\mu\nu}\bar T^\mu{}_\sigma ,
\end{align}
and for $\bar T^\mu{}_\nu=-\rho_\Lambda\delta^\mu{}_\nu$ the last two terms cancel. (c) Perturbing
$T^\mu{}_\nu=\rho\,u^\mu u_\nu+p\,\gamma^\mu{}_\nu+\dots$, the terms that contain $\delta u^\mu$ carry the factor
$\bar\rho_{\rm x}+\bar p_{\rm x}=0$. For the same reason a change of frame $u^\mu\to u^\mu+v^\mu$ changes the energy
flux by $-(\bar\rho_{\rm x}+\bar p_{\rm x})v^\mu=0$. Gauge invariance follows from (a).
\end{proof}

Equations~(\ref{eq:encov}) and (\ref{eq:momcov}) are the projections of Eq.~(\ref{eq:L}) with the tensor
(\ref{eq:dTx}), obtained with $\bar\nabla_\mu\bar u_\nu=\tfrac13\bar\Theta\bar\gamma_{\mu\nu}$.

\subsection{Kodama--Sasaki form}

In the harmonic formulation of Kodama and Sasaki~\cite{KodamaSasaki1984} (see also Ref.~\cite{MalikWands2009}) the
action of the gauge group is explicit. For scalar-type perturbations with harmonics $Y$,
$Y_j=-k^{-1}\partial_jY$ and $Y_{ij}$, a first-order gauge transformation is the infinitesimal diffeomorphism
\begin{equation}
\xi^0=\zeta(\tau)\,Y,\qquad \xi^j=\lambda(\tau)\,Y^j ,
\label{eq:KSxi}
\end{equation}
with two arbitrary functions of time per mode. For a component $A$ with background tensor
$\bar T_A{}^\mu{}_\nu={\rm diag}(-\bar\rho_A,\bar p_A\delta^i{}_j)$ and energy transfer
$\bar Q_{A\,\mu}=(-a\bar Q_A,0)$, we write
\begin{align}
\delta T_A{}^0{}_0&=-\delta\rho_A\,Y,\qquad \delta T_A{}^0{}_j=\mathcal{J}_A\,Y_j,\nonumber\\
\delta T_A{}^i{}_j&=\delta p_A\,Y\delta^i{}_j+\Pi_A\,Y^i{}_j,\nonumber\\
\delta Q_{A\,0}&=-a\,\delta q_A\,Y,\qquad \delta Q_{A\,j}=a\,f_A\,Y_j .
\end{align}
Evaluating the Lie derivatives in Eq.~(\ref{eq:gauge}) with the generator (\ref{eq:KSxi}) gives
\begin{align}
\delta\rho_A&\to\delta\rho_A+\bar\rho_A'\,\zeta, &
\mathcal{J}_A&\to\mathcal{J}_A+k(\bar\rho_A+\bar p_A)\,\zeta,\nonumber\\
\delta p_A&\to\delta p_A+\bar p_A'\,\zeta, &
\Pi_A&\to\Pi_A, \label{eq:KST}\\
a\,\delta q_A&\to a\,\delta q_A+(a\bar Q_A\zeta)', &
f_A&\to f_A+k\,\bar Q_A\,\zeta , \label{eq:KSQ}
\end{align}
and $\lambda$ does not appear. Every shift is proportional to $\bar\rho_A'$, $\bar p_A'$, $\bar\rho_A+\bar p_A$,
$\bar Q_A$ or $\bar Q_A'$. For the component x of the frozen model with $w=-1$ all of them vanish, so that
$\delta\rho_{\rm x}$, $\mathcal{J}_{\rm x}=\bar\rho_{\rm x}q/k$, $\delta p_{\rm x}$, $\Pi_{\rm x}$, $\delta q$ and $f$
are unchanged by every $\zeta$ and $\lambda$. The gauge freedom does not act on the dark sector of the model. In
particular, $q=0$ and $\delta Q=0$ are not gauge conditions and are not consequences of covariance; they are
physical restrictions on the perturbations.

\subsection{Consistency for any background and any closure}

\begin{prop}\label{prop:consist}
Let $(\bar g,\bar T_A,\bar Q_A)$ be any Friedmann solution with
$\bar\nabla_\nu\bar T_A{}^{\mu\nu}=\bar Q_A^\mu$ and $\sum_A\bar Q_A^\mu=0$, where the transfers $\bar Q_A$ may or may
not vanish. Let $(\delta T_A,\delta Q_A)$ be first-order fields that satisfy
$\delta(\nabla_\nu T_A{}^{\mu\nu})=\delta Q_A^\mu$ and $\sum_A\delta Q_A^\mu=0$, closed by arbitrary constitutive
relations, and let
$\delta E_{\mu\nu}\equiv\delta\mathcal{G}_{\mu\nu}-8\pi G_{\rm N}\sum_A\delta T^{(A)}_{\mu\nu}$. If the constraints
$\delta E^0{}_\nu=0$ hold on an initial hypersurface and $\delta E^i{}_j=0$ holds at all times, then
$\delta E^0{}_\nu=0$ at all times.
\end{prop}
\begin{proof}
The proof has four steps and uses three distinct objects, which we define first. For
each member of the one-parameter family of configurations, labeled by $\lambda$, we define on the physical
spacetime the tensor
\begin{equation}
E_{\mu\nu}(\lambda)\equiv\mathcal{G}_{\mu\nu}[g(\lambda)]-8\pi G_{\rm N}\sum_AT^{(A)}_{\mu\nu}(\lambda),
\label{eq:Edef}
\end{equation}
which measures the failure of the configuration to satisfy the Einstein equations; it is \emph{not} assumed to
vanish. Its background value, $\bar E_{\mu\nu}=\bar{\mathcal{G}}_{\mu\nu}-8\pi G_{\rm N}\sum_A\bar T^{(A)}_{\mu\nu}$, is
built from the unperturbed metric and the unperturbed tensors $\bar T^{(A)}_{\mu\nu}$, and it vanishes because the
background is a solution:
\begin{equation}
\bar E_{\mu\nu}=0 .
\label{eq:Ebar}
\end{equation}
Its first-order part, $\delta E_{\mu\nu}=dE_{\mu\nu}/d\lambda$ at $\lambda=0$, is a tensor on $\bar M$ built from
$\delta g_{\mu\nu}$ and from the perturbations $\delta T^{(A)}_{\mu\nu}$; the statement to be proved concerns its
components.

\emph{Step 1: the divergence of $E$ on the physical spacetime.} The contracted Bianchi identity,
$\nabla_\mu\mathcal{G}^\mu{}_\nu=0$, holds for every metric and therefore for every $\lambda$. Taking the divergence of
Eq.~(\ref{eq:Edef}) with the connection $\nabla_\mu$ of $g(\lambda)$,
\begin{equation}
\nabla_\mu E^\mu{}_\nu=-8\pi G_{\rm N}\sum_A\nabla_\mu T^{(A)\mu}{}_\nu .
\label{eq:divE}
\end{equation}
This is an identity: it uses no field equation.

\emph{Step 2: its first-order part.} We now expand Eq.~(\ref{eq:divE}) in $\lambda$. At zeroth order its right-hand
side is $-8\pi G_{\rm N}\sum_A\bar Q_{A\,\nu}=0$. At first order it is
$-8\pi G_{\rm N}\sum_A\delta(\nabla_\mu T^{(A)\mu}{}_\nu)=-8\pi G_{\rm N}\sum_A\delta Q_{A\,\nu}$, where
$\delta Q_{A\,\nu}=\bar g_{\nu\mu}\delta Q_A^\mu+\delta g_{\nu\mu}\bar Q_A^\mu$; both terms cancel in the sum over the
components, by the hypotheses $\sum_A\delta Q_A^\mu=0$ and $\sum_A\bar Q_A^\mu=0$. Hence
\begin{equation}
\delta\!\left(\nabla_\mu E^\mu{}_\nu\right)=0 .
\label{eq:ddivE}
\end{equation}
The constitutive relations enter the proof only here, through the conservation of the total energy-momentum tensor.

\emph{Step 3: from $\nabla_\mu$ to $\bar\nabla_\mu$.} The left-hand side of Eq.~(\ref{eq:ddivE}) is the perturbation
of a covariant derivative, in which both the tensor and the connection are perturbed. For any tensor,
\begin{align}
\delta\!\left(\nabla_\mu E^\mu{}_\nu\right)&=\bar\nabla_\mu\,\delta E^\mu{}_\nu
+\delta\Gamma^\mu_{\mu\sigma}\,\bar E^\sigma{}_\nu-\delta\Gamma^\sigma_{\mu\nu}\,\bar E^\mu{}_\sigma ,
\label{eq:dnabla}
\end{align}
where $\delta\Gamma^\sigma_{\mu\nu}$ is the first-order difference between the connections of $g$ and $\bar g$, a
tensor on $\bar M$. The first term contains the perturbation $\delta E$ and the background connection; the last two
contain the perturbed connection and the background tensor $\bar E$. By Eq.~(\ref{eq:Ebar}) the last two terms
vanish, whatever $\delta g_{\mu\nu}$ is. For the same reason $\delta E^\mu{}_\nu=\bar g^{\mu\sigma}\delta E_{\sigma\nu}$,
since the term $\delta g^{\mu\sigma}\bar E_{\sigma\nu}$ vanishes, so the indices of $\delta E$ are moved with
$\bar g_{\mu\nu}$ without ambiguity. Equations~(\ref{eq:ddivE}) and (\ref{eq:dnabla}) then give an equation on the
background alone,
\begin{equation}
\bar\nabla_\mu\,\delta E^\mu{}_\nu=0 .
\label{eq:consE}
\end{equation}
Note that this step uses $\bar E_{\mu\nu}=0$ and does not use $\delta E_{\mu\nu}=0$.

\emph{Step 4: propagation of the constraints.} We write Eq.~(\ref{eq:consE}) in the conformal coordinates of the
background, in which the nonvanishing connection coefficients are $\bar\Gamma^0_{00}=\Hc$,
$\bar\Gamma^0_{ij}=\Hc\delta_{ij}$ and $\bar\Gamma^i_{0j}=\Hc\delta^i{}_j$, and we use
$\delta E^i{}_0=-\delta E^0{}_i$, which follows from the symmetry of $\delta E_{\mu\nu}$. Its time and space components
are
\begin{align}
(\delta E^0{}_0)'+3\Hc\,\delta E^0{}_0-\partial_i\delta E^0{}_i-\Hc\,\delta E^i{}_i&=0,\nonumber\\
(\delta E^0{}_j)'+4\Hc\,\delta E^0{}_j+\partial_i\delta E^i{}_j&=0 .
\end{align}
If the evolution equations hold, $\delta E^i{}_j=0$, the second equation gives $\delta E^0{}_j\propto a^{-4}$, which
vanishes at all times if it vanishes initially; the first one then gives $\delta E^0{}_0\propto a^{-3}$, which
vanishes for the same reason. The constraints are therefore preserved~\cite{Wald1984}.
\end{proof}

It is useful to state precisely what the proof uses. Of the background it uses two facts: that its metric is the
Friedmann metric of Sec.~\ref{sec:background}, from which the connection coefficients of Step~4 are computed, and
that it solves the Einstein equations, $\bar E_{\mu\nu}=0$, which removes the perturbed connection in Step~3. It does
not use the matter content of the background: which components are present, their equations of state, or whether
the background transfers $\bar Q_A$ vanish. Of the first-order closure it uses only the conservation of the total
energy-momentum tensor, in Step~2, and not the form of the constitutive relations. Three consequences follow. First,
the same closure is consistent on different backgrounds: the exchange (\ref{eq:exchcov}) extends without change to a
component with constant $w\neq-1$, because $D_\mu\rho_{\rm x}$ and $D_\mu\Theta_T$ vanish on every Friedmann
background, and the closure (\ref{eq:closurecov}) extends with $c_s^2-w$ in place of $1+c_s^2$; the result is
Eqs.~(\ref{eq:dc})--(\ref{eq:q}). Second, different closures are consistent on the same background: on that of \LCDM,
$\Sigma$ labels a one-parameter family of first-order theories, and the inhomogeneous vacuum~\cite{Wands2012} is
another one. Third, the correspondence between backgrounds and closures is many-to-many, which is the freedom by
which dark-energy models with identical expansion histories differ in their sound speed or anisotropic
stress~\cite{KunzSapone2006,Kunz2009}.

\subsection{Regularity at $w=-1$ and the necessity of perturbations}

\begin{prop}\label{prop:reg}
In the variables $(\delta_{\rm x},q)$, the coefficients of the linear system formed by
Eqs.~(\ref{eq:dc})--(\ref{eq:q}) and the Einstein equations, and the initial data (\ref{eq:ic}), are real-analytic
functions of $w$ in a neighborhood of $w=-1$, for every $a>0$ and $\Sigma>-26/3$. The solution is therefore an
analytic function of $w$ there.
\end{prop}
\begin{proof}
The factor $1+w$ enters Eqs.~(\ref{eq:dc})--(\ref{eq:q}) polynomially, $c_a^2=w$, and the only denominator is that
of Eq.~(\ref{eq:thetaT}), $1+w+r^{-1}$, which is positive near $w=-1$. The component sources the Einstein equations
through $\bar\rho_{\rm x}\delta_{\rm x}$, $\bar\rho_{\rm x}q$ and $\delta p_{\rm x}$. The denominator of
Eq.~(\ref{eq:ic}) equals $26+3\Sigma$ at $w=-1$ and $c_s^2=1$. Analytic dependence on a parameter is a standard
property of linear ordinary differential equations.
\end{proof}

The velocity divergence $\theta_{\rm x}=q/(1+w)$ is unbounded as $w\to-1$ at fixed $q$. This has no consequence,
because $\theta_{\rm x}$ does not enter the system; it reflects the fact that a tensor proportional to
$\delta^\mu{}_\nu$ has every timelike vector as eigenvector and selects no rest frame~\cite{Wands2012,KunzSapone2006}.

\begin{prop}\label{prop:trivial}
Let $\Sigma\neq0$ and $w=-1$. If $\delta T_{\rm x}{}^\mu{}_\nu=0$ solves the first-order equations of the frozen
model, then $D_\mu\Theta=0$ and $\bar D_\mu(\delta\rho+3\delta p)_{\rm tot}=0$ at all times.
\end{prop}
\begin{proof}
Apply $\bar D_\mu$ to Eq.~(\ref{eq:encov}). With $\delta\rho_{\rm x}=\delta p_{\rm x}=0$ and $J^\mu=0$ the left-hand
side vanishes, so $\bar D_\mu\delta Q=0$, and Eq.~(\ref{eq:exchcov}) reduces to $\Sigma\,\rho_\Lambda D_\mu\Theta=0$.
For the geodesic DM congruence the gradient of the Raychaudhuri equation is, to first order,
\begin{equation}
(D_\mu\Theta)^{\displaystyle\cdot}+\bar\Theta\,D_\mu\Theta
=-4\pi G_{\rm N}\,\bar D_\mu(\delta\rho+3\delta p)_{\rm tot},
\label{eq:raych}
\end{equation}
and with $D_\mu\Theta=0$ it gives $\bar D_\mu(\delta\rho+3\delta p)_{\rm tot}=0$.
\end{proof}

In the synchronous gauge $D_\mu\Theta=0$ reads $h'=0$, and Eq.~(\ref{eq:raych}) is
$h''+\Hc h'=-8\pi G_{\rm N}a^2(\delta\rho+3\delta p)_{\rm tot}$. An unperturbed component x is therefore compatible
with the equations only in a universe without density perturbations. Since $h'$ is fixed by the Einstein equations
and does not vanish in the presence of the primordial adiabatic mode, the contrapositive holds: the equations
require $\delta T_{\rm x}{}^\mu{}_\nu\neq0$, in agreement with Eq.~(\ref{eq:L}), by which a nonvanishing
$\delta Q_\nu$ requires a nonvanishing $\delta T_{\rm x}{}^\mu{}_\nu$. The explicit solution on sub-horizon scales
is Eq.~(\ref{eq:qqs}).

\subsection{Relation with a vacuum component}

\begin{prop}\label{prop:char}
To first order, $T_{\rm x}{}^\mu{}_\nu=-V\delta^\mu{}_\nu$ for some scalar $V=\rho_\Lambda+\delta V$ if and only if
conditions (\ref{eq:vacuum}) hold and the anisotropic stress vanishes.
\end{prop}
\begin{proof}
Compare $\delta T_{\rm x}{}^\mu{}_\nu=-\delta V\,\delta^\mu{}_\nu$ with Eq.~(\ref{eq:dTx}).
\end{proof}

Under conditions (\ref{eq:vacuum}), Eq.~(\ref{eq:L}) becomes
\begin{equation}
\delta Q_\nu=\partial_\nu\delta V,\qquad
\delta Q=-\delta\dot V,\qquad F_\mu=\bar D_\mu\delta V ,
\label{eq:vacQ}
\end{equation}
so that the energy and the momentum transfer derive from one scalar.

\begin{prop}\label{prop:vac}
Let $T_{\rm x}^{\mu\nu}=-Vg^{\mu\nu}$ and $F^\mu=0$. Then $\bar D_\mu\delta V=0$. If in addition $\bar Q=0$, then
$\delta Q=0$.
\end{prop}
\begin{proof}
By Eq.~(\ref{eq:vacQ}), $F^\mu=0$ gives $\bar D_\mu\delta V=0$. Then
$\bar D_\mu\delta Q=-(\bar D_\mu\delta V)^{\displaystyle\cdot}-\tfrac13\bar\Theta\,\bar D_\mu\delta V=0$. A
first-order scalar with vanishing spatial gradient is a function of time only, that is, a change of the background
value $\bar Q$, which is zero.
\end{proof}

Proposition~\ref{prop:vac} is the geodesic interacting-vacuum scenario~\cite{Wands2012,Martinelli2019}: the vacuum
is homogeneous on the DM rest spaces and the whole exchange resides in the background. The hypothesis that leads to
$\delta Q=0$ is $F^\mu=0$ together with conditions (\ref{eq:vacuum}), and not covariance: with $\bar V$ constant,
$\delta V$ is a gauge-invariant function by Lemma~\ref{lem:SW}(ii), and an inhomogeneous vacuum with
$\delta V\neq0$ is covariant and exchanges energy and momentum at first order~\cite{Wands2012}. In the geodesic
vacuum class the transfer is slaved to a background function: $Q$ depends on the DM proper time alone, and in the
language of Eq.~(\ref{eq:KSQ}) its perturbation has the pure-gauge form $a\,\delta q=(a\bar Q\zeta)'$,
$f=k\bar Q\zeta$, generated by the time shift to the DM-comoving slicing; it is proportional to $\bar Q$ and
vanishes with it. The frozen model has no such link between the two orders. Its transfer perturbation is intrinsic,
that is, gauge invariant and not of the form (\ref{eq:KSQ}), and it exists with $\bar Q=0$, because the component
does not satisfy conditions (\ref{eq:vacuum}): by Eq.~(\ref{eq:closurecov}) with $c_s^2=1$ and
Proposition~\ref{prop:trivial},
\begin{equation}
\bar D_\mu\delta p_{\rm x}=\bar D_\mu\delta\rho_{\rm x}-2\bar\Theta J_\mu\neq-\bar D_\mu\delta\rho_{\rm x},
\qquad J^\mu\neq0 .
\end{equation}
In terms of the counting of Sec.~\ref{sec:pertstruct}, the vacuum form and the frozen model are two different sets of
constitutive relations for the same gauge-invariant functions, and both are covariant.

\subsection{Propagation speed}

\begin{prop}\label{prop:speed}
For $w=-1$, the characteristic speeds of the perturbations of the component x in the DM frame are
$\pm c_{\rm eff}$, with $c_{\rm eff}^2=c_s^2\,\eta$ and $\eta=1+\Sigma\,\bar\rho_{\rm x}/\bar\rho_{\rm c}$.
\end{prop}
\begin{proof}
The characteristic speeds of a system of evolution equations are fixed by its terms with the highest number of
derivatives, its principal part. We isolate it in five steps, first covariantly and then in the gauge of
Sec.~\ref{sec:sync}.

\emph{Step 1: variables and commutation rule.} We use the spatial gradient of the density perturbation and the
divergence of the energy flux,
\begin{equation}
X_\mu\equiv\bar D_\mu\delta\rho_{\rm x},\qquad Z\equiv\bar D_\nu J^\nu ,
\end{equation}
both of first order. For a first-order scalar $f$ on the Friedmann background, whose DM congruence is geodesic and
shear free, time and space derivatives commute up to the expansion,
\begin{equation}
\bar D_\mu\dot f=(\bar D_\mu f)^{\displaystyle\cdot}+\tfrac13\bar\Theta\,\bar D_\mu f .
\label{eq:commute}
\end{equation}

\emph{Step 2: the gradient of the energy equation.} We apply $\bar D_\mu$ to Eq.~(\ref{eq:encov}), use
Eq.~(\ref{eq:commute}), and insert the energy transfer (\ref{eq:exchcov}). Since $\bar\rho_{\rm c}$ is homogeneous,
$D_\mu\Theta_T=D_\mu\Theta+\bar D_\mu Z/\bar\rho_{\rm c}$, and we obtain
\begin{align}
\dot X_\mu=&-\bar D_\mu Z-\Sigma\,\frac{\rho_\Lambda}{\bar\rho_{\rm c}}\,\bar D_\mu Z-\Sigma\,\rho_\Lambda D_\mu\Theta\nonumber\\
&-\left(\tfrac43+\Sigma\right)\bar\Theta\,X_\mu-\bar\Theta\,\bar D_\mu\delta p_{\rm x} .
\label{eq:Xdot}
\end{align}
The first term is the divergence of the energy flux in the conservation law. The second has the same structure and
comes from the energy transfer, which depends on the expansion $\Theta_T$ of the frame of the total dark momentum and
therefore on the flux itself. Together they give $-\eta\,\bar D_\mu Z$: this is the origin of the factor $\eta$.

\emph{Step 3: the momentum equation.} With $F^\mu=0$ and the closure (\ref{eq:closurecov}),
Eq.~(\ref{eq:momcov}) becomes
\begin{equation}
\dot J_\mu=-c_s^2\,X_\mu-\left(\tfrac13-c_s^2\right)\bar\Theta\,J_\mu ,
\label{eq:Jdot}
\end{equation}
and the same closure turns the last term of Eq.~(\ref{eq:Xdot}) into terms proportional to $X_\mu$ and $J_\mu$,
without spatial derivatives.

\emph{Step 4: wave equation.} Taking the divergence of Eq.~(\ref{eq:Jdot}), then its time derivative, and using
Eq.~(\ref{eq:Xdot}), the two first-order equations combine into a single second-order equation for the divergence of
the flux,
\begin{equation}
\ddot Z-c_s^2\,\eta\,\bar D^\mu\bar D_\mu Z=c_s^2\,\Sigma\,\rho_\Lambda\,\bar D^\mu D_\mu\Theta+\dots ,
\label{eq:waveZ}
\end{equation}
where the dots stand for terms with at most one derivative of $Z$, multiplied by $\bar\Theta$. The left-hand side
is a wave operator with speed $c_s\sqrt{\eta}$.

\emph{Step 5: the metric term does not change the principal part.} The right-hand side of Eq.~(\ref{eq:waveZ})
contains the metric through $D_\mu\Theta$, and it must be checked that it does not feed back two derivatives of
$Z$. By Eq.~(\ref{eq:raych}), $D_\mu\Theta$ is the time integral of the gradients of the density and pressure
perturbations of all components; for the component x it involves $X_\mu$ and $\bar D_\mu\delta p_{\rm x}$ with no
further spatial derivative, and by Eq.~(\ref{eq:Jdot}) $X_\mu$ is itself a first time derivative of $J_\mu$. The term
$\bar D^\mu D_\mu\Theta$ therefore contributes to Eq.~(\ref{eq:waveZ}) terms with no more than the divergence of
$J_\mu$, that is, $Z$ without derivatives. It acts as a source and as a lower-order coupling, and the principal part
is the left-hand side of Eq.~(\ref{eq:waveZ}).

The same result is explicit in the synchronous gauge. For $w=-1$, Eqs.~(\ref{eq:dx}) and (\ref{eq:q}) with
$\theta_T=r\,q$ give, after eliminating $\delta_{\rm x}$,
\begin{equation}
q''+(4+3\Sigma)\Hc\,q'+c_s^2\,\eta\,k^2q=-c_s^2\,\Sigma\,k^2\,\frac{h'}{2}+\mathcal{O}(\Hc^2)\,q ,
\label{eq:qwave}
\end{equation}
a damped oscillator of frequency $c_s\sqrt{\eta}\,k$ forced by the metric. The response of $h'$ to $q$ is suppressed
by $(\Hc/k)^2$, because $h'$ is sourced by density perturbations and not by their gradients, so it does not modify
the term $c_s^2\eta k^2q$. For $k\gg\Hc$ the dispersion relation is $\omega^2=c_s^2\,\eta\,k^2$, and the stationary
solution of Eq.~(\ref{eq:qwave}) is the quasi-static flux of Eq.~(\ref{eq:qqs}), $q=-(\Sigma/\eta)\,h'/2$.
\end{proof}

\begin{figure}[tbp]
\includegraphics[width=\columnwidth]{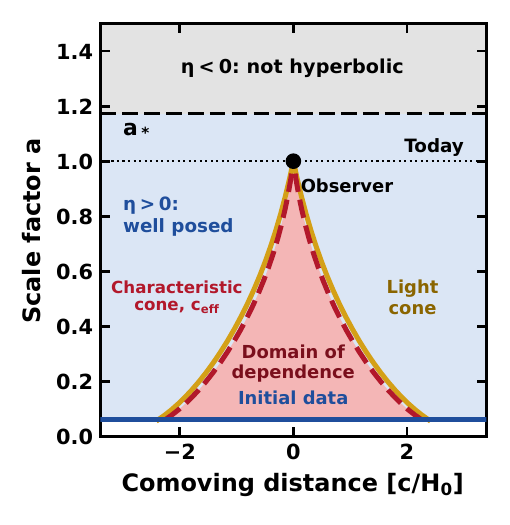}
\caption{Initial-value problem of the frozen model for $\Sigma=-0.24$, in a flat \LCDM\ background with
$\Omega_{\rm m}=0.31$. The horizontal axis is the comoving distance from the observer in units of $c/H_0$, and the
vertical axis the scale factor. Initial data are given on the hypersurface at the bottom (blue line). The dashed
lines are the characteristic cone of the dark-energy perturbations that reach the observer today, with speed
$c_{\rm eff}=c_s\sqrt{\eta}$ and $c_s=1$, and the shaded area inside it is the domain of dependence of the
observer; the solid lines are the past light cone. The system is hyperbolic and the problem is well posed in the
blue region, $\eta>0$, which extends to $a_*=1.17$; hyperbolicity is lost in the gray region, which lies in the
future of the observer and does not influence any observable.}
\label{fig:cauchy}
\end{figure}

For $\eta>0$ the system is hyperbolic; for $\eta<0$ it is elliptic, with a gradient instability of rate
$k\sqrt{c_s^2|\eta|}$. Equation~(\ref{eq:astar}) follows from $\eta(a)=1+\Sigma r_0a^3$. Figure~\ref{fig:cauchy}
illustrates the resulting initial-value problem. Initial data for $(\delta_{\rm x},q)$ and for the other components,
subject to the constraints of the Einstein equations, are given on a hypersurface deep in the radiation era. As long
as $\eta>0$, the perturbations of the component x propagate along characteristic cones of finite speed
$c_{\rm eff}$, the solution at any event depends only on the data inside its past characteristic cone, and it depends
continuously on them. For a negative coupling this holds on $0<a<a_*$, a region that contains the observer and the
whole past light cone, on which all observables are defined.

\begin{figure}[tbp]
\includegraphics[width=\columnwidth]{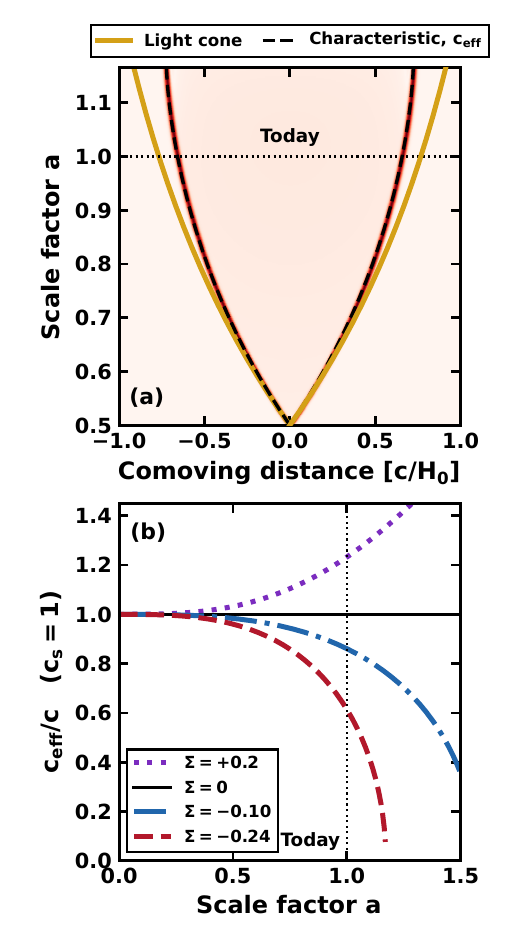}
\caption{Propagation of a localized perturbation of the dark-energy fluid. (a) Numerical solution of the wave
equation of the energy flux, the homogeneous part of Eq.~(\ref{eq:qwave}), for a narrow pulse released at $a=0.5$
with $\Sigma=-0.24$ and $c_s=1$; the color scale is the amplitude normalized to its maximum at each time. The two
fronts follow the characteristics (dashed lines), which lie inside the light cone (solid lines) and become vertical
as $a\to a_*=1.17$, where the effective speed vanishes. (b) Effective speed $c_{\rm eff}=c_s\sqrt{\eta}$ in units of the
speed of light, for $c_s=1$ and four values of the coupling. For $\Sigma<0$ it is subluminal and vanishes at $a_*$;
for $\Sigma>0$ it exceeds the speed of light unless $c_s^2\leq1/\eta$.}
\label{fig:pulse}
\end{figure}

Figure~\ref{fig:pulse} shows this propagation directly. We solved the homogeneous part of Eq.~(\ref{eq:qwave}) in
one spatial dimension for a narrow pulse of the energy flux released at $a=0.5$, with $\Sigma=-0.24$ and $c_s=1$.
The pulse splits into two fronts that follow the characteristics, $x=\pm\int c_{\rm eff}\,d\tau$, to better than
$0.3\%$, and they remain inside the light cone: at $a=1$ the fronts have traveled $0.66\,c/H_0$, against $0.77\,c/H_0$
for light. The effective speed decreases with time, from $c_{\rm eff}=0.96\,c$ at the release to $0.62\,c$ today, and
it tends to zero at $a_*$, where the characteristics become vertical in the diagram (panel a). Panel (b) compares
the effective speed for several couplings: it is smaller than the speed of light at all times for negative
couplings, and larger for positive couplings if $c_s=1$.

\subsection{The two signs of the coupling}
\label{app:signs}

The loss of hyperbolicity is specific to negative couplings. We now compare the initial-value problem and the
dynamical evolution for the two signs of $\Sigma$, in order to show which properties of the model depend on the sign.

For $\Sigma>0$, $\eta(a)=1+\Sigma r_0a^3$ is larger than unity and grows monotonically. It never vanishes, so no
critical scale factor $a_*$ exists: by Proposition~\ref{prop:speed} the system is hyperbolic for all $a>0$, and the
initial-value problem is well posed at all times, including the whole future. The evolution is also bounded. On
sub-horizon scales the energy flux is $q=-(\Sigma/\eta)\,h'/2$ [Eq.~(\ref{eq:qqs})], whose magnitude in units of $h'/2$
never exceeds $\Sigma$ and decreases toward zero as $\eta$ grows, and the dark-matter density responds to the flow
with the factor $1/\eta<1$ [Eq.~(\ref{eq:eps})], so that its growth is slowed and its contrast eventually freezes. For
$\Sigma<0$ the same two quantities, $\Sigma/\eta$ and $1/\eta$, diverge at $a_*$, and this divergence precedes the
gradient instability of the region $\eta<0$.

The price of a positive coupling is the propagation speed. Since $c_{\rm eff}^2=c_s^2\eta$ grows with $\eta$, with
$c_s^2=1$ the perturbations of the dark-energy fluid propagate faster than light in the dark-matter frame, and for any
constant $c_s<1$ they do so after the scale factor
\begin{equation}
a_{\rm c}=\left[\frac{c_s^{-2}-1}{\Sigma\,r_0}\right]^{1/3}.
\end{equation}
Subluminal propagation up to today requires $c_s^2\leq1/\eta_0$, and at all times it requires a sound speed that
decreases as $1/\eta$. This affects neither the well-posedness of the problem nor the observables, which are
insensitive to $c_s$ (Appendix~\ref{app:numcheck}). The two signs are therefore not equivalent: a negative coupling is
limited in the future by $a_*$, and a positive one by the requirement of subluminal propagation, which constrains the
sound speed and not the coupling.

\begin{figure}[tbp]
\includegraphics[width=\columnwidth]{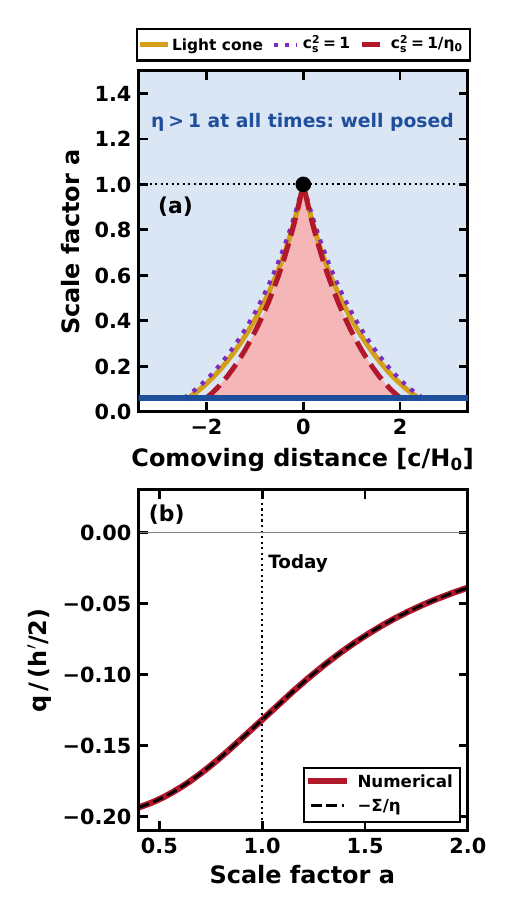}
\caption{Positive coupling, $\Sigma=+0.2$. (a) Initial-value problem, as in Fig.~\ref{fig:cauchy}: $\eta>1$ at all
times, the system is hyperbolic everywhere and no critical scale factor exists. The solid lines are the past light
cone of the observer, the dotted lines the characteristic cone for $c_s^2=1$, which lies outside it, and the dashed
lines the characteristic cone for $c_s^2=1/\eta_0$, which lies inside it and bounds the shaded domain of dependence.
(b) Energy flux of the dark-energy fluid in units of $h'/2$ at $k=0.1\,{\rm Mpc}^{-1}$, followed into the future
through the density ratio $\bar\rho_{\rm x}/\bar\rho_{\rm c}$ in a universe with a smaller dark-matter density (see
text): the numerical solution (thick line) follows $-\Sigma/\eta$ (dashed), remains bounded and tends to zero.}
\label{fig:positive}
\end{figure}

Figure~\ref{fig:positive} illustrates these statements for $\Sigma=+0.2$. Panel (a) is the analog of
Fig.~\ref{fig:cauchy}: there is no region without hyperbolicity, the characteristic cone for $c_s^2=1$ lies slightly
outside the light cone, and the one for $c_s^2=1/\eta_0$ lies inside it. To follow the evolution into the future
with the Boltzmann solver, we use the fact that at $w=-1$ the coupling enters only through the product
$\Sigma\,\bar\rho_{\rm x}/\bar\rho_{\rm c}$: a universe with a smaller dark-matter density, $\omega_{\rm c}=0.02$, has
$r_0=20.5$, the value that the density ratio of our Universe reaches at $a=2.0$. For $\Sigma=+0.2$ this gives
$\eta_0=5.1$; the solution is regular, the ratio of the computed flux to Eq.~(\ref{eq:qqs}) stays within $10^{-4}$ of
unity at $k=0.05$ and $0.1\,{\rm Mpc}^{-1}$, $|q|$ never exceeds $0.18\,h'/2$, and $\sigma_8$ is $11\%$ lower than for
$\Sigma=0$ (panel b). For $\Sigma=-0.2$ the same universe crosses $\eta=0$ at $a=0.62$, and $|q|$ then grows by $52$
and $105$ orders of magnitude at the two wave numbers, at a rate proportional to $k$, as expected for a gradient
instability.

The data do not select the sign. With cosmic shear included the coupling is consistent with zero, and the posterior
probability of $\Sigma>0$ is $0.72$ for \Pl+shear and $0.39$ for the full combination, against $0.05$ and $0.02$ for
\Pl\ and \Pl+\DK+\DESI. A sizable part of the allowed region therefore corresponds to models for which the
initial-value problem is well posed at all times.

For the remaining part of the posterior, the relevant question is how far in the future the critical scale factor
lies. Table~\ref{tab:astar} gives the posterior distribution of $a_*$ and of the cosmic time $t_*-t_0$ between today
and $a_*$, computed from the same samples as the constraints, with the points of positive coupling counted as
$a_*=\infty$. With all data, $a_*>1.44$ at the $95\%$ level, which is more than $5.5$~Gyr in the future, and the median
is $a_*=2.6$, about $15$~Gyr from now; the values obtained with Planck PR4 are nearly the same. The critical scale
factor approaches the present epoch only for the combinations without cosmic shear, where the median is
$a_*=1.11$--$1.16$, between $1.5$ and $2.2$~Gyr from now. This is the region of strong coupling that we interpret
with caution in Sec.~\ref{sec:planck}, and the shear data exclude it.

\begin{table}[tbp]
\caption{Posterior of the critical scale factor $a_*$ of Eq.~(\ref{eq:astar}) and of the cosmic time between today
and $a_*$: probability that no critical scale factor exists ($\Sigma\geq0$), lower limit at the $95\%$ level and
median. A dash indicates that the median lies in the region $\Sigma\geq0$. The CMB data are those
of Planck 2018, except in the last row.}
\label{tab:astar}
\begin{ruledtabular}
\begin{tabular}{lccccc}
 & & \multicolumn{2}{c}{$a_*$} & \multicolumn{2}{c}{$t_*-t_0$ [Gyr]}\\
Data & $P(\Sigma\geq0)$ & $95\%$ & Median & $95\%$ & Median\\
\hline
CMB & $0.05$ & $>1.03$ & $1.16$ & $>0.4$ & $2.2$\\
CMB+SN+BAO & $0.02$ & $>1.02$ & $1.11$ & $>0.3$ & $1.5$\\
CMB+shear & $0.72$ & $>1.63$ & --- & $>7.6$ & ---\\
All data & $0.39$ & $>1.44$ & $2.6$ & $>5.5$ & $15$\\
All data (PR4) & $0.41$ & $>1.46$ & $2.8$ & $>5.7$ & $17$\\
\end{tabular}
\end{ruledtabular}
\end{table}

A finite domain of validity in the future is a common property of effective descriptions of the dark sector. A
constant equation of state $w<-1$ leads to a divergence of the scale factor at a finite time~\cite{Caldwell2003}, and
it is nevertheless used routinely to describe the observed epoch, because the data constrain the expansion only up to
today. The frozen model with a constant negative coupling is in the same situation, with two differences in its
favor. First, the background is that of \LCDM\ and is regular at all times; only the perturbations of the dark sector
lose hyperbolicity at $a_*$. Second, the solution on $a\leq1$ is determined by the initial data alone and is
independent of any continuation beyond $a_*$, as shown above. The existence of $a_*$ therefore limits the
extrapolation of the model to the future, and it does not affect any of the predictions that we compare with the
data. A complete description, in which the product $\Sigma\,r$ saturates before reaching $-1$, would remove the
critical scale factor while leaving these predictions unchanged for moderate couplings.

The case of a positive coupling, including the role of the sound
speed and its phenomenology, will be studied in more detail in a separate paper.

\subsection{Numerical verification}
\label{app:numcheck}

\begin{figure}[tbp]
\includegraphics[width=\columnwidth]{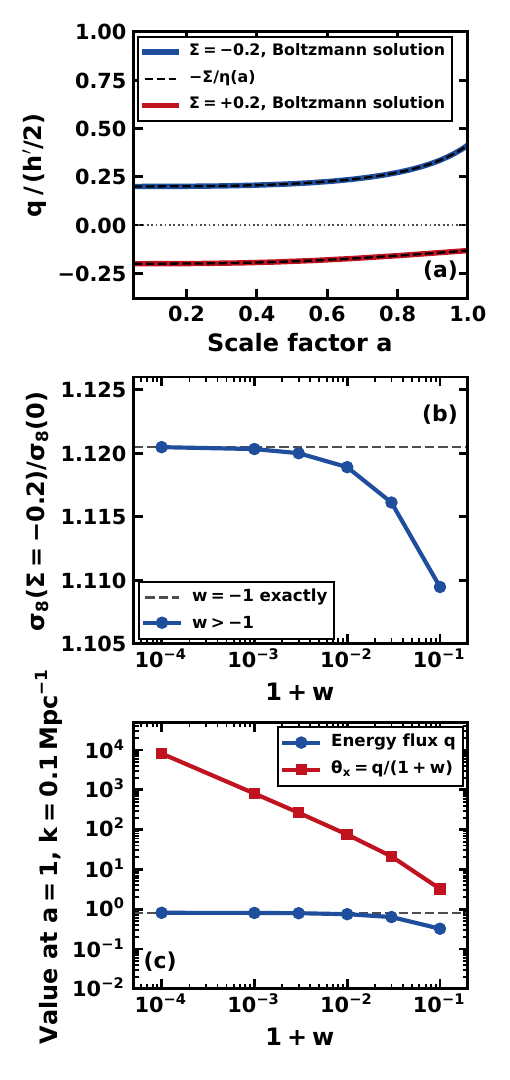}
\caption{Energy flux of the dark-energy fluid and regularity of the limit $w\to-1$. (a) Energy flux $q$ in units of
$h'/2$ from the Boltzmann solution at $k=0.1\,{\rm Mpc}^{-1}$ and $w=-1$ (thick lines), compared with the
quasi-static prediction $-\Sigma/\eta$ of Eq.~(\ref{eq:qqs}) (dashed). (b) Ratio $\sigma_8(\Sigma=-0.2)/\sigma_8(0)$
as $w\to-1$; the dashed line is the value at $w=-1$ exactly. (c) Energy flux $q$ and velocity divergence
$\theta_{\rm x}=q/(1+w)$ at $a=1$ and $k=0.1\,{\rm Mpc}^{-1}$: the flux converges to its value at $w=-1$ (dashed
line), while $\theta_{\rm x}$, which enters no equation, grows as $(1+w)^{-1}$.}
\label{fig:R1}
\end{figure}

We have verified these results with the Boltzmann solver used in the paper, for the parameters of
Fig.~\ref{fig:pheno}. Figure~\ref{fig:R1} concerns the energy flux and the limit $w\to-1$. For $\Sigma=\pm0.2$,
$w=-1$ and $a>0.3$, the ratio of the computed $q$ to Eq.~(\ref{eq:qqs}) lies in $[0.9995,1.0000]$ at
$k=0.1\,{\rm Mpc}^{-1}$, equals unity to four digits at $k=0.5\,{\rm Mpc}^{-1}$, and lies in $[0.961,1.001]$ at
$k=0.01\,{\rm Mpc}^{-1}$, where the sub-horizon expansion is less accurate; the relation
$\eta\,\delta_{\rm c}'=-h'/2$ holds to $0.2\%$ for $k\geq0.1\,{\rm Mpc}^{-1}$. The limit is regular, as stated in
Proposition~\ref{prop:reg}: $\sigma_8(\Sigma=-0.2)/\sigma_8(0)=1.1095$, $1.1189$, $1.1203$, $1.12048$ and $1.12049$
for $w=-0.9$, $-0.99$, $-0.999$, $-0.9999$ and $-1$, and $q$ converges in the same way, while $\theta_{\rm x}=q/(1+w)$
grows as $(1+w)^{-1}$. The component is not of the vacuum form: at $a=1$ and $k\geq0.1\,{\rm Mpc}^{-1}$,
$\delta p_{\rm x}/\delta\rho_{\rm x}=+6.0$ for $\Sigma=-0.2$ and $-0.90$ for $\Sigma=+0.2$, against $-1$ for a
vacuum; on sub-horizon scales this ratio is $(q'+4\Hc q)/(q'-2\Hc q)$, independent of $k$.

\begin{figure}[tbp]
\includegraphics[width=\columnwidth]{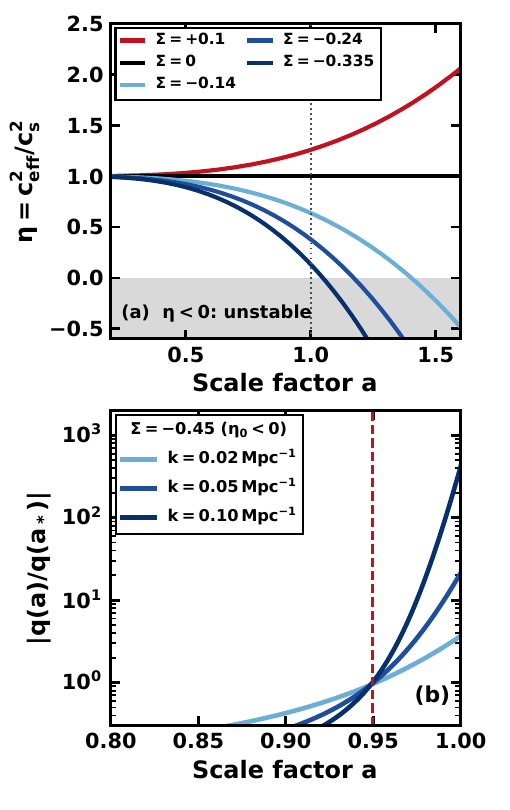}
\caption{Stability boundary of the model. (a) The parameter $\eta(a)=c_{\rm eff}^2/c_s^2$ for several couplings,
extended to the future; for $\Sigma<0$ it vanishes at the scale factor $a_*$ of Eq.~(\ref{eq:astar}), beyond which
the perturbations of the dark-energy fluid are unstable (shaded region). The dotted line marks the present epoch.
(b) Growth of the energy flux after $\eta$ changes sign at $a_*$ (dashed line), for $\Sigma=-0.45$ and three wave
numbers; the growth rate is proportional to $k$, as expected for a gradient instability.}
\label{fig:R2}
\end{figure}

Figure~\ref{fig:R2} concerns Proposition~\ref{prop:speed}. For $\Sigma=-0.45$, a value outside the prior of the
analysis for which $\eta$ changes sign at $a_*=0.95$, the logarithmic growth rate of $q$ after $a_*$ is
$(0.92$--$1.00)\,k\sqrt{|\eta|}$, and $|q|$ grows between $a_*$ and $a=1$ by factors $3.7$, $21$ and $420$ at
$k=0.02$, $0.05$ and $0.1\,{\rm Mpc}^{-1}$. For $\Sigma=+0.1$, lowering $c_s^2$ from $1$ to $0.5$ changes
$\sigma_8$, $C_L^{\phi\phi}$ at $L=100$ and $C_\ell^{TT}$ at $\ell=1000$ by less than $10^{-5}$, and $C_\ell^{TT}$
by $0.8\%$ at $\ell=2$ and $0.03\%$ at $\ell=10$, so the condition $c_s^2\leq1/\eta_0$ can be imposed without
changing the results.

\section{Cosmic shear in \LCDM\ and in the frozen model}
\label{app:wl}

This appendix explains how the shear data enter our analysis, compares the shear signal of the two models with the
published source distributions of DES Year 3 and KiDS-Legacy, and quantifies how well the effective amplitude of
Eq.~(\ref{eq:s8eff}) summarizes the frozen model for a survey analyzed under \LCDM.

\subsection{Tomographic shear spectra}

In the Limber approximation, the cross-spectrum of the shear (or convergence) between tomographic bins $i$ and
$j$ is
\begin{equation}
C_\ell^{ij}=\int_0^{\chi_{\rm H}}\frac{d\chi}{\chi^2}\,q_i(\chi)\,q_j(\chi)\,
P_{\rm m}\!\left(k=\frac{\ell+1/2}{\chi},z(\chi)\right),
\label{eq:limber}
\end{equation}
where $\chi$ is the comoving distance, $P_{\rm m}$ the nonlinear matter power spectrum, and the lensing efficiency
of bin $i$, with normalized source distribution $n_i(z)$, is
\begin{equation}
q_i(\chi)=\frac{3}{2}\Omega_{\rm m}H_0^2\,\frac{\chi}{a(\chi)}
\int_\chi^{\chi_{\rm H}}d\chi'\,n_i(\chi')\,\frac{\chi'-\chi}{\chi'}.
\end{equation}
The real-space correlation functions measured by DES are the Hankel transforms
\begin{equation}
\xi_\pm^{ij}(\theta)=\int_0^\infty\frac{\ell\,d\ell}{2\pi}\,C_\ell^{ij}\,J_{0,4}(\ell\theta),
\end{equation}
with $J_0$ for $\xi_+$ and $J_4$ for $\xi_-$. In \LCDM\ the amplitude of these spectra scales approximately as
$S_8^{2.5}$--$S_8^{3}$ on the scales used by current surveys, which is why $S_8$ is the best-measured combination.

\begin{figure}[tbp]
\includegraphics[width=\columnwidth]{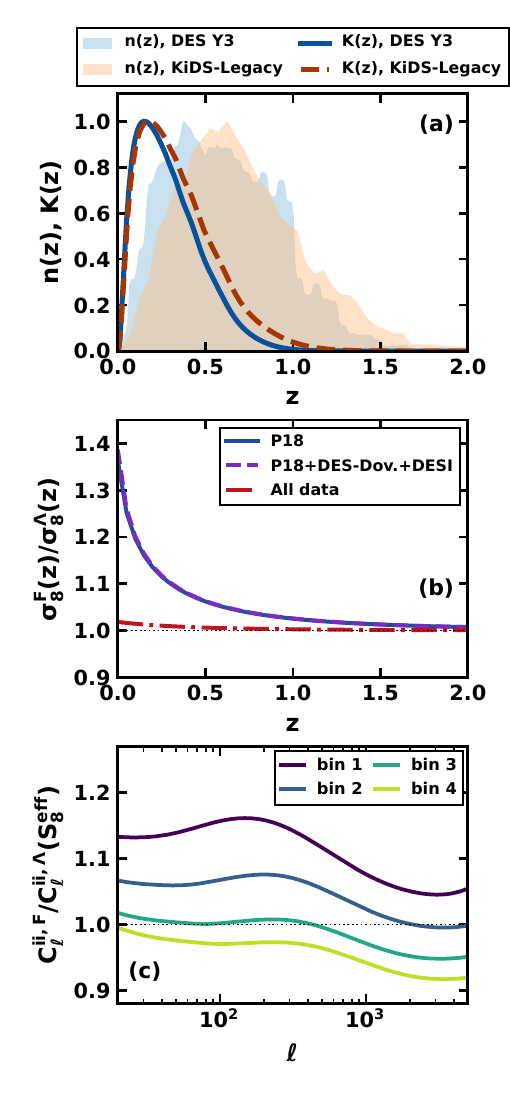}
\caption{(a) Published source distributions of DES Year 3 (blue area) and KiDS-Legacy (orange area), summed over
the tomographic bins and normalized to their maxima, together with the redshift kernels $K(z)$ of the shear power at
$\ell=300$ of the two surveys (solid and dashed lines). (b) Ratio of $\sigma_8(z)$ in the frozen model and in \LCDM\ with
the same parameters, at the best fits of three data combinations. (c) Auto-spectra of the four DES Year 3 bins of
the frozen best fit to \Pl+\DK+\DESI, divided by those of a \LCDM\ model with the same parameters and the amplitude
rescaled to $S_8^{\rm eff}$: the low-redshift bins are enhanced and the high-redshift bins suppressed relative to
\LCDM, a tomographic signature of the late growth of the frozen model.}
\label{fig:shearapp}
\end{figure}

The frozen model changes Eq.~(\ref{eq:limber}) only through $P_{\rm m}$, since the distances and the lensing
efficiencies depend on the background, which is that of \LCDM. On linear scales the change is a rescaling of the
amplitude at each redshift,
\begin{equation}
P_{\rm m}^{\rm F}(k,z)=\left[\frac{\sigma_8^{\rm F}(z)}{\sigma_8^{\Lambda}(z)}\right]^2P_{\rm m}^{\Lambda}(k,z),
\label{eq:pkratio}
\end{equation}
because the growth modification is scale independent on the relevant scales: the ratio of the density growth rate
of the frozen model to that of \LCDM\ is constant to better than $0.2\%$ for $0.05<k<0.5\,h\,{\rm Mpc}^{-1}$ at
$z\ge0.1$, even close to the boundary ($\Sigma=-0.3$). On larger scales, around $k=0.01\,h\,{\rm Mpc}^{-1}$, the
dark-energy perturbations with $c_s^2=1$ introduce a scale dependence of up to $3\%$ at $z=0.1$ for
$\eta_0=0.2$, which is irrelevant for the shear and cluster amplitudes. On nonlinear scales we apply the same
halo-model correction as for the CMB lensing potential to the linear spectrum of each model. The ratio in
Eq.~(\ref{eq:pkratio}) is shown in Fig.~\ref{fig:shearapp}b for the best fits: it is close to unity at $z\gtrsim1$
and grows steeply at low redshift, reaching $1.36$ and $1.39$ today at the best fits of \Pl\ and \Pl+\DK+\DESI, respectively.

\subsection{Effective amplitude}

Since the shear signal of a survey is dominated by the redshifts where its kernel peaks, a model whose amplitude
evolves differently from \LCDM\ is seen by a \LCDM\ analysis as an effective $S_8$. We define it through the
redshift kernel of the convergence power at $\ell=300$, near the middle of the scales that carry most of the
signal,
\begin{equation}
K(z)\propto\frac{1}{H(z)\chi^2}\sum_iq_i^2(\chi)\,P_{\rm m}\!\left(\frac{300.5}{\chi},z\right),
\end{equation}
which leads to Eq.~(\ref{eq:s8eff}). In the fits we tabulate $G(z)$ with the Boltzmann solver on a grid of $30$
values of $\ln\eta_0\in[\ln0.05,\ln1.8]$ and five values of $\Omega_{\rm m}\in[0.27,0.35]$, which is sufficient
because at fixed background the late-time growth depends on $\eta_0$ alone [the combination $\eta_0-1=\Sigma r_0$
of Eq.~(\ref{eq:eta})], and interpolate it with bicubic splines; the effective amplitudes agree with a direct
computation to $0.2\%$ over the posterior, an order of magnitude below the errors of the shear measurements. For
the kernels we approximated the source distributions by Gaussian tomographic bins at the published mean redshifts,
which for the four bins of DES Year 6 are $0.41$, $0.54$, $0.85$ and $1.16$~\cite{DESY6shear2026}. The resulting
kernels have mean redshifts $0.33$ (DES Year 6) and $0.36$ (KiDS-Legacy). The dependence on the kernel is weak: at
the posterior of \Pl+\DK+\DESI\ the effective amplitude is $0.868$ with the kernel of KiDS-Legacy, $0.872$ with that
of DES Year 6 and $0.877$ with that of DES Year 3, whose mean redshift is $0.29$.

We tested this approximation with the published source distributions of DES Year 3 and KiDS-Legacy, shown in
Fig.~\ref{fig:shearapp}a; we use DES Year 3 for this test and for the simulated analysis below because its source
distributions, data vector and covariance are public. Their
kernels have mean redshifts $\KzD$ (DES Year 3) and $\KzK$ (KiDS-Legacy), and at the best fit of \Pl+\DK+\DESI\ they
give $S_8^{\rm eff}=\SeffnzDAEH$ and $\SeffnzKAEH$, against $\SeffgDAEH$ and $\SeffgKAEH$ with the Gaussian bins,
for a model with $S_8=\SbfFAEH$. The two kernels thus agree to $0.5\%$, at the most extreme point of the posterior.

\subsection{What would a \LCDM\ analysis infer?}

A more demanding test is to ask what a \LCDM\ analysis of the shear data would infer if the frozen model were true.
We computed the DES Year 3 correlation functions $\xi_\pm$ of the frozen best fit of \Pl+\DK+\DESI\ for all ten
bin pairs, kept the $\npxi$ data points with $\theta>10'$ for $\xi_+$ and $\theta>40'$ for $\xi_-$, and fitted
them with \LCDM\ spectra that share all the cosmological parameters of the frozen model except the amplitude,
using the published covariance of the survey. The fit gives $S_8=\SinfAEH$, in agreement with
$S_8^{\rm eff}=\SeffnzDAEH$ and far from the true $S_8=\SbfFAEH$ of the model; for the best fit of \Pl\
alone the analogous numbers are $\SinfA$, $\SeffnzDA$ and $\SbfFA$. The effective amplitude is therefore what a
\LCDM\ analysis measures, and our comparison of the frozen model with the published $S_8$ values is consistent.

The fit is not perfect, and the residual has a definite pattern. Relative to the rescaled \LCDM\ model, the frozen model
enhances the auto-spectrum of the lowest DES bin by $\binoneD\%$ and suppresses that of the highest one by
$\binfourD\%$ at $\ell=300$ (Fig.~\ref{fig:shearapp}c); for KiDS-Legacy the lowest bin is enhanced by $\binoneK\%$
and the highest suppressed by $\binsixK\%$. This differential growth across the bins is a signature of
the model. With the DES Year 3 covariance it costs only $\Delta\chi^2=\chiinfAEH$ for $\npxi$ data points, and it
would in practice be partly absorbed by the photometric-redshift and intrinsic-alignment parameters of a full
analysis. Since $\Delta\chi^2$ grows in proportion to the inverse covariance, a survey with errors several times
smaller, such as Euclid or the Rubin Observatory LSST, could detect it for the couplings favored by \Pl+\DESI\ if
the nuisance parameters are well controlled.

\begin{figure}[tbp]
\includegraphics[width=\columnwidth]{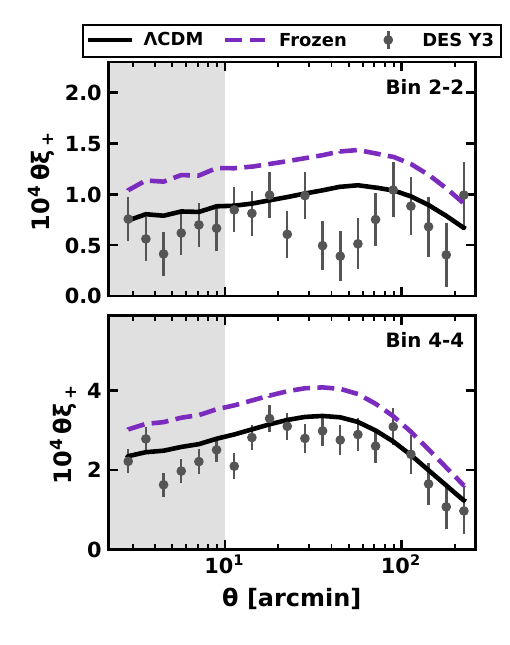}
\caption{DES Year 3 measurements of $\theta\xi_+(\theta)$ for the auto-correlations of the second and the fourth
source bins, with the predictions of the \LCDM\ best fit of \Pl+\DK+\DESI\ (solid) and of the frozen best fit of the
same combination (dashed). The frozen best fit of all data overlaps with \LCDM\ and is not drawn. The predictions do
not include
intrinsic alignments, baryonic effects or survey nuisance parameters; the shaded region marks small scales that
are most affected by them.}
\label{fig:xip}
\end{figure}

Figure~\ref{fig:xip} shows the DES Year 3 measurements of $\xi_+$ for the auto-correlations of the second and the
fourth source bins together with the predictions of the best fits of \Pl+\DK+\DESI. The predictions do not include intrinsic alignments, baryonic feedback or the
nuisance parameters of the survey, so the figure is meant only to illustrate the size of the effect: the frozen best
fit lies above the data in all four bins, while the best fit of the full combination, which is not drawn, is
indistinguishable from \LCDM. The excess of the frozen best fit is largest in the lowest-redshift bin, about $50\%$
in $\xi_+$ near $\theta=30'$, and decreases to about $20\%$ in the highest one, in line with the tomographic
pattern of Fig.~\ref{fig:shearapp}c and with the steep rise of the growth enhancement at $z\lesssim0.5$.

\subsection{ISW amplitude}

The same tabulation gives the ISW amplitude. Since the Poisson equation holds with the total matter density,
the Weyl potential scales as $\delta_{\rm m}/a$ and its decay rate is $f_\delta-1$, with
$f_\delta=d\ln\delta_{\rm m}/d\ln a$ the growth rate of the density. The ratio of the ISW--galaxy and
lensing--galaxy cross-spectra in a narrow redshift bin, which is independent of galaxy bias and of the amplitude of
the fluctuations~\cite{Zhang2006}, is proportional to this decay rate, and its value relative to \LCDM,
\begin{equation}
A_{\rm ISW}(z)=\frac{1-f_\delta(z)}{1-f_\delta^{\Lambda}(z)},
\label{eq:aisw}
\end{equation}
agrees with the full Limber computation of the two cross-spectra for galaxy samples at $0.3\le z\le2.1$ to
$0.01$--$0.03$.

\section{Cluster abundances in \LCDM\ and in the frozen model}
\label{app:clusters}

The abundance of massive halos is exponentially sensitive to the amplitude of the linear fluctuations, which makes
cluster counts a natural probe of a coupling that changes the late growth. This appendix summarizes the standard
formalism, describes how we apply it to the frozen model, and quantifies the accuracy of the effective amplitude
used in Sec.~\ref{sec:data}.

\subsection{Formalism}

The number of clusters per unit redshift and solid angle above a mass threshold $M_{\rm lim}(z)$ is
\begin{equation}
\frac{dN}{dz\,d\Omega}=\frac{c\,\chi^2(z)}{H(z)}\int_{M_{\rm lim}(z)}^{\infty}dM\,\frac{dn}{dM}(M,z),
\label{eq:dndz}
\end{equation}
where, in a real survey, the sharp threshold is replaced by the selection function in the observable and by the
scaling relation between the observable and the mass. The halo mass function is written as
\begin{equation}
\frac{dn}{d\ln M}=f(\sigma)\,\frac{\bar\rho_{\rm m}}{M}\,\left|\frac{d\ln\sigma}{d\ln M}\right|,
\end{equation}
with the variance of the linear density field smoothed with a top-hat filter of radius
$R=(3M/4\pi\bar\rho_{\rm m})^{1/3}$,
\begin{equation}
\sigma^2(M,z)=\frac{1}{2\pi^2}\int_0^\infty dk\,k^2P_{\rm lin}(k,z)\,W^2(kR),
\end{equation}
where $W(x)=3(\sin x-x\cos x)/x^3$,
and the multiplicity function of Ref.~\cite{Tinker2008},
\begin{equation}
f(\sigma)=A\left[\left(\frac{\sigma}{b}\right)^{-a}+1\right]e^{-c/\sigma^2},
\end{equation}
whose parameters depend on the overdensity that defines the halo mass and on redshift. We use masses
$M_{500{\rm c}}$, defined with respect to $500$ times the critical density, which corresponds to an overdensity
$500/\Omega_{\rm m}(z)$ with respect to the mean, and interpolate the published parameters in this overdensity.

\subsection{Application to the frozen model}

The linear power spectrum of the frozen model differs from that of \LCDM\ only by the scale-independent factor of
Eq.~(\ref{eq:pkratio}), so $\sigma(M,z)$ is that of \LCDM\ multiplied by $\sigma_8^{\rm F}(z)/\sigma_8^{\Lambda}(z)$.
The multiplicity function, on the other hand, has been calibrated with $N$-body simulations of \LCDM-like cosmologies~\cite{Tinker2008}, in which the mass of the dark matter is conserved. In the frozen model the exchange converts vacuum energy into dark matter where matter
converges, which modifies the spherical collapse and may change the collapse threshold and the universality of
$f(\sigma)$. We use the \LCDM\ form of $f$, which is the assumption implicit in any comparison with the published
cluster constraints, and we regard the results as indicative.

\begin{figure}[tbp]
\includegraphics[width=\columnwidth]{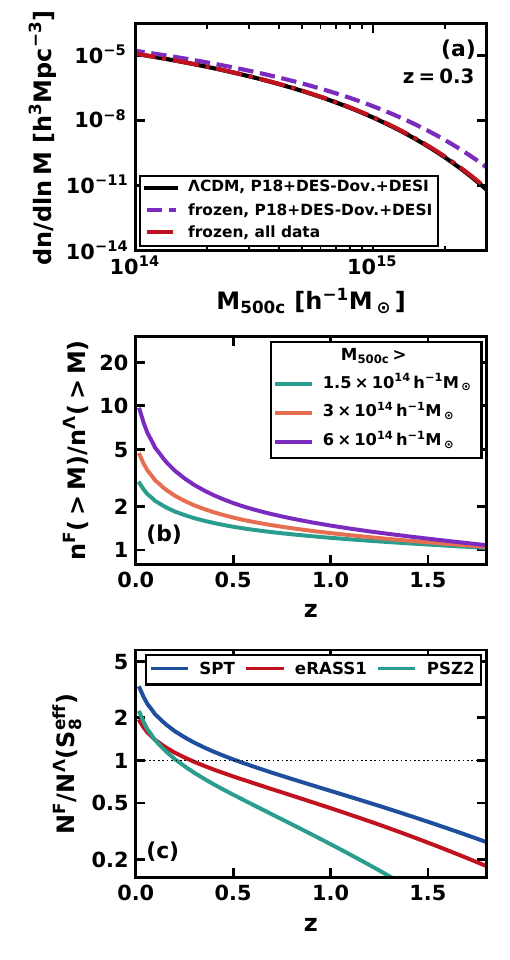}
\caption{Cluster abundances in \LCDM\ and in the frozen model. (a) Mass function at $z=0.3$ for the best fits of
\LCDM\ and of the frozen model to \Pl+\DK+\DESI, and of the frozen model to all data. (b) Ratio of the comoving
number densities of halos above three mass thresholds in the frozen model and in \LCDM, for the best fits of
\Pl+\DK+\DESI. (c) Ratio of the counts per unit redshift of the frozen best fit and of a \LCDM\ model with the same
parameters and the amplitude rescaled to the effective amplitude $S_8^{\rm eff}$ of each cluster sample, for the
mass thresholds $3\times10^{14}$ (SPT and PSZ2) and $1.5\times10^{14}\,h^{-1}M_\odot$ (eRASS1). The thresholds are
representative of the mass range of each sample and are not meant to reproduce its selection function; the
kernel-weighted ratios quoted in the text are close to unity, while the redshift dependence of the ratio shows that
the frozen model changes the shape of the redshift distribution of the counts, and not only their amplitude.}
\label{fig:clusters}
\end{figure}

Figure~\ref{fig:clusters}a compares the mass function at $z=0.3$ for the best fits of \LCDM\ and of the frozen model.
For \Pl+\DK+\DESI\ the frozen model predicts $\nratioc$ times more halos above $3\times10^{14}\,h^{-1}M_\odot$ at
$z=0.3$ than \LCDM, and $\nratiocsix$ times more above $6\times10^{14}\,h^{-1}M_\odot$. The excess grows toward low
redshift, where the growth is enhanced, and toward high masses, where the mass function is steepest
(Fig.~\ref{fig:clusters}b). For the best fit of the full combination the two models are indistinguishable.

\subsection{Accuracy of the effective amplitude}

To test the compression used in the main text, we compare the counts of the frozen best fit of \Pl+\DK+\DESI\ with
those of a \LCDM\ model that has the same parameters and an amplitude rescaled to the effective amplitude of each
sample. With representative mass thresholds and the redshift kernels of the three samples, the kernel-weighted
counts of the frozen model are $\NFRspt$ (SPT), $\NFReras$ (eRASS1) and $\NFRpsz$ (PSZ2) times those of the
rescaled \LCDM\ model. Since the counts of these samples scale approximately as $S_8^{\alpha}$ with $\alpha=\slopespt$,
$\slopeeras$ and $\slopepsz$, respectively, these differences correspond to shifts of $\dSspt\%$, $\dSeras\%$ and
$\dSpsz\%$ in the effective amplitude, comparable to the error of the most precise sample and smaller than the
others. The compression is therefore adequate for the consistency check of Sec.~\ref{sec:late}.

The redshift distribution of the counts is a different matter. Figure~\ref{fig:clusters}c shows that, relative to
the rescaled \LCDM\ model, the frozen model predicts more clusters at $z\lesssim0.3$ and fewer at higher redshift,
by factors of two or more at the extremes of the samples. A full analysis of the counts in bins of redshift, which
all the cluster surveys perform, would therefore be considerably more sensitive to the coupling than the
compressed amplitude we use, once a mass function appropriate for a dark matter whose mass is not conserved
becomes available.

\section{Tension estimators}
\label{app:estimators}

Let $P_A$, $P_X$ and $P_{AX}$ be the posteriors of \Pl, of the external data set and of their combination, with
means $\bm\mu$ and covariances $\mathcal{C}$, and $\mathcal{C}_\Pi$ the prior covariance.

\paragraph{Effective number of parameters.}
$N_{\rm eff}=N-{\rm tr}(\mathcal{C}_\Pi^{-1}\mathcal{C})$ counts the parameters constrained beyond the
prior~\cite{Raveri2019}.

\paragraph{Difference in means.} In the parameters $\bm p$ constrained by both data sets,
\begin{equation}
Q_{\rm DM}=(\bm\mu_A-\bm\mu_X)^{\rm T}(\mathcal{C}_A+\mathcal{C}_X)^{-1}(\bm\mu_A-\bm\mu_X),
\end{equation}
distributed as $\chi^2$ with ${\rm rank}(\mathcal{C}_A+\mathcal{C}_X)$ degrees of freedom.

\paragraph{Update difference in means.}
\begin{equation}
Q_{\rm UDM}=(\bm\mu_{AX}-\bm\mu_A)^{\rm T}(\mathcal{C}_A-\mathcal{C}_{AX})^{-1}(\bm\mu_{AX}-\bm\mu_A),
\end{equation}
restricted to the directions in which the update improves the constraint~\cite{Raveri2020}.

\paragraph{Goodness-of-fit loss.} $Q_{\rm DMAP}=\chi^2_{AX}-\chi^2_A-\chi^2_X$ at the respective best fits, with
$N_{\rm eff}^A+N_{\rm eff}^X-N_{\rm eff}^{AX}$ degrees of freedom~\cite{Raveri2019}.

\paragraph{Parameter shift.} With $\Delta\bm p=\bm p_A-\bm p_X$ drawn from the product of the two posteriors, the
shift probability is
\begin{equation}
P_{\rm shift}=\int_{P(\Delta\bm p)>P(\bm0)}P(\Delta\bm p)\,d\Delta\bm p ,
\end{equation}
estimated with a kernel density estimate evaluated by fast Fourier transform~\cite{Raveri2021}.

\paragraph{Evidence ratio, information and suspiciousness.}
\begin{align}
\ln R&=\ln Z_{AX}-\ln Z_A-\ln Z_X,\\
D_i&=\langle\ln\mathcal{L}\rangle_{P_i}-\ln Z_i,\qquad \ln I=D_A+D_X-D_{AX},\\
\ln S&=\ln R-\ln I=\langle\ln\mathcal{L}_{AX}\rangle_{P_{AX}}\nonumber\\
&\quad-\langle\ln\mathcal{L}_A\rangle_{P_A}-\langle\ln\mathcal{L}_X\rangle_{P_X},\\
d_i&=2\left[\langle(\ln\mathcal{L})^2\rangle_{P_i}-\langle\ln\mathcal{L}\rangle_{P_i}^2\right],\\
d&=d_A+d_X-d_{AX},
\end{align}
and the tension probability is $P(\chi^2_d>d-2\ln S)$~\cite{Handley2019a,Handley2019b}. $R$ depends on the prior
volume of the parameters that $X$ does not constrain; $S$ does not.

\paragraph{Deviance information criterion.} ${\rm DIC}=\chi^2_{\rm MAP}+2p_D$ with
$p_D=\langle\chi^2\rangle_P-\chi^2_{\rm MAP}$, and $\Delta{\rm DIC}={\rm DIC}_{AX}-{\rm DIC}_A-{\rm DIC}_X$;
$\Delta{\rm DIC}>0$ indicates disagreement~\cite{Spiegelhalter2002,Joudaki2017}.

\section{Sampling and validation}
\label{app:sampling}

\begin{table}[tbp]
\caption{Importance sampling. $N$ is the total number of exact likelihood evaluations, accumulated over all the rounds of the sampling and shared by all data combinations, $N_{\rm eff}$ the effective sample size
after Pareto smoothing, $\hat k$ the Pareto diagnostic and $\ln Z$ the log-evidence with its Monte Carlo
error. The evidences of \DK\ and \DESI\ alone, from nested sampling, are $\ln Z=-819.60$ and $-13.90$ in \LCDM\ and $-819.73$ and $-14.13$ in the frozen model.}
\label{tab:sampling}
\begin{ruledtabular}
\footnotesize
\begin{tabular}{lccc}
Data & $N_{\rm eff}$ & $\hat k$ & $\ln Z$ \\
\hline
\multicolumn{4}{l}{\LCDM, $N=3298$} \\
\Pl & $1129$ & $-0.24$ & $-535.66\pm0.02$ \\
$+$Dov. & $881$ & $-0.31$ & $-1351.87\pm0.03$ \\
$+$DESI & $974$ & $-0.27$ & $-544.22\pm0.03$ \\
$+$Dov.$+$DESI & $1087$ & $-0.19$ & $-1361.73\pm0.02$ \\
$+$shear & $908$ & $-0.23$ & $-537.57\pm0.03$ \\
$+$all & $906$ & $-0.28$ & $-1362.35\pm0.03$ \\
\hline
\multicolumn{4}{l}{Frozen model, $N=5966$} \\
\Pl & $1369$ & $0.05$ & $-534.94\pm0.02$ \\
$+$Dov. & $992$ & $-0.19$ & $-1351.35\pm0.03$ \\
$+$DESI & $963$ & $0.21$ & $-542.58\pm0.03$ \\
$+$Dov.$+$DESI & $1104$ & $0.36$ & $-1360.24\pm0.03$ \\
$+$shear & $634$ & $-0.01$ & $-538.72\pm0.04$ \\
$+$all & $694$ & $-0.06$ & $-1363.78\pm0.04$ \\
\end{tabular}
\end{ruledtabular}
\end{table}

Table~\ref{tab:sampling} summarizes the importance samples. Each model was sampled with a single adaptive importance-sampling run (Sec.~\ref{sec:sampling}), whose successive
rounds refine the proposal and accumulate draws: \LCDM\ in four rounds, with $\NdrawsL$ exact likelihood evaluations
in total, and the frozen model in five rounds, with $\NdrawsF$; every target has at least
$\ESSmin$ effective samples and a Pareto diagnostic $\hat k\le\khatmax$, well below the threshold $0.7$ above which
the weights become unreliable~\cite{Vehtari2015}. The raw and Pareto-smoothed evidences agree to better than
$0.002$. In the frozen model $6.2\%$ of the draws fall beyond the physical boundary; they are rejected before the
perturbations are computed and cost a negligible fraction of the time. No evaluation failed inside the prior and the boundary.

\paragraph{Validation.} With $\Sigma=0$ the fluid is unperturbed and the code reduces to \LCDM\ exactly. Our
\LCDM\ results reproduce the companion analysis of the same data, obtained with a different version of the
Boltzmann solver: the evidences of \Pl, \Pl+\DK\ and \Pl+\DK+\DESI\ agree to $0.08$, those of \DK\ and \DESI\
alone to $0.01$, and the tension between \Pl+\DK\ and \DESI\ is $2.3$--$2.5\sigma$ in both. Evaluations of the same
points on two computers with different processor architectures agree to $10^{-3}$ in $\ln\mathcal{L}$.

\paragraph{Best fits.} The surrogate refinement of Sec.~\ref{sec:sampling} improves the best exact draw by
$0.2$--$1.3$ in $\chi^2$ and converges in one to five iterations. In the frozen model the best fits of the
combinations without shear lie at $\eta_0=0.08$--$0.12$, inside the allowed region; the coordinate
$\ln(\eta_0-\eta_{\rm min})$ keeps every trial point inside it.

\section{Full posterior of the joint analysis}
\label{app:global}

\begin{figure*}[t]
\includegraphics[width=\textwidth]{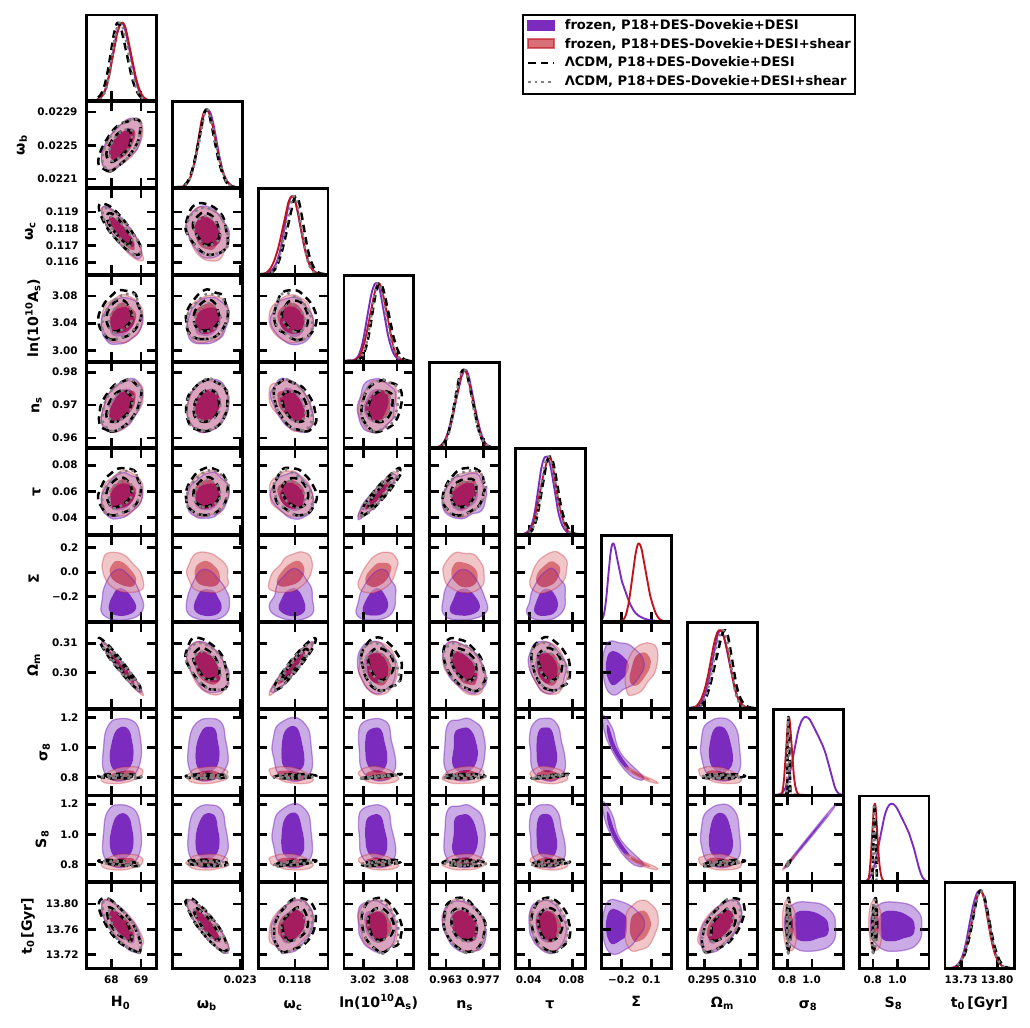}
\caption{Marginalized posteriors (68\% and 95\% regions) of the cosmological parameters for \Pl+\DK+\DESI\ and
\Pl+\DK+\DESI+shear in the frozen model (filled) and in \LCDM\ (dashed and dotted lines). The last three
parameters are derived: the matter density, the amplitude of the fluctuations and the age of the Universe.}
\label{fig:global}
\end{figure*}

For completeness, Fig.~\ref{fig:global} shows the marginalized posteriors of all the sampled cosmological
parameters and of the main derived ones for the two largest data combinations, in both models. The parameters
shared with \LCDM\ are almost unaffected by the coupling: the only appreciable correlations of $\Sigma$ are with
$\sigma_8$ and $S_8$, which it controls directly, and, more weakly, with the amplitude $\ln(10^{10}A_{\rm s})$ and
the optical depth, which adjust the height of the acoustic peaks once the lensing smoothing is provided by the late
growth. Adding cosmic shear removes the region of large $\sigma_8$ without moving any of the other parameters, and
the two models then give practically identical posteriors.

\bibliography{references}

\end{document}